\documentclass[12pt, a4paper]{article}

\usepackage[english]{babel}
\usepackage{xcolor}
\usepackage{amsmath}
\usepackage{xcolor}
\usepackage{amssymb,amsmath}
\usepackage{bm}
\usepackage{color}
\usepackage{graphicx}
\usepackage{empheq}
\usepackage{appendix}
\usepackage[hidelinks]{hyperref}
\usepackage[left=2cm,right=2cm,top=2cm,bottom=2cm]{geometry} 
\usepackage[algorithme]{algorithm}
\usepackage{algorithmic}
\usepackage{listings}
\usepackage{siunitx}
\usepackage{cite}
\usepackage{tabularx}
\usepackage{array}
\numberwithin{equation}{section}
\usepackage{physics} 
\usepackage{arydshln}
\usepackage{adjustbox}
\usepackage{booktabs}
\usepackage{lineno}
\usepackage{subfig}
\usepackage{todonotes}
\usepackage{graphicx}
\usepackage{makecell} 
\usepackage{textcomp}
\usepackage{geometry}
\usepackage{xspace}
\usepackage{orcidlink}

\newcommand{\aunit}[1]{\ensuremath{\text{\,#1}}}   
\newcommand{\tev}{\aunit{Te\kern -0.1em V}\xspace}
\newcommand{\gev}{\aunit{Ge\kern -0.1em V}\xspace}
\newcommand{\mev}{\aunit{Me\kern -0.1em V}\xspace}
\newcommand{\kev}{\aunit{ke\kern -0.1em V}\xspace}
 
\def\rad  {\aunit{rad}\xspace}

\def\cm   {\aunit{cm}\xspace}

\def\mm   {\aunit{mm}\xspace}

\def\mum  {\aunit{\textmu{m}}\xspace}

\def\deriv{\ensuremath{\mathrm{d}}}

\mathchardef\PLambda="7103
\renewcommand{\Lambda}{\PLambda}
\newcommand{\Lc}{\ensuremath{\Lambda_c^+}\xspace}
\newcommand{\Dtoppnn}{\ensuremath{D^0\to\pi^+\pi^-\nu\bar{\nu}}\xspace}
\newcommand{\Lctopmm}{\ensuremath{\Lc\to p \mu^+\mu^-}\xspace}

\begin{document}

\def\papertitle{New opportunities for rare charm from $Z\to c\bar{c}$ decays}
\def\paperauthors{ 
Angelo Di Canto$^{\,a}$, Tabea Hacheney$^{\,b}$, Gudrun Hiller$^{\,b,c}$, Dominik Stefan Mitzel$^{\,b}$, St{\'e}phane Monteil$^{\,d}$, Lars R\"{o}hrig$^{\,b,d}$, Dominik Suelmann$^{\,b}$
}

\begin{titlepage}
\pagenumbering{roman}


\vspace*{3.0cm} 

\begin{center}
{\normalfont\Large
\papertitle
}
\vspace*{1.5cm} 

\paperauthors
\vspace*{1.0cm} 

\footnotesize \it
\noindent
${}^a$Physics Department, Brookhaven National Laboratory, Upton NY, United States \\
${}^b$Fakult\"at Physik, Technische Universit\"at Dortmund, Dortmund, Germany \\
${}^c$Theoretical Physics Department, CERN 1211, Geneva 23, Switzerland\\
${}^d$Universit\'e Clermont Auvergne, CNRS/IN2P3, LPC, Clermont-Ferrand, France \\
\vspace{1.0cm}

\end{center}
\vspace*{0.5cm}

\begin{abstract}
  \noindent
We analyze the potential of rare charm decays as probes of new physics at a high-luminosity flavor facility operating at the $Z$ pole,
such as the FCC-ee or CEPC. In particular, we identify clean null-test observables in $D^0 \to \pi^+ \pi^- \nu\bar{\nu}$ and in polarized $\Lc \to p \ell^+ \ell^-$ decays with $\ell=e, \mu$.
Complementarity with the LHC and HL-LHC flavor programs arises from the characteristic features of a Tera-$Z$ environment: the capability to study missing-energy modes and charm production with significant polarization.
We improve the theoretical description of $D^0 \to \pi^+ \pi^- \nu\bar{\nu}$ decays and work out the phenomenology of polarization-induced null-test observables in $\Lc \to p \ell^+ \ell^-$ decays.
In regions of dilepton mass near the $\phi$ resonance, polarization asymmetries can reach $O(5 \%)$ for muons and $O(14 \%)$ for electrons times the
\Lc polarization.
We also point out synergies between the dineutrino and the dilepton modes using the SMEFT framework of heavy new physics.
Using the IDEA detector concept at FCC-ee, we find in simulation studies that dineutrino branching fractions as low as $\sim 2 \times 10^{-7}$ can be probed, which reaches well into the parameter space of
new physics, and also allows for discrimination of lepton flavor structures. Furthermore, the measurement of asymmetries in $\Lc \to p \mu^+ \mu^-$ at $O(1 \%)$ will be possible. Similar sensitivities are expected for dielectron final states, although robust predictions will require further dedicated studies.
\end{abstract}

\end{titlepage}

\renewcommand{\thefootnote}{\arabic{footnote}}
\setcounter{footnote}{0}

\pagestyle{plain} 
\setcounter{page}{1}
\pagenumbering{arabic}


\tableofcontents

\newpage


\section{Introduction}

For decades, the Standard Model of Particle Physics (SM) has withstood numerous precision tests at collider experiments and remains the most successful description of fundamental interactions. Nevertheless, its limitations motivate experimental input from the next generation of particle accelerators.

A future collider combining high-energy electron–positron collisions with high intensity, such as the Future Circular Collider in its electron–positron mode (FCC-ee) at CERN or the Circular Electron–Positron Collider (CEPC) in China, would provide new opportunities to address these open questions and to search for phenomena beyond the SM (BSM), often referred to as New Physics (NP)~\cite{FCC:2025lpp,FCC:2025uan,FCC:2025jtd,CEPCStudyGroup:2018rmc,CEPCStudyGroup:2018ghi}. While high-precision measurements of the Higgs boson properties are a central objective of their physics programs, these facilities also offer unique opportunities for precise studies in the quark-flavor sector when operated at the $Z$ pole, as highlighted by several sensitivity studies in $b$-hadron decays (\textit{e.g.}, see Refs.~\cite{b_to_snunu_FCC,AlvarezCartelle:2025mtx,b_to_taunu_FCC, b_to_stautau_FCC,Bordone:2025cde,Kwok:2025fza}). However, despite the enormous yield of expected $Z\to c\bar{c}$ events to be produced, about $7 \times 10^{11}$ at FCC-ee, the opportunities at such facilities in the field of charm physics remain largely unexplored~\cite{FCCee_ESPPU,Ai:2024nmn}.

Charm physics plays a unique role in searches for NP in the up-type quark sector and is a central part of the physics program at current flavor experiments such as LHCb~\cite{LHCb-DP-2008-001,LHCb:2018roe}, Belle II~\cite{Belle-II:2010dht,Belle2_Physics_Book}, BESIII~\cite{Asner:2008nq,BESIII:2009fln}, their upgrades and other proposed projects such as the Super Tau–Charm Factory (STCF)~\cite{Achasov:2023gey}. Here, we focus on rare charmed hadron decays into semileptonic final states, which are sensitive to flavor-changing neutral current (FCNC) processes and constitute particularly powerful probes of BSM physics~\cite{Gisbert:2020vjx}. 
Since charm FCNC processes are induced by quantum loops of down-type quarks, which are essentially degenerate in mass compared to the much heavier $W$ boson, they are strongly suppressed. This suppression, resulting from the Glashow-Iliopoulos-Maiani (GIM) mechanism, is a distinctive feature of the SM's flavor sector and breaks down in general in BSM models. The GIM-protected null-test observables are therefore ideal targets for dedicated BSM searches in charm, despite the otherwise sizable theoretical uncertainties from non-perturbative quantum chromodynamics (QCD) \cite{DeBoer:2018pdx}. Importantly, probing flavor in the up-type quark sector with charm FCNCs provides complementary sensitivity to kaon and $b$ hadron decays, accessing otherwise untested couplings~\cite{Gisbert:2024kob}. 

Current facilities benefit either from high charm-production rates in proton–proton collisions (LHCb) or from clean experimental environments at $e^+e^-$ colliders (Belle II, BESIII). Future projects such as FCC-ee and CEPC could combine these advantages. They would allow high-precision searches for rare decays with missing particles in the final state, such as semileptonic dineutrino decays sensitive to $c \to u \nu \bar{\nu}$ transitions. Semileptonic dineutrino modes are particularly well motivated, as the absence of intermediate hadronic resonances in the dilepton system suppresses SM backgrounds and enhances sensitivity to BSM effects~\cite{Bause:2020xzj}. At present, the only experimental search for semileptonic dineutrino decays has been performed by BESIII, which set an upper limit on the branching fraction of $D^0 \to \pi^0 \nu\bar{\nu}$ decays of $2.1 \times 10^{-4}$ at 90\% confidence level (CL)~\cite{BESIII:2021slf}. Belle also set an upper limit of $9.4 \times 10^{-5}$ at 90 $\%$ CL on the branching fraction of $D^0$ decays to invisible final states~\cite{Belle:2016qek}, which is a complementary decay process with sensitivity to additional light and invisible degrees of freedom and to different NP models.

Moreover, the sizable polarization of charmed baryons produced in $e^+e^-$ collisions at the $Z$ pole would allow access to additional angular observables in $c \to u \ell^+ \ell^-$ ($\ell=e,\mu$) transitions, providing powerful null tests of the SM with negligible theoretical uncertainties. The LHCb collaboration has already performed the first measurement of angular observables in $D^0 \to P^+P^- \mu^+\mu^-$ ($P=\pi,K$) and $\Lambda_c^+ \to p \mu^+ \mu^-$ decays~\cite{LHCb:2018qsd,LHCb:2021yxk,LHCb:2025bfy}. However, the $\Lambda_c^+$ baryons studied in Ref.~\cite{LHCb:2025bfy} originate directly from the primary $pp$ interaction and are therefore largely unpolarized, restricting the range of accessible observables. In $pp$ collisions, samples of polarized $\Lambda_c^+$ can be obtained by selecting those produced in $b$-hadron decays, but this strategy comes at the cost of severely reduced signal yields. In contrast, at an $e^+e^-$ collider the polarization of $\Lambda_c^+$ baryons can be inferred from the underlying charm-quark polarization. At the $Z$ pole, this is predicted to be $P_{c}^{Z\text{-pole}}\simeq -0.65$, which implies a sizable charmed baryon polarization of similar size $P_{\Lambda_c}^{Z\text{-pole}}\simeq -0.44\pm0.02$ 
\cite{Falk:1993rf,deBoer:2017que}.

In this article, we explore the unique NP discovery potential of future $e^+e^-$ colliders in two benchmark channels. The first is $D^0 \to \pi^+ \pi^- \nu \bar{\nu}$ as representative of the $c \to u \nu \bar{\nu}$ transition with the experimental advantage of a displaced vertex reconstructed from two charged pions. The second channel is $\Lambda_c^+ \to p \mu^+ \mu^-$, which is one of the most promising candidates for null-test measurements in rare decays of polarized charm baryons.
We go beyond existing  works by the following: using a data-driven approach for the $D^0 \to \pi^+ \pi^- \nu \bar{\nu}$ transversity form factors and comparing them to predictions based on heavy hadron chiral perturbation theory, and detailing the phenomenology of the
angular observables in $\Lambda_c^+ \to p \ell^+ \ell^-$ that are induced by finite $\Lambda_c^+$ polarization. In addition, we work out correlations between the two modes in concrete BSM models depending on lepton flavor pattern.
Finally, we estimate the sensitivity to these observables at a Tera-Z facility, using the FCC-ee IDEA detector with $6 \times 10^{12}$ produced $Z$ bosons as a case study.

The paper is organized as follows. In Sec.~\ref{sec:theory} we discuss the theory of rare charm decays. We introduce the effective field theory (EFT) formalism to describe $c \to u \nu \bar{\nu}$ and $c \to u \ell^+ \ell^-$ transitions and make predictions for null test observables.
In Sec.~\ref{sec:exp}, we  complement the phenomenology work with sensitivity studies for both channels at a future $e^+e^-$ collider operating at the $Z$ pole using simulated samples. The studies strengthen the physics case for a future accelerator and show the unique capability of $e^+e^-$ colliders at the $Z$ pole. We conclude in Sec.~\ref{sec:con}. Auxiliary information on $\Lc\to p \ell^+ \ell^-$  form factors and helicity amplitudes is given in App.~\ref{app:formfactors} and App.~\ref{app:helicity_amps}, respectively.

\section{Theory of rare charm decays \label{sec:theory}}

We present the theory framework to compute rare charm decay observables in and beyond the SM.
We briefly review the underlying EFT set up in Sec.~\ref{sec:EFT}, including the standard model effective theory (SMEFT), allowing to harvest
correlations manifest in the unbroken electroweak theory, and the low energy weak effective theory (WET).
We discuss $D^0 \to \pi^+ \pi^- \nu \bar \nu$ decays and $\Lambda^+_c \to p \ell^+ \ell^-$ decays with polarized $\Lambda^+_c$ in Sec.~\ref{sec:pipinunu_theo} and Sec.~\ref{sec:l2pmumu_theo}, respectively. 
The BSM reach of the observables is worked out in Sec.~\ref{sec:nulltest_pred}. We present correlations in NP scenarios between the two decay modes in Sec.~\ref{sec:NP}.

\subsection{Effective Field Theory Set Up}
\label{sec:EFT}
We utilize the SMEFT to harness the symmetries of 
the SM as a guideline for NP and to link different observables. For the low energy observables in rare charm decays we employ the WET and give the matching between the EFTs.

The SMEFT is an effective theory  framework, which uses the SM degrees of freedom 
to describe particle interactions and 
includes heavy NP model-independently 
consistent with Lorentz and $SU(3)_C\times SU(2)_L \times U(1)_Y$ gauge invariance.
The Lagrangian of the SMEFT reads
\begin{equation}
\label{eq:SMEFT}
    \mathcal{L}_{\mathrm{SMEFT}} = \mathcal{L}_{\mathrm{SM}} + \sum_{d}^{\infty} \sum_i \frac{\mathcal{C}^{(d)}_i}{\Lambda^{d-4}} \mathcal{Q}_i^{(d)} \,,
\end{equation}
where the Wilson coefficients $\mathcal{C}_i^{(d)}$ parametrize the NP contributions to the $d$-dimensional operators $\mathcal{Q}_i^{(d)}$. The  NP scale $\Lambda$, which is assumed to be sufficiently above the electroweak scale  $v=(\sqrt{2}\,G_F)^{-1/2} \simeq 246\,\mathrm{GeV}$, with Fermi's constant $G_F$.
To study $c\to u\nu\overline{\nu}$ and $c\to u\ell^+\ell^-$ transitions  at leading order in the SMEFT we consider the 
semileptonic four-fermion operators with dimension six  in the Warsaw basis \cite{Grzadkowski:2010es} 
\begin{equation}
    \begin{aligned}
        \mathcal{Q}_{\ell q}^{(1)} &= \overline{Q} \gamma_\mu Q\, \overline{L} \gamma^\mu L \:,&\quad \mathcal{Q}_{\ell q}^{(3)} &= \overline{Q} \gamma_\mu \tau^a Q \,\overline{L} \gamma^\mu \tau^a L \:,\\
        \mathcal{Q}_{\ell u} &= \overline{U} \gamma_\mu U \,\overline{L} \gamma^\mu L \:,&\quad \mathcal{Q}_{\ell d} &= \overline{D} \gamma_\mu D\, \overline{L} \gamma^\mu L \:,\\
        \mathcal{Q}_{e u} &= \overline{U} \gamma_\mu U \,\overline{E} \gamma^\mu E \:,&\quad \mathcal{Q}_{e d} &= \overline{D} \gamma_\mu D\, \overline{E} \gamma^\mu E\:,\\
        \mathcal{Q}_{q e} &= \overline{Q} \gamma_\mu Q \,\overline{E} \gamma^\mu E \:,&\quad & \\
    \end{aligned}
\end{equation}
where $Q$ is the left-handed quark $SU(2)_L$-doublet, $U$, $D$ denote the right-handed up-type and down-type quark singlets respectively, $L$ is the lepton-doublet and $E$  the lepton singlet. The Pauli-matrices are 
denoted as $\tau^a$.
To avoid clutter we suppress quark and lepton flavor indices.
The operators $\mathcal{Q}_{\ell d}$ and $\mathcal{Q}_{e d}$ do not contribute to rare charm decays.
We give them here because 
they contribute to the strongly constrained  rare kaon decays where they interfere with operators that appear in both $|\Delta c|=|\Delta u|=1$ and $|\Delta s|=|\Delta d|=1$ transitions, and hence are  relevant for the NP reach in charm.

The $c\to u\nu\overline{\nu}$ and $c\to u\ell^+\ell^-$ transitions 
are  linked by  $SU(2)_L$  since
the  left-handed charged leptons and the neutrinos live in the  same $SU(2)_L$-doublet $L$. As observed
in Refs.\cite{Bause:2020xzj,Bause:2020auq} this connection allows to obtain limits on the 
branching fractions of the dineutrino modes using experimental constraints on semileptonic four-fermion operators with charged lepton modes of various flavors. 
Tree level contributions to dineutrino and dilepton modes are also induced by $Z$ penguins from dimension six 
operators with two Higgs fields and a covariant derivative.
These are constrained by electroweak and top observables, or mixing~\cite{Efrati:2015eaa,Brivio:2019ius}, and subleading as the
dipole contributions.

Low-energy processes are expressed in terms of the WET which is obtained by integrating out the Higgs, $Z$, $W$-bosons and the top quark.
The relevant WET Lagrangian for $q_n\to q_m \nu_i \overline{\nu}_j$ and $q_n\to q_m \ell^+_i\ell^-_j$ FCNC transitions is given as 
\begin{equation}
    \label{eq:WET}
    \begin{aligned}        
    \mathcal{L}_{\mathrm{WET}} \supset  \frac{4 G_F}{\sqrt{2}} \frac{\alpha_e}{4\pi} \sum_{X=\{U,\,D\}} \:\sum_{n,m,i,j}\bigg(&\mathcal{C}_L^{X_{nm}\,ij}\mathcal{Q}_L^{nmij} + \mathcal{C}_R^{X_{nm}\,ij}\mathcal{Q}_R^{nmij} \\
    &+ \sum_{k=7,9,10} \left(\mathcal{K}_k^{X_{nm}\,ij} \mathcal{O}_k^{nmij} + \mathcal{K}_k^{\prime\,X_{nm}\,ij} \mathcal{O}_k^{\prime\,nmij}\right) \bigg) +\mathrm{H.c.}\:,
    \end{aligned}
\end{equation}
where $\alpha_e$ denotes the fine structure constant.  The indices $i$, $j$ indicate lepton flavor and $n$, $m$ are quark flavor indices. $X=U,D$ indicates up-type or down-type transitions. We denote Wilson coefficients for charged leptons by $\mathcal{K}$ and those of the dineutrinos with $\mathcal{C}$.   Unless mentioned otherwise, omitted flavor indices correspond to $c\to u$ transitions with  $X=U_{12}$.
The dimension six  operators are given as  
\begin{equation}
    \begin{aligned}        
        \mathcal{Q}_{L(R)}^{nmij} &= \left( \overline{q}_{m\,L(R)} \gamma_\mu q_{n\,L(R)} \right) \left( \overline{\nu}_{j\,L} \gamma^\mu \nu_{i\,L}\right)\,, & \quad
        \mathcal{O}_7^{nm} &= \frac{m_c}{e} \left(\overline{q}_{m\,L} \sigma_{\mu\nu} q_{n\,R} \right) F^{\mu\nu}\,, \\
        \mathcal{O}_9^{nmij} &= \left(\overline{q}_{m\,L} \gamma_\mu q_{n\,L} \right) \left(\overline{\ell}_j \gamma^\mu \ell_i \right)\,, &\quad
        \mathcal{O}_{10}^{nmij} &= \left(\overline{q}_{m\,L} \gamma_\mu q_{n\,L} \right) \left(\overline{\ell}_j \gamma^\mu \gamma_5 \ell_i \right) \,.
    \end{aligned}
\end{equation}
Here $F^{\mu\nu}$ is the electromagnetic field strength tensor, $L(R) = \frac{1}{2}\left(1\mp \gamma_5\right)$ 
denote the chiral projectors and $m_c$ the mass of the charm quark.
The electromagnetic coupling $e$ is related to the fine structure constant via $\alpha_e = e^2/(4\pi)$.
The primed operators $\mathcal{O}^\prime$ for the charged lepton modes are obtained from the $\mathcal{O}$  by 
swapping  chiral projectors $L(R)\to R(L)$. 

In the following we give the 
tree-level matching conditions for the semileptonic four-fermion operators from SMEFT \eqref{eq:SMEFT} onto WET.
To correlate charged lepton and dineutrino transitions~\cite{Bause:2020xzj} one needs to consider the  operators 
with lepton doublets, 
\begin{alignat}{4} \nonumber
    C_L^{D}= K_L^{U} &=K_9^{U} -K_{10}^{U} &=&k\cdot \left(\mathcal{C}_{\ell q}^{(1)}-\mathcal{C}_{\ell q}^{(3)}\right) \, , \\
    C_L^{U}=K_L^{D} &= K_9^{D} -K_{10}^{D} &=& k\cdot \left(\mathcal{C}_{\ell q}^{(1)}+\mathcal{C}_{\ell q}^{(3)}\right) \, , \label{eq:LHleptons}\\
    C_R^U = K_R^{U} &=K_9^{\prime U} -K_{10}^{\prime U} &=& k\cdot \mathcal{C}_{\ell u} \, ,  \nonumber \\
    C_R^D = K_R^{D} &= K_9^{\prime D} -K_{10}^{\prime D} &=& k\cdot \mathcal{C}_{\ell d} \, ,  \nonumber
    \end{alignat} 
    where
 \begin{alignat}{4}    
    k&= \frac{\sqrt{2}\pi}{\alpha_e G_F \Lambda^2} &\approx& \, 0.5 \left( \frac{10 \, \text{TeV}}{\Lambda} \right)^2\, .  
\end{alignat} 
Here we introduced chiral Wilson coefficients  $K_{L,R}^{U,D}$, and
ignored RG-induced contributions or those from other operators that do not induce dineutrino modes or are subleading.
Note in the upper two equations in (\ref{eq:LHleptons})  for the left-handed quark currents the interplay between the up and down sector:
left-handed $|\Delta c|=|\Delta u |=1$ dineutrino modes are subject of constraints from $|\Delta s|=|\Delta d|=1$
transitions with charged leptons. Likewise,  constraints on $K \to \pi \nu \bar \nu$ decays impose
constraints on 
left-handed $|\Delta c|=|\Delta u|=1$ transitions with charged leptons.
 Charged current processes involving a charged lepton of specific flavor and a neutrino are probing the
triplet operator $\mathcal{Q}_{\ell q}^{(3)}$. 
In rather global analyses, \textit{e.g.} Refs.~\cite{Grunwald:2023nli,Fajfer:2023nmz}, an endeavor beyond the scope of this work, these provide complementary constraints in addition to $\mathcal{C}_{\ell q}^{(1)} \pm \mathcal{C}_{\ell q}^{(3)}$.
Note, unlike in the dineutrino modes, factors from the Pontecorvo-Maki-Nakagawa-Sakata (PMNS) matrix do not drop out.

We also give the matching for operators with right-handed leptons as they contribute to 
FCNCs with charged leptons
\begin{align} \label{eq:RHleptons}
    K_9^U + K_{10}^U &= k\cdot \mathcal{C}_{qe}  \, , ~~
    K_9^{\prime U} + K_{10}^{\prime U} = k\cdot \mathcal{C}_{eu} \, , ~~
    K_9^{\prime D} + K_{10}^{\prime D} = k\cdot  \mathcal{C}_{ed}  \, . 
    \end{align}
The dipole operators $\mathcal{O}_7^{(\prime)}$ are induced at tree-level by electroweak dipole operators
after electroweak symmetry breaking.

The operators in (\ref{eq:SMEFT}) are given in the flavor basis. The coefficients $C,K$ in (\ref{eq:LHleptons})  and (\ref{eq:RHleptons}) therefore correspond to this basis.
Rotation to the mass basis gives the (calligraphic) coefficients $\mathcal{C}, \mathcal{K}$ 
\begin{align} \nonumber
     C_{L(R)}^U &= \mathcal{C}_{L(R)}^U \\
    C_{L(R)}^D &= V\mathcal{C}_{L(R)}^D V^\dagger \\
    K_{L(R)}^D &= W^\dagger V\mathcal{K}_{L(R)}^D V^\dagger W  \nonumber\\
    K_{L(R)}^U &= W^\dagger \mathcal{K}_{L(R)}^U  W  \nonumber
\end{align} 
using $Q_\alpha = \left(u_{L\,\alpha},\:V_{\alpha\beta} d_{L\,\beta}\right)$ and $L_i = \left(\nu_{L\,i},\:W^\ast_{ki}\ell_{L\,k}\right)$ 
with the Cabibbo-Kobayashi-Maskawa (CKM) and PMNS matrices $V$ and $W$, respectively.
We obtain the relations for WET operators with vector and axial-vector couplings to leptons (\ref{eq:WET})
\begin{equation}
    \begin{aligned}
        \mathcal{C}_L^U &= W^\dagger V \mathcal{K}_L^D  V^\dagger W \,,\\
           \mathcal{C}_R^U &= W^\dagger \mathcal{K}_R^{U}   W\,, \\
                  \end{aligned} \label{eq:WET_correlations}
\end{equation}
where
\begin{equation}
    \begin{aligned}
        \mathcal{K}_{L}^{X} &= \mathcal{K}_9^X - \mathcal{K}_{10}^X\:,\quad \mathcal{K}_R^{X} =\mathcal{K}_9^{\prime\,X} - \mathcal{K}_{10}^{\prime\,X} \:, \quad X=U,D \, .
    \end{aligned} \label{eq:KL910}
\end{equation}
Notice that the unitary matrix $W$ drops out in observables after summing over lepton flavors in the dineutrino modes.

The relations \eqref{eq:WET_correlations} link 
different charged lepton down-type quark transitions with dineutrino up-type transitions for $\mathcal{C}_L^U$, and 
charged lepton and dineutrino modes for the same quark transition with $\mathcal{C}_R^{U}$. The latter implies
a one-to-one BSM correlation between the branching fraction of $D^0\to\pi^+\pi^-\nu\overline{\nu}$ and null-test observables of 
$\Lambda^+_c\to p\ell^+\ell^-$ decays, as worked out in Sec.~\ref{sec:NP}. In the general case with both $\mathcal{C}_{L,R}^{U}$ present flavor pattern specific upper limits on dineutrino modes can be obtained from charged dilepton modes~\cite{Bause:2020xzj}.
Upper limits on dineutrino modes depend on the lepton flavor structure, which can therefore be tested in correlations.
We consider 
lepton universal (LU), charged lepton flavor conserving (cLFC), democratic, general and lepton-specific flavor patterns, defined as
\begin{equation}
    \mathcal{K}_{L,R}^{X\,ij}\bigg|_{\text{LU}} = \begin{pmatrix}
        k & 0 & 0 \\
        0 & k & 0 \\
        0 & 0 & k \\
    \end{pmatrix}_{ij}\:,\quad 
    \mathcal{K}_{L,R}^{X\,ij}\bigg|_{\text{cLFC}} = \begin{pmatrix}
        k_e & 0 & 0 \\
        0 & k_\mu & 0 \\
        0 & 0 & k_\tau \\
    \end{pmatrix}_{ij}\:,\quad 
    \mathcal{K}_{L,R}^{X\,ij}\bigg|_{\text{democratic}} = \begin{pmatrix}
        \tilde{k} & \tilde{k} & \tilde{k} \\
        \tilde{k} & \tilde{k} & \tilde{k} \\
        \tilde{k} & \tilde{k} & \tilde{k} \\
    \end{pmatrix}_{ij} \:,
\end{equation}
with "general" meaning arbitrary coefficients and "lepton-specific" for the case with only one coefficient of 
a specific lepton transition  switched on. 
The parameters $k,k_\ell,\tilde{k}$ are constrained by experimental data.

\subsection{\texorpdfstring{$D^0\to\pi^+\pi^-\nu\overline{\nu}$}{D0 -> pi pi nu nubar} decays}\label{sec:pipinunu_theo}
The triple differential branching fraction of $D^0\to\pi^+\pi^-\nu\bar{\nu}$ can be written as 
\begin{equation}
    \label{eq:diffbranching_dineutrino}
    \frac{\mathrm{d^3}\mathcal{B}(D^0\to\pi^+\pi^-\nu\bar{\nu})}{\mathrm{d}q^2\,\mathrm{d}p^2\,\mathrm{d}\!\cos\theta_\pi} =  b_{+}(q^2,p^2,\theta_\pi) \,x_U^+ +  b_{-}(q^2,p^2,\theta_\pi) \, x_U^- \:,
\end{equation}
where $q^2$ is the invariant mass squared of the dineutrino pair and $p^2$ is the invariant mass squared of the dipion pair. The functions $b_\pm$ are defined as
\begin{align}
    b_-(q^2,p^2,\theta_{\pi}) &= \frac{\tau_{D^0}}{6} \left(\left|\mathcal{F}_0\right|^2 + \sin^2\theta_{\pi}\left|\mathcal{F}_\parallel\right|^2\right) \:, \\
    b_+(q^2,p^2,\theta_{\pi}) &= \frac{\tau_{D^0}}{6} \sin^2\theta_{\pi}\left|\mathcal{F}_\perp\right|^2 \:,
\end{align}
where $\tau_{D^0}$ is the lifetime of the $D^0$ meson and $\theta_{\pi}$ is the angle between the $\pi^+$-momentum 
and negative direction of flight of the $D^0$ meson in the dipion rest frame. 
The dependence of the transversity form factors $\mathcal{F}_i$, $i=0,\perp,\parallel$ 
on the kinematic variables $q^2$, $p^2$ and $\cos\theta_\pi$ is omitted for brevity.
As the neutrinos are not tagged, the distribution Eq.~\eqref{eq:diffbranching_dineutrino} includes an implicit 
sum over the neutrino flavors, encoded in the short-distance coefficients $x_U^\pm$,
\begin{equation} \label{eq:XU}
    x_U^\pm = \sum_{i,j}\left|\mathcal{C}_L^{Uij}\pm \mathcal{C}_R^{Uij}\right|^2 \:.
\end{equation}
The differential distributions with respect to $q^2$ or $p^2$ can be obtained via 
\begin{align}
    \frac{\mathrm{d}\mathcal{B}(D^0\to\pi^+\pi^-\nu\bar{\nu})}{\mathrm{d}q^2} &= \int_{4m_\pi^2}^{(m_D-\sqrt{q^2})^2}\mathrm{d}p^2\int_{-1}^{1}\mathrm{d}\cos\theta_{\pi}\,\frac{\mathrm{d^3}\mathcal{B}(D^0\to\pi^+\pi^-\nu\bar{\nu})}{\mathrm{d}q^2\,\mathrm{d}p^2\,\mathrm{d}\!\cos\theta_\pi}\:, \\ 
    \frac{\mathrm{d}\mathcal{B}(D^0\to\pi^+\pi^-\nu\bar{\nu})}{\mathrm{d}p^2} &= \int_{0}^{(m_D-\sqrt{p^2})^2}\mathrm{d}q^2\int_{-1}^{1}\mathrm{d}\cos\theta_{\pi}\,\frac{\mathrm{d^3}\mathcal{B}(D^0\to\pi^+\pi^-\nu\bar{\nu})}{\mathrm{d}q^2\,\mathrm{d}p^2\,\mathrm{d}\!\cos\theta_\pi}  \:,
\end{align}
where $m_D$ ($m_\pi$)~\cite{PDG2024} denotes the mass of the $D^0$ meson (pion). Additionally, a differential distribution with respect to the missing momentum $\vec{q}_{\text{miss}}$ in the rest frame of the $D^0$ meson can be obtained from
\begin{equation}
    \frac{\mathrm{d^2}\mathcal{B}(D^0\to\pi^+\pi^-\nu\bar{\nu})}{\mathrm{d}\left|\vec{q}_{\text{miss}}\right|\mathrm{d}p^2} = 2m_D \frac{\sqrt{\lambda(m_D^2,p^2,q^2)}}{m_D^2+q^2-p^2}\frac{\mathrm{d^2}\mathcal{B}(D^0\to\pi^+\pi^-\nu\bar{\nu})}{\mathrm{d}q^2\mathrm{d}p^2} \:,
\end{equation}
where $\lambda(a,b,c)=a^2+b^2+c^2-2(ab+ac+bc)$ is the 
Källén function.

The transversity form factors
 ${\cal{F}}_i$, $i=0, \perp, \parallel$ can be written as
\begin{align} \nonumber 
{\cal{F}}_0 &= \frac{{\cal N}_{\rm nr}}{2}  \bigg[  \lambda^{1/2 }w_+(q^2,p^2,\cos \theta_{\pi})-\frac{1}{p^2}(m_D^2-q^2-p^2) \lambda^{1/2}_{p}\cos \theta_{\pi} w_-(q^2,p^2,\cos \theta_{\pi}) \bigg]\,,\\
\label{eq:Fi}
{\cal{F}}_\parallel &= {\cal N}_{\rm nr}  \sqrt{ \lambda_p \frac{q^2}{p^2}} \, w_-(q^2,p^2,\cos \theta_{\pi})\,, \qquad
{\cal{F}}_\perp = \frac{{\cal N}_{\rm nr}}{2}\sqrt{ \lambda \lambda_p \frac{  q^2}{p^2}} \, h(q^2,p^2,\cos \theta_{\pi})\,, \\
 & \quad \quad  \quad \quad   {\cal{N}}_\text{nr}=\frac{G_F\alpha_e}{2^7\pi^4m_D}\sqrt{\pi\frac{\sqrt{  \lambda \lambda_p }}{m_Dp^2}} \, .   \nonumber
\end{align} 
with $\lambda=\lambda(m_D^2,q^2,p^2)$ and $\lambda_p=\lambda(p^2,m_{\pi}^2,m_{\pi}^2)$. 
The  $D^0 \to \pi^+\pi^-$ transition form factors  $w_\pm^{(\prime)}$, $r^{(\prime)}$, $h^{(\prime)}$ are defined as
\begin{align}
\langle \pi^+(p_1)\pi^-(p_2)|\bar u\gamma_\mu(1-\gamma_5)c|
D^0(p_D)\rangle& 
		=i \left[w_+p_\mu+w_-P_\mu+rq_\mu+ih\epsilon_{\mu\alpha\beta\gamma}p_D^\alpha p^\beta P^\gamma\right]\,, \label{eq::FFLLW} \\
       \langle \pi^+(p_1)\pi^-(p_2)|
\bar u i q^\nu \sigma_{\mu \nu}(1+\gamma_5)   c|D^0(p_D)\rangle &=
- i m_D \! \left[ w_+' p_{ \mu} + w_-' P_{\mu} + r' q_{\mu} \!
       + i h' \varepsilon_{\mu\alpha\beta\gamma}p_D^\alpha p^\beta
       P^\gamma \right]  \, , \label{eq::FFLLW2}
\end{align}
where the dependence on $q^2,p^2$ and $\cos \theta_{\pi}$ in the form factors is suppressed. 
Here, $q^\mu=p_+^\mu+p_-^\mu$, $p^\mu=p_1^\mu+p_2^\mu=p_D^\mu-q^\mu$ and $P^\mu=p_1^\mu-p_2^\mu$.
Contributions to the decay distribution from  $r$ and $r^\prime$ vanish  for massless leptons.

The form factors  are calculated in the following sections using different methods. 
We utilize heavy hadron chiral perturbation theory (HH$\chi$PT)~\cite{Lee:1992ih,Bause:2020xzj}, 
HH$\chi$PT with vector mesons added~\cite{Fajfer:1998dv,Adolph:2020ema},
and a data driven approach by relating the decay $D^+\to\pi^+\pi^-e^+\nu_e$ to 
the dineutrino mode and using the measurements by BESIII~\cite{BESIII:2018qmf}.

\subsubsection{HH\texorpdfstring{$\chi$}{chi}PT form factors} 
The non-resonant form factors $w_\pm,h,r$ entering (\ref{eq::FFLLW}) are calculated using HH$\chi$PT in Ref.~\cite{Lee:1992ih}.
HH$\chi$PT is applicable if the pion momenta are soft, a condition that is reasonably well satified in most region of phase space in $D$-meson decays.
Expanding formally to lowest order,
\begin{align} \label{eq:nrFF}
 &w_\pm=\pm\frac{\hat gf_D}{2f_{\pi}^2}\frac{m_D}{v\cdot p_{1}+\Delta}\,,\quad h=\frac{\hat g^2f_D}{2f_{\pi}^2}\frac1{(v\cdot p_{1}+\Delta)(v\cdot p+\Delta)}\,,
\end{align}
with the mass splitting  $\Delta=(m_{{D^*}^0}-m_{D^0})=0.1421\,\text{GeV}$, and the decay constants  $f_D=0.21\,\text{GeV}$ and $f_\pi=0.13\,\text{GeV}$ of the $D$ and $\pi$ meson respectively. $ \hat g=0.570\pm0.006$ \cite{BaBar:2013thi} and the scalar products are given as $v\cdot p_{1}=((m_D^2-q^2+p^2)-\sqrt{\lambda(m_D^2,q^2,p^2)(1-4m_{\pi}^2/p^2)}\cos\theta_{\pi})/(4m_D)$ and $v\cdot p=(m_D^2-q^2+p^2)/(2m_D)$. 

\subsubsection{HH\texorpdfstring{$\chi$}{chi}PT with \texorpdfstring{$\rho^0$}{rho } resonance} 

The form factors for $D^0\to\pi^+\pi^- \gamma$ decays
have been worked out within HH$\chi$PT with vector resonances~\cite{Adolph:2020ema}.
The form factors of the (axial) vector currents can be 
obtained from the tensor ones using form factor relations, here at lowest order,
$w_\pm^\prime =w_\pm$, $h^\prime = h$ ~\cite{Das:2014sra}.
The transition form factors then read  
  \begin{align}
        w_+(q^2,p^2,\theta_\pi) & \simeq \frac{\hat{g}f_D}{2m_Df_\pi^2 (v\cdot p_1 + \Delta)}\big(m_D^2 -p\cdot p_D - 2 p_1\cdot p_D\big) \nonumber\\&+\frac{\hat{g}^2f_D}{2m_Df_\pi^2 (v\cdot p_1 + \Delta)}\frac{\left((v\cdot p_2 - v\cdot p_1) (p\cdot q) \right)}{(v\cdot p+\Delta)}\nonumber \\
      &- {\frac{\alpha_1(p_2\cdot q - p_1\cdot q)}{f_\pi^2 m_D^{3/2}}}\left[{1} + {\frac{f_\pi^2 m_\rho^4}{g_\rho^2} BW_\rho(p^2)}\right] \\
      &+ {\frac{\lambda f_D g_v}{\sqrt{2}m_D (v\cdot p+\Delta)}\frac{m_\rho^2}{g_\rho}\bigg((v\cdot q)(p_1\cdot q-p_2\cdot q) + q^2 (v\cdot p_1 - v\cdot p_2)\bigg)BW_\rho(p^2)} \:, \nonumber\\
        w_-(q^2,p^2,\theta_\pi) & \simeq  -\frac{\hat{g}f_D}{2m_Df_\pi^2 (v\cdot p_1 + \Delta)}\big(m_D^2 -p\cdot p_D \big)\nonumber\\
        &+\frac{\hat{g}^2f_D}{2f_\pi^2 (v\cdot p_1 + \Delta)}\frac{\left((v\cdot q)(v\cdot p) - p\cdot q\right)}{(v\cdot p+\Delta)} \nonumber\\
      &+ {\frac{\alpha_1(v\cdot q )}{f_\pi^2 m_D^{1/2}}}\left[{1} + {\frac{f_\pi^2 m_\rho^4}{g_\rho^2} BW_\rho(p^2)}\right] \\
      &+ \frac{\lambda f_D g_v}{\sqrt{2} (v\cdot p+\Delta)}\frac{m_\rho^2}{g_\rho}\bigg((v\cdot q)(v\cdot p)-p\cdot q \bigg)BW_\rho(p^2) \:, \nonumber\\
      h(q^2,p^2,\theta_\pi) & \simeq \frac{\hat{g}f_D}{2f_\pi^2 m_D (v\cdot p_1 + \Delta)} \left[1 + \frac{\hat{g}\,(m_D - v\cdot p)}{(v\cdot p+\Delta)}\right] + {\frac{\alpha_1}{f_\pi^2 m_D^{\frac{3}{2}}}}\left[{1} + {\frac{f_\pi^2 m_\rho^4}{g_\rho^2} BW_\rho(p^2)}\right] \nonumber\\
      &+ {\frac{\lambda f_D g_v\left(v\cdot q\right)}{\sqrt{2}m_D(v\cdot p+\Delta)}\frac{m_\rho^2}{g_\rho}BW_\rho(p^2)} \:,
  \end{align}
  with $m_\rho$ the mass of the $\rho$ meson, $v = p_D/m_D$, $g_v = 5.9$, $g_\rho\approx m_\rho f_\rho$, $\alpha_1 = 0.188\,\sqrt{\mathrm{GeV}}$, $\lambda = -0.49\,\mathrm{GeV}^{-1}$ and the Breit-Wigner function 
  $BW_{\rho}(p^2) = 1/(p^2-m_\rho^2+im_\rho \Gamma_\rho)$ with the total decay width $\Gamma_\rho$. 
$\alpha_1$ describes weak currents between heavy meson and a light meson, and the coupling $\lambda$ parametrizes the odd-parity light quark-photon interaction.

  A few comments are in order. 
  Since the relations between axial(vector) and tensor form factors are only at lowest order in the chiral and heavy quark expansion, the results for $w_\pm,h$ are an incomplete beyond-lowest order result.
  We find, \textit{cf.} Sec.~\ref{sec:comp}, that these partial higher order terms matter  for the decay distributions, and result in suppression of the form factors relative to the ones at lowest order, Eq.~\eqref{eq:nrFF}.
  Yet, more realistic line-shapes of the $\rho$ could be explored in the future, as well as further resonances, such as in the data driven approach, see Sec.~\ref{sec:dda}.
The form factors obtained here serve as a model to estimate sensitivities. They are at best a sensible approximation. At present,  no first principle determination of the $D \to \pi \pi$ form factors is available, but clearly desirable. 
We note progress on the lattice on two-pion states\cite{Briceno:2015tza},
 in particular the study of $B \to \pi \pi \ell \bar \nu$ decays \cite{Leskovec:2025gsw}, which demonstrates the feasibility of such computations towards the physical regime. Further study, also specifically for charm, is highly encouraged.

\subsubsection{Data driven approach \label{sec:dda}} 

For the data driven approach we utilize 
the measurements of the decay $D^+\to\pi^+\pi^-e^+\nu_e$ by BESIII~\cite{BESIII:2018qmf},
where different resonance components are fitted to the angular distribution with five kinematic variables.

\begin{table}
    \renewcommand{\arraystretch}{1.2}
    \centering
    \caption{Measured branching fractions contributing to $D^+\to\pi^+\pi^-e^+\nu_e$~\cite{BESIII:2018qmf}. 
      $^\dagger$ Extracted from~\cite{BESIII:2018qmf} using $\mathcal{B}(\omega\to\pi^+\pi^-)=(1.53\pm0.12)\%$~\cite{PDG2024}. }
    \label{tab:BRs_BESIII}
    \begin{tabular}{llll}
         \hline
         Signal mode & $\mathcal{B}\:\:\left[10^{-3}\right]$  \\
         \hline \hline
         $D^+\to \pi^+\pi^- e^+\nu_e$ (Total) & $2.449 \pm 0.074 \pm 0.073$ \\ \hline 
         $D^+\to \rho^0 e^+\nu_e$, $\rho^0\to\pi^+\pi^-$ & $1.860\pm 0.070 \pm 0.061$ \\
         $D^+\to \omega e^+ \nu_e$, $\omega\to\pi^+\pi^-$ & $0.031 \pm 0.011^\dagger$ \\
         $D^+\to f_0(500) e^+\nu_e,\,f_0(500) \to \pi^+\pi^-$ & $0.630\pm 0.043 \pm 0.032$ \\
         \hline
    \end{tabular}
\end{table}

The $D^+\to \pi^+\pi^- e^+\nu_e$ five-fold differential 
distribution can be written as~\cite{BaBar:2010vmf}  
\begin{equation}
    \mathrm{d}^5 \Gamma = \frac{2 G_F^2 \left|V_{cd}\right|^2p_{\pi\pi}p^\ast}{(4\pi)^6 m_D^2\,\sqrt{p^2}}     \mathcal{I}(p^2,q^2,\theta_{\pi},\theta_e,\chi) \,\mathrm{d}p^2\mathrm{d}q^2\mathrm{d}\cos\theta_{\pi}\mathrm{d}\cos\theta_e\mathrm{d}\chi
\end{equation}
with $q^2$ here the invariant mass squared of the $e^+\nu_e$ system, $p_{\pi\pi} = \frac{\sqrt{\lambda(m_D^2,q^2,p^2)}}{2m_D}$ and 
$p^\ast = \frac{\sqrt{\lambda(p^2,m_\pi^2,m_\pi^2)}}{2\sqrt{p^2}}$.
The angular distribution can be written as 
\begin{equation}
    \mathcal{I} = \mathcal{I}_1 + \mathcal{I}_2\cos(2\theta_e) + \dots
\end{equation}
with 
\begin{align}
    \mathcal{I}_1 &= \frac{\sqrt{p^2}}{8 p_{\pi\pi}\, m_D \,p^\ast}  \left\{ \left|\mathcal{F}_0\right|^2 + \frac{3}{2}\sin^2\theta_{\pi} \left( \left|\mathcal{F}_\parallel\right|^2 + \left|\mathcal{F}_\perp\right|^2 \right) \right\}  \:,\\
    \mathcal{I}_2 &= -\frac{\sqrt{p^2}}{8 p_{\pi\pi}\, m_D\, p^\ast}  \left\{ \left|\mathcal{F}_0\right|^2 - \frac{1}{2}\sin^2\theta_{\pi} \left( \left|\mathcal{F}_\parallel\right|^2 + \left|\mathcal{F}_\perp\right|^2 \right) \right\} \:,
\end{align}
and 
\begin{align}
    \mathcal{F}_0         &=  2p_{\pi\pi} m_D \frac{p^\ast}{\sqrt{p}}\left( N_{f_0(500)}\,\mathcal{F}_{10} + N_{\rho^0}\,\mathcal{F}_{11}\cos\theta_{\pi}\right) \:,\nonumber\\
    \mathcal{F}_\parallel &=  \sqrt{2}p_{\pi\pi} m_D \frac{p^\ast}{\sqrt{p}}  N_{\rho^0}\,\mathcal{F}_{21}  \:,\\
    \mathcal{F}_\perp     &=  \sqrt{2}p_{\pi\pi} m_D \frac{p^\ast}{\sqrt{p}} N_{\rho^0}\,\mathcal{F}_{31} \:,\nonumber
\end{align}
where the coefficients $N_{f_0(500)/\rho^0}$ are normalization constants.
The form factors $\mathcal{F}_{i1}$, $i=1,2,3$ are related 
to the helicity amplitudes $H_{0,+,-}$ via 
\begin{eqnarray}
    {\cal F}_{11}  &=&  2 \sqrt{2 q^2} \,H_0 \:,\nonumber\\
    {\cal F}_{21}  &=& 2  \, \sqrt{q^2}\, \left (H_+ + H_- \right )\label{eq:Fdeux} \:,\\
    {\cal F}_{31}  &=& 2  \, \sqrt{q^2}\, \left (H_+ - H_- \right ) \:,\nonumber
\end{eqnarray}
and the helicity amplitudes are in turn related to the vector form factor 
$V(q^2)$ and the axial-vector form factors $A_{1,2}(q^2)$ of the $D^0\to\rho^0$ 
transition by 
\begin{eqnarray}
 H_0 &=&\frac{1}{2 \sqrt{p^2}\, \sqrt{q^2}} \left[ \left (
m_D^2-p^2-q^2 \right ) \left ( m_D+\sqrt{p^2} \right ) A_1(q^2,p^2)\right .\nonumber\\ 
 &-& \left . 4 \frac{m_D^2\,p_{\pi\pi}^2}{m_D+\sqrt{p^2}} A_2(q^2,p^2) \right]\:, \\
 H_{\pm}& =&\left (m_D + \sqrt{p^2} \right ) A_1(q^2,p^2)
\mp \frac{2m_D\, p_{\pi\pi}}{m_D+\sqrt{p^2}} V(q^2,p^2)\:.\nonumber
\label{eq:FF}
\end{eqnarray}
Here, the form factors $V,A_1,A_2$ are factorized into  $q^2$ 
and $p^2$-dependent components 
\begin{equation} 
    (V,A_1,A_2)(q^2,p^2) = (V,A_1,A_2)(q^2)\times \mathcal{A}(p^2) \:,
\end{equation} 
with the lineshape given as 
\begin{equation}
    \mathcal{A}(p^2) =  \frac{1}{\sqrt{\pi} } \frac{p^\ast(p^2)}{p^{\ast}_0} \frac{B(p^*) }{ B(p_0^*) } \frac{1}{P_{\rho^0}(p^2)} \left( 1 + a_{p^2,\omega} \, e^{i \delta_{p^2,\omega}} \: p^2\,\text{BW}_\omega (p^2) \right) \,.\label{eq:lineshapeBESIII}
\end{equation}
$B(p) = 1/\sqrt{1+r_{\text{BW}}^2\,p^2}$ is the Blatt-Weisskopf damping factor with the barrier factor $r_{BW}=3\:\mathrm{GeV}^{-1}$ and $P_{\rho^0}(p^2)$ is defined in the appendix of Ref.~\cite{Fajfer:2023tkp}.
The $q^2$ dependent functions are fitted in Ref.~\cite{BESIII:2018qmf} using a single pole ansatz
\begin{eqnarray}
    V(q^2)&=&\frac{V(0)}{1-\frac{q^2}{m_{V}^2}} \:,\nonumber\\
    A_{1,2}(q^2)&=&\frac{A_{1,2}(0)}{1-\frac{q^2}{m_A^2}}\:, \label{eq:ffdef}
\end{eqnarray}
with $m_V \simeq 2.01\,\mathrm{GeV}$, $m_A \simeq 2.42\,\mathrm{GeV}$. 
The results of the fit $r_V = V(0)/A_1(0) = 1.695\pm 0.083\pm 0.051$ and $r_2 = A_2(0)/A_1(0) = 0.845\pm 0.056\pm 0.039$ ~\cite{BESIII:2018qmf} 
are consistent with theoretical results from Ref.~\cite{Melikhov:2000yu, Lin:2025cmn}.

The $S$-wave contribution stemming from the $f_0(500)$ resonance is given as
\begin{equation}
    \mathcal{F}_{10} = p_{\pi\pi} m_D \frac{e^{i\phi_S} \mathcal{A}_S(p^2)}{1-\frac{q^2}{m_A^2}}
\end{equation}
with the lineshape $\mathcal{A}_S(p^2)$ provided in the appendix of Ref.~\cite{Fajfer:2023tkp}.
The relative phases are given as $\phi_S=3.4044\pm 0.0738$ and $\delta_{p^2,\omega} = 2.93\pm 0.17$ in Ref.~\cite{Fajfer:2023tkp}. 
The parameter $a_{p^2,\omega}$ is fitted together with the normalization constants $N_{f_0(500)/\rho^0}$ to the 
components of the branching fraction in Table~\ref{tab:BRs_BESIII}.  
In agreement with Ref.~\cite{Fajfer:2023tkp} we find $a_{p^2,\omega} \simeq 0.006$.

The contribution of the left-handed current $\mathcal{O}_L^{U_{12}\,ee}$ to the decay rate of $D^0\to\pi^+\pi^-\nu\overline{\nu}$ is  related  by isospin symmetry to decay rate of $D^+\to\pi^+\pi^-e^+\nu_e$ as
\begin{equation}
  \label{eq:relation_dineutrino_decaywidths}
  \mathrm{d}^5\Gamma (D^+\to \pi^+\pi^-e^+\nu_e) \simeq \left|\frac{\frac{\alpha_e}{4\pi} \mathcal{C}_{L}^{U_{12}\,ee}}{V_{cd}}\right|^2 \times \mathrm{d}^5\Gamma(D^0\to \pi^+\pi^-\nu\overline{\nu})\bigg|_{\mathcal{O}_L^{U_{12}\,ee}}  \, .
\end{equation}
This allows us to extract the transversity form factors $\mathcal{F}_{0,\perp,\parallel}$ for the dineutrino mode 
assuming that BSM effects are negligible in the semileptonic decay. The total and differential branching fractions of charged current 
$c \to d e^+ \nu$ mediated decays $D^+\to \pi^0 e^+\nu_e$, $D^0\to \pi^-e^+\nu_e$ and $\Lambda_c^+ \to n e^+ \nu_e$ are in good agreement with the SM and NP contributions are at most at the $\sim 10$\% level \cite{FermilabLattice:2022gku, BESIII:2015tql, BESIII:2017ylw, BESIII:2024mgg}.

\subsubsection{Comparison of dineutrino spectra \label{sec:comp}}

Here we compare the decay distributions resulting from the hadronic form factors in the different approaches. As the dineutrino decays are SM null tests we need a BSM benchmark to illustrate the height and shape of the decay distributions. 
To be specific, we choose $x_U^+=x_U^-=392$, corresponding to the maximal cLFC scenario. Different  values would change 
the height of the curves, i.e. the branching fractions,
and  $x_U^+ \neq x_U^-$ would in addition change the shapes as it would reweight different transversities, see (\ref{eq:diffbranching_dineutrino}).

\begin{figure}
    \centering
    \includegraphics[width=0.49\textwidth]{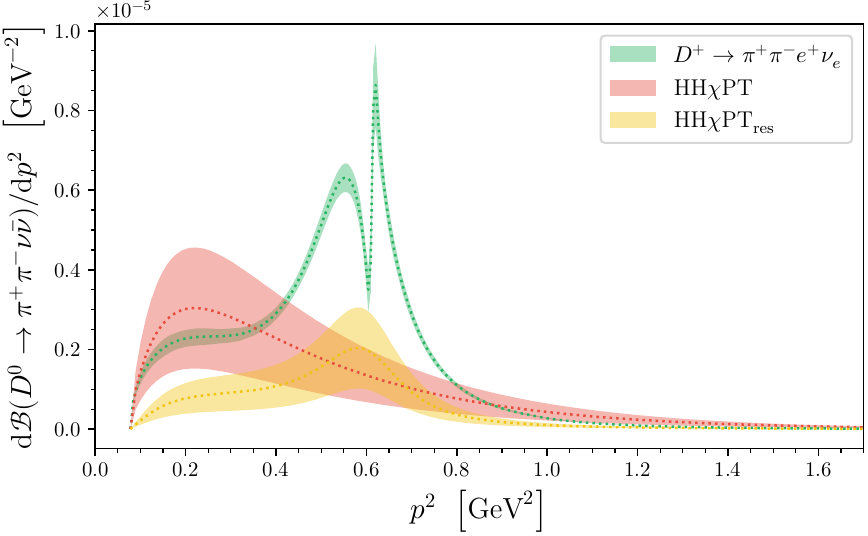}
    \includegraphics[width=0.49\textwidth]{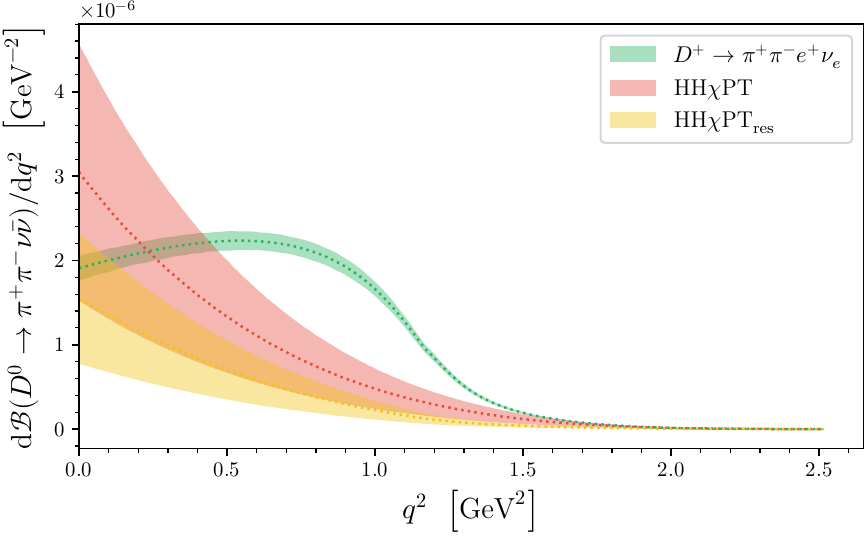}
    \caption{Differential branching fraction of $D^0\to\pi^+\pi^-\nu\overline{\nu}$ decays for BSM benchmark $x_U^+=x_U^-=392$ (\ref{eq:XU}) against  $q^2$ (right panel) and $p^2$ (left panel) for different hadronic models, HH$\chi$PT (red), HH$\chi$PT$_\text{res}$ (yellow) and data driven (green). For the HH$\chi$PT-based models the bands illustrate $50\%$ uncertainty in the branching fractions, while for the data driven one they give  the $1 \sigma$ uncertainties 
    resulting from the fit, which are below 10\%.}
    \label{fig:dBR_dx2_dineutrino}
\end{figure}

In Fig.~\ref{fig:dBR_dx2_dineutrino} we show the differential $D^0\to\pi^+\pi^-\nu\overline{\nu}$ branching fraction as a function of $q^2$ (right panel) and  $p^2$ (left panel) in the HH$\chi$PT based approach (red), HH$\chi$PT$_\text{res}$ amended by the  $\rho^0$  (yellow) and the data driven approach using data on $D^+\to\pi^+\pi^- e^+\nu_e$ decays (green). 

\begin{figure}
    \centering
    \includegraphics[width=0.49\textwidth]{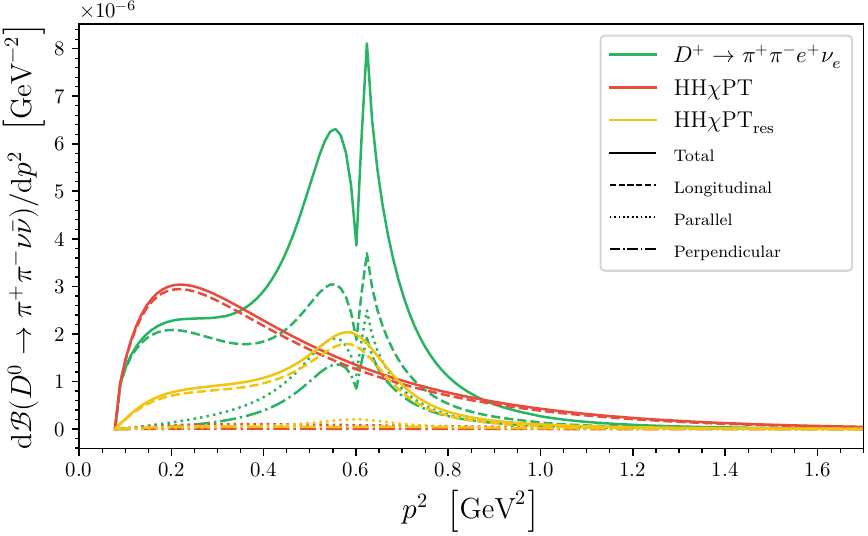}
    \includegraphics[width=0.49\textwidth]{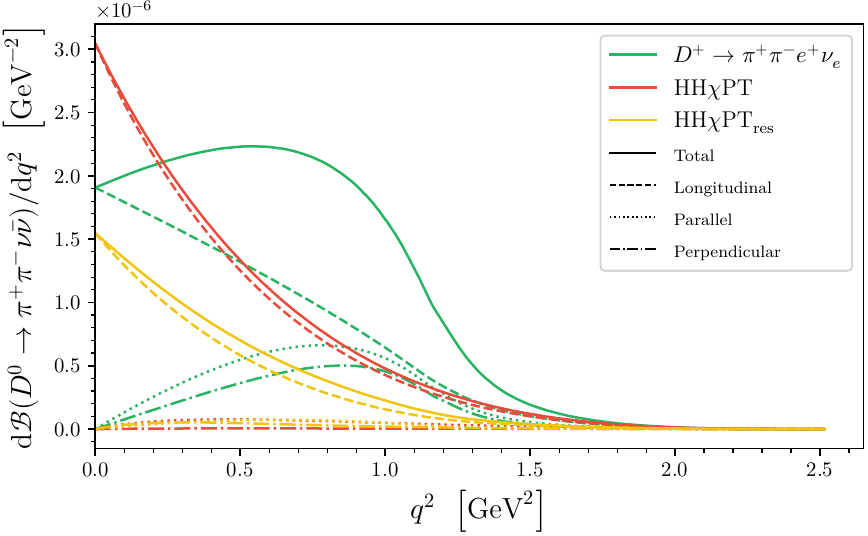}
    \caption{Breakdown of transversity contributions to the differential branching fraction of $D^0\to\pi^+\pi^-\nu\overline{\nu}$ decays as in Fig.~\ref{fig:dBR_dx2_dineutrino} for central values showing the
    total, i.e., the sum (solid), longitudinal (dashed), parallel (dotted) and perpendicular (dash dotted) contributions.
    }
    \label{fig:dBR_dx2_dineutrino_contribution}
\end{figure}

The bands display $50\%$ uncertainty in the branching fractions using HH$\chi$PT-based form factors for which  $1/m_D$ corrections are to be expected, which are sizable. For the differential branching fraction based on the data driven form factors  we compute the uncertainties from the fit, and find they are smaller,
between  five and ten percent.
In Fig.~\ref{fig:dBR_dx2_dineutrino_contribution} we show the breakdown of the transversity components, total (solid), longitudinal (dashed), parallel (dotted) and perpendicular (dash dotted), to the differential branching fraction.

We observe several differences between the decay distributions:
For $d\mathcal{B}/dp^2$, the non-resonant form factors from HH$\chi$PT (red) do not exhibit
resonance structure in $\pi^+ \pi^-$, such as the $\rho^0$. The latter is included  in the HH$\chi$PT$_\text{res}$ model with $\rho^0$ (yellow). As the $\omega$ is very narrow and predominantly decays to three pions -- decays to two pions arise from isospin breaking --
it gives a spike that is restricted to a narrow region close to and above the $\rho^0$. This can be seen in the data driven model (green), which
includes the $\rho^0,\omega$ and $f_0(500)$. The latter is however very broad, and does not give a visible peak.

For $d\mathcal{B}/dq^2$ noticeable differences are between the shape of the data-driven model 
and the HH$\chi$PT ones. In particular,  the latter grow towards the low $q^2$ end point. All models feature the longitudinal component induced by $\mathcal{F}_0$ as the largest, and the perpendicular one as the smallest, 
 as expected on general grounds \cite{Hiller:2013cza}, however hierarchies differ quantitatively between
the data driven model and the HH$\chi$PT ones,
see Fig.~\ref{fig:dBR_dx2_dineutrino_contribution}. 
 We stress, however, that despite the apparent differences, all methods that 
range from pure theory, to theory supplemented by resonances, to a data-driven phenomenological ansatz, result in branching fractions that agree within a factor of a few, see Table~\ref{tab:BR_upper_limits}.

\subsection{Polarized \texorpdfstring{$\Lambda^+_c$}{Lambdac} decays}\label{sec:l2pmumu_theo}
The four-differential distribution of polarized $\Lambda^+_c \to p\ell^+\ell^-$ decays 
can be written as 
\begin{align}
    \frac{\deriv^{4}\Gamma(\Lambda^+_c\to p \ell^+\ell^-)}{\deriv q^2\,\deriv\!\cos\theta\,\deriv\!\cos\theta_\ell\,\deriv\phi_\ell} &= \frac{3}{8\pi}\bigg(K_{1ss}(q^2)\sin^2\theta_\ell + K_{1cc}(q^2)\cos^2\theta_\ell + K_{1c}(q^2)\cos\theta_\ell  \nonumber\\
    \label{eq:d4G}
    &\quad +\left( K_{11}(q^2)\sin^2\theta_\ell + K_{12}(q^2)\cos^2\theta_\ell + K_{13}(q^2)\cos\theta_\ell \right)\cos\theta \vphantom{\bigg)}\\ 
    &\quad +\left( K_{21}(q^2)\sin\phi_\ell + K_{23}(q^2)\cos\phi_\ell  \right)\cos\theta_\ell\sin\theta \vphantom{\bigg)}\nonumber \\ 
    &\quad +\left( K_{22}(q^2)\sin\phi_\ell + K_{24}(q^2)\cos\phi_\ell  \right)\sin\theta_\ell\sin\theta \bigg) \:,\nonumber
\end{align}
where $q^2$ is the invariant mass squared of the dileptons and $\theta_\ell$ is the angle between the positively charged lepton and the $\Lambda^+_c$ baryon 
in the dilepton rest frame. The angles $\theta$ and $\phi_\ell$ are defined in respect to the normal 
vector $\hat{n}$ spanned by the $\Lambda^+_c$ baryon in the lab-frame and the beam-axis of the experiment ($\hat{n}=\hat{p}_{\Lambda_c}\times \hat{p}_{\text{beam}}$).
The angle $\theta$ is the angle between the normal vector and decay plane spanned by the $\hat{p}_p$ and $\hat{p}_{\ell\bar{\ell}}$. 
The angle $\phi_\ell$ is the angle between the dilepton plane and the vector $\hat{x}_{\ell\bar{\ell}}$ of the 
coordinate system constructed with $\hat{z}_{\ell\bar{\ell}} = \hat{p}_{\ell\bar{\ell}}$ and $\hat{y}_{\ell\bar{\ell}} = \hat{n} \times \hat{p}_{\ell\bar{\ell}}$ 
all defined in the $\Lambda^+_c$ rest-frame. For more details on the angles see \cite{Blake:2017une}.
The physical regions of the angles and invariant mass are given by
\begin{equation}
    0 \leq \phi_\ell < 2\pi \,,\quad -1\leq \cos\theta < 1  \,,\quad -1\leq \cos\theta_\ell < 1 \,,\quad 4m_\ell^2 \leq q^2 \leq (m_{\Lambda_c}-m_p)^2 \:.
\end{equation}
The NP Wilson coefficients enter into the angular coefficients $K_i=K_i(q^2)$. Resonance 
contribution of $\Lambda^+_c \to p R(\to \ell^+\ell^-)$,\,$R=\rho,\omega,\phi$, without any better 
alternative, need to be fitted to data. For the current precision it is sufficient to use a pragmatic approach, we  add a $q^2$-dependent resonance contribution, $\mathcal{K}_9^{R}$,  to the coefficient of $\mathcal{O}_9$,
\begin{equation}
    \mathcal{K}_9^R(q^2) = \frac{a_\rho \mathrm{e}^{i\delta_\rho}}{q^2-m_\rho^2 + im_\rho \Gamma_\rho} + \frac{a_\omega \mathrm{e}^{i\delta_\omega}}{q^2-m_\omega^2 + im_\omega \Gamma_\omega} + \frac{a_\phi \mathrm{e}^{i\delta_\phi}}{q^2-m_\phi^2 + im_\phi \Gamma_\phi} \,,
    \label{eq:K9R}
\end{equation}
with the resonance masses $m_R$, decay widths $\Gamma_R$ and resonance parameters $a_R$ and $\delta_R$. 
For the parameters $a_R$ we use the values fitted to data from Ref.~\cite{Gisbert:2024kob}, while for the 
strong phases we use the isospin limit $\delta_\rho-\delta_\omega=\pi$ and vary other phases between $0$ and $2\pi$. This 
will be the main source of uncertainty for calculating observables. We use the same parameter values for both $\Lambda^+_c\to p e^+ e^-$ and $\Lambda^+_c\to p\mu^+\mu^-$ decays. We refrain from including the $\eta$ and $\eta^\prime$ resonances here as they are very narrow and will not effect our results in Sec.~\ref{sec:ellell}. Their inclusion in a similar manner as Eq.~\eqref{eq:K9R} via a resonant pseudo-scalar contribution is however straightforward and has been done in Ref.~\cite{Golz:2021imq}.

Integration of the angular distribution over the angle $\theta$ and $\phi_\ell$ leaves one with the 
identical angular distribution as for unpolarized $\Lambda^+_c\to p\ell^+\ell^-$ decays and the expressions for 
the angular coefficients $K_i$ with $i=\left\{1ss,1cc,1c\right\}$ can be found in Ref.~\cite{Golz:2021imq}. 
We compute here the additional angular coefficients that appear in the polarized case using the helicity formalism, following~\cite{Blake:2017une,Golz:2021imq,Golz:2022alh}.
The additional $q^2$-dependent angular coefficients $K_i$ in (\ref{eq:d4G}) are given as 
\begin{equation}    
\begin{aligned}
    K_{11} &= \frac{1}{2}\,P_{\Lambda_c}\,q^2\bigg( -2(\beta_\ell^2-1) S_{P}^{22} + \beta_\ell^2 \left(P^{11}-P^{22}\right) + 2\beta_\ell^2 L_P^{22} + 2\left(L_P^{11}-P^{11}\right)\bigg)\:,\\
    K_{12} &= P_{\Lambda_c}\,q^2\bigg(
        -(\beta_\ell^2-1)\,S_P^{22}
        -\beta_\ell^2\,P^{11+22}
        +L_P^{11}-P^{11}
    \bigg)\:,\\
    K_{13} &= 2\beta_\ell\, P_{\Lambda_c}\,q^2\,U^{12} \:,\\
    K_{21} &= - \sqrt{2} \beta_\ell^2\, P_{\Lambda_c}\,q^2\,J_{4P}^{11+22} \:,\\
    K_{22} &= - \sqrt{2} \beta_\ell \, P_{\Lambda_c}\,q^2\,J_{3P}^{12} \:,\\
    K_{23} &= - \sqrt{2} \beta_\ell^2\, P_{\Lambda_c}\,q^2\,J_{2P}^{11+22} \:,\\
    K_{24} &= \sqrt{2} \beta_\ell\, P_{\Lambda_c}\,q^2\,J_{1P}^{12} \:,
\end{aligned} \label{eq:K_i}
\end{equation}
with the short-hand notation $X^{11+22} = X^{11}+X^{22}$ for $X=P,\,J_{2P},\,J_{4P}$, and $\beta_\ell=\sqrt{1-\frac{4m_\ell^2}{q^2}}$ with the lepton mass $m_\ell$. See Appendix~\ref{app:helicity_amps} for further details.
Our results are in agreement with \cite{Blake:2017une,Golz:2022alh}\footnote{The right plot of Figure 2 in Ref.~\cite{Golz:2022alh} erroneously shows $\frac{3}{2} A_{\text{FB}}^\ell$ instead of $A_{\text{FB}}^\ell$. We thank Tom Magorsch for confirmation.}.

We identify four null tests in the distribution polarized $\Lambda^+_c$ decays, (\ref{eq:d4G}), which 
are proportional to $K_{1c}$, $K_{13}$, $K_{22}$, $K_{24}$. Here, $K_{1c}$ corresponds to the forward-backward asymmetry in the leptons, $A_\text{FB}^\ell$ that can be accessed also without polarization, and has been measured by LHCb for muons~\cite{LHCb:2025bfy}.
The dependence on Wilson coefficients and form factors in $K_{13}$ is identical to the combined asymmetry in the leptons and hadrons $A_\text{FB}^{\ell H}$, which arises in baryonic decays with a self-analyzing secondary baryon \cite{Golz:2022alh}. 
The resonance enhanced contributions are sensitive to the following combinations of coefficients
\begin{equation}
\begin{aligned}
    K_{1c} &\propto \mathrm{Re}\left\{ \mathcal{K}_9^{\mathcal{R}} \mathcal{K}_{10}^\ast  \right\} \:,\\
    K_{13} &\propto \mathrm{Im}\left\{ \mathcal{K}_9^{\mathcal{R}} \left(\mathcal{K}_{10} \pm \mathcal{K}_{10}^\prime \right)^\ast  \right\} \:,\\
    K_{22} &\propto \mathrm{Im}\left\{ \mathcal{K}_9^{\mathcal{R}} \mathcal{K}_{10}^{\prime,\ast}  \right\} \:,\\
    K_{24} &\propto \mathrm{Re}\left\{ \mathcal{K}_9^{\mathcal{R}} \left(\mathcal{K}_{10} \pm \mathcal{K}_{10}^\prime \right)^\ast  \right\} \:. 
\end{aligned}
\end{equation}
We learn that $K_{13},K_{24}$ probe both $\mathcal{K}_{10} $ and $\mathcal{K}_{10}^\prime$ while  $K_{1c} $ tests  only the former and $K_{22}$ only the primed coefficient.

Using an angular decomposition we define the null-test asymmetries as 
\begin{equation}
\begin{aligned}
    \langle A_{\mathrm{FB}}^\ell\rangle &= \frac{1}{\Gamma} \int_{-1}^{1}\deriv\!\cos\theta\, \left(\int_{0}^{1}-\int_{-1}^{0}\right)\deriv\!\cos\theta_\ell\,\int_{0}^{2\pi}\deriv\phi_\ell \int \mathrm{d}q^2 \frac{\mathrm{d}^4 \Gamma}{\deriv q^2\,\deriv\!\cos\theta\,\deriv\!\cos\theta_\ell\,\deriv\phi_\ell}   \\ 
    &= \frac{3}{2} \frac{1}{\Gamma} \int \mathrm{d}q^2 K_{1c} (q^2)  \:,\\
    \langle A_{K_{13}}^\ell\rangle &= \frac{1}{\Gamma}  \left(\int_{0}^{1}-\int_{-1}^{0}\right)\deriv\!\cos\theta\, \left(\int_{0}^{1}-\int_{-1}^{0}\right)\deriv\!\cos\theta_\ell\,\int_{0}^{2\pi}\deriv\phi_\ell \int \mathrm{d}q^2 \frac{\mathrm{d}^4 \Gamma}{\deriv q^2\,\deriv\!\cos\theta\,\deriv\!\cos\theta_\ell\,\deriv\phi_\ell}  \\  
    &= \frac{3}{4} \frac{1}{\Gamma} \int \mathrm{d}q^2 K_{13} (q^2)  \:,\\
    \langle A_{K_{22}}^\ell\rangle &= \frac{1}{\Gamma} \int_{-1}^{1}\deriv\!\cos\theta\, \int_{-1}^{1} \deriv\!\cos\theta_\ell\,\left(\int_{0}^{\pi}-\int_{\pi}^{2\pi}\right)\deriv\phi_\ell \int \mathrm{d}q^2 \frac{\mathrm{d}^4 \Gamma}{\deriv q^2\,\deriv\!\cos\theta\,\deriv\!\cos\theta_\ell\,\deriv\phi_\ell}  \\
    &= \frac{3\pi}{8} \frac{1}{\Gamma} \int \mathrm{d}q^2 K_{22} (q^2) \:, \\
    \langle A_{K_{24}}^\ell\rangle &= \frac{1}{\Gamma} \int_{-1}^{1}\deriv\!\cos\theta\, \int_{-1}^{1} \deriv\!\cos\theta_\ell\,\left(\int_{0}^{\frac{\pi}{2}}-\int_{\frac{\pi}{2}}^{\frac{3\pi}{2}}+\int_{\frac{3\pi}{2}}^{2\pi} \right)\deriv\phi_\ell \int \mathrm{d}q^2 \frac{\mathrm{d}^4 \Gamma}{\deriv q^2\,\deriv\!\cos\theta\,\deriv\!\cos\theta_\ell\,\deriv\phi_\ell}  \\
    &= \frac{3\pi}{8} \frac{1}{\Gamma} \int \mathrm{d}q^2 K_{24} (q^2)  \:. 
\end{aligned} \label{eq:afbs}
\end{equation}
An additional  asymmetry that arises from finite $\Lambda^+_c$-polarization is
\begin{equation}\label{eq:AK21}
\begin{aligned}
    \langle A_{K_{21}}^\ell\rangle &= \frac{1}{\Gamma} \int_{-1}^{1}\deriv\!\cos\theta\, \left(\int_{0}^{1}-\int_{-1}^{0}\right)\deriv\!\cos\theta_\ell\,\left(\int_{0}^{\pi}-\int_{\pi}^{2\pi}\right)\deriv\phi_\ell \int \mathrm{d}q^2 \frac{\mathrm{d}^4 \Gamma}{\deriv q^2\,\deriv\!\cos\theta\,\deriv\!\cos\theta_\ell\,\deriv\phi_\ell} \\
    &= \frac{3}{4} \frac{1}{\Gamma} \int \mathrm{d}q^2 K_{21} (q^2) \:,
\end{aligned}
\end{equation}
which is driven by dipole couplings $K_{21} \propto \mathrm{Im}\left\{ \mathcal{K}_9^{\mathcal{R}} \left(\mathcal{K}_{7} \pm \mathcal{K}_{7}^\prime \right)^\ast  \right\}$.
Since the dipole coefficients in the SM are finite, yet small \cite{Greub:1996wn,deBoer:2017que}, $K_{21}$ is
not as strongly GIM-protected as the null tests in Eq.~(\ref{eq:afbs}), which vanish for vanishing  leptonic axial-vector currents \cite{DeBoer:2018pdx}.
Moreover, since NP effects in the dipoles are stronger constrained than those in the four-fermion operators,  see Sec.~\ref{sec:nulltest_pred}, maximum values of $\langle A_{K_{21}}^\ell\rangle $ are generically expected to be smaller than for the other asymmetries.

\subsection{Upper limits on NP and null tests}\label{sec:nulltest_pred}

We give phenomenological predictions for rare charm decays into dineutrinos in 
Sec.~\ref{sec:nunu} and for the dileptons in Sec.~\ref{sec:ellell}.

\subsubsection{\texorpdfstring{$c \to u \nu \bar \nu$}{c -> u nu nubar} in SMEFT  \label{sec:nunu}}

We compute upper limits on the total branching fraction of $D^0\to\pi^+\pi^-\nu\overline{\nu}$ decays, employing the  hadronic models for the form factors
of Sec.~\ref{sec:pipinunu_theo}.
We utilize bounds on the Wilson coefficients, following Ref.~\cite{Bause:2020xzj} using  the $SU(2)_L$ link in the SMEFT and  high-$p_T$ data. 
 We distinguish between NP scenarios with contributions with  $\mathcal{K}_{L}$ and $\mathcal{K}_R$,  only $\mathcal{K}_L$, and  only $\mathcal{K}_R$. We consider different lepton flavor patterns,  LU, democratic, cLFC and a general structure as detailed in Sec.~\ref{sec:EFT}. 
Results are shown in Table~\ref{tab:BR_upper_limits}.  The Drell-Yan data are advantageous as they provide a single source of constraints for all accessible flavors as well as conservative limits unaffected by cancellations. If one neglects the possibility of cancellations from different operators in the amplitudes, stronger limits from the combination with LFV decays~\cite{Angelescu:2020uug}, \textit{e.g.}~$K \to e \mu$, $\tau\to \ell K^\ast$, $\tau\to\ell\rho$ and $\mu \to e$ conversion exist.
We show these stronger limits, available for the democratic and general structures, in parentheses.

The flavor hierarchies $ \mathcal{B}_\text{LU}^{\text{max}}<\mathcal{B}_\text{democratic}^{\text{max}}<\mathcal{B}^{\text{max}}_\text{cLFC}<\mathcal{B}^{\text{max}}$ are apparent in  the limits  using only Drell-Yan data. The additional constraints from low-energy LFV data affect mostly the democratic scenario, which then becomes the one with the smallest maximum branching fraction,
$ \mathcal{B}_\text{democratic}^{\text{max}}<\mathcal{B}_\text{LU}^{\text{max}}$.
As expected from the spectra in Fig.~\ref{fig:dBR_dx2_dineutrino} the branching fractions in the data-based form factor approach are the largest, about a factor or three larger than the smallest ones, obtained at lowest order HH$\chi$PT.
The NP scenario with only $\mathcal{K}_{R}$, which corresponds for example to $S_2$ or $\tilde{V}_2$ lepto-quark 
models~\cite{Bause:2020xzj}\footnotemark,  gives roughly an order of magnitude stronger bounds on the branching fractions than the scenario with both $
\mathcal{K}_{L}$ \&  $\mathcal{K}_R$ present.

The general upper limit derived in SMEFT is more than an order of magnitude stronger than the one obtained in WET using only $\mathcal{Q}_L$ or $\mathcal{Q}_R$ and the experimental upper limit $\mathcal{B}(D^0 \to \pi^0 \nu \bar \nu) < 2.1 \cdot 10^{-4}$~\cite{BESIII:2021slf}.
An observation above the SMEFT limits would imply NP light degrees of freedom.

\begin{table}
  \caption[Caption for BRs table]{Upper limits on the branching fraction of $D^0\to\pi^+\pi^-\nu\overline{\nu}$ decays for different 
  flavor scenarios,  LU, democratic, cLFC and general, the latter denoted by $\mathcal{B}^{\text{max}}$. The bounds are extracted using high-$p_T$ data following Ref.~\cite{Bause:2020xzj} for different hadronic form factors 
  and allowing for NP either in left-handed FCNCs $\mathcal{K}_{L}$, or right-handed ones $\mathcal{K}_{R}$, or both. Limits including additionally constraints from lepton-flavor violating low-energy processes~\cite{Angelescu:2020uug}, neglecting large cancellations, are shown in parentheses. See text for details.}
  \label{tab:BR_upper_limits}
  \renewcommand{\arraystretch}{1.5}
\begin{adjustbox}{max width=\textwidth}
\begin{tabular}{lllllcllllcllll}
\hline
      Form Factors                & \multicolumn{4}{c}{$\mathcal{K}_L$}            &\phantom{For}& \multicolumn{4}{c}{$\mathcal{K}_R$}            &\phantom{For}& \multicolumn{4}{c}{$\mathcal{K}_L\, \&  \,\mathcal{K}_R$}        \\
\cmidrule[0.4pt](lr{0.4em}){2-5}\cmidrule[0.4pt](lr{0.4em}){7-10}\cmidrule[0.4pt](lr{0.4em}){12-15}
    & $\mathcal{B}^{\text{max}}_{\mathrm{LU}}$   &  $\mathcal{B}^{\text{max}}_{\mathrm{democratic}}$   & $\mathcal{B}^{\text{max}}_{\mathrm{cLFC}}$   & $\mathcal{B}^{\text{max}}$   && $\mathcal{B}^{\text{max}}_{\mathrm{LU}}$   & $\mathcal{B}^{\text{max}}_{\mathrm{democratic}}$   &$\mathcal{B}^{\text{max}}_{\mathrm{cLFC}}$   & $\mathcal{B}^{\text{max}}$   && $\mathcal{B}^{\text{max}}_{\mathrm{LU}}$   & $\mathcal{B}^{\text{max}}_{\mathrm{democratic}}$   &  $\mathcal{B}^{\text{max}}_{\mathrm{cLFC}}$   & $\mathcal{B}^{\text{max}}$  \\
    & $\left[10^{-7}\right]$   & $\left[10^{-7}\right]$   & $\left[10^{-7}\right]$   & $\left[10^{-6}\right]$   && $\left[10^{-8}\right]$   & $\left[10^{-7}\right]$   & $\left[10^{-7}\right]$   & $\left[10^{-7}\right]$   && $\left[10^{-7}\right]$   & $\left[10^{-6}\right]$   & $\left[10^{-6}\right]$   & $\left[10^{-6}\right]$  \\
\hline
data driven                     & $1.7$    & $5.2(0.08)$  & $9.9$        & $3.7(1.0)$          && $4.8$   &  $1.4$   & $2.7$        & $9.3$           && $4.4$  &  $1.3(0.31)$     & $2.5$         & $9.2(3.9)$                     \\
HH$\chi$PT                      & $1.1$    & $3.4(0.05)$  & $6.4$        & $2.4(0.7)$          && $3.1$   & $0.93$   & $1.8$        & $6.0$           && $2.9$   &  $0.86(0.21)$    & $1.6$         & $6.0(2.5)$                     \\
HH$\chi$PT with $\rho^0$ & $0.56$   &  $1.7(0.03)$   & $3.2$        & $1.2(0.3)$          && $1.6$   & $0.47$   & $0.90$        & $3.0$           && $1.4$   & $0.43(0.10)$    & $0.83$         & $3.0(1.3)$                     \\
\hline
\end{tabular}
\end{adjustbox}
\end{table}
\footnotetext{\label{footnote}The bounds with only  $\mathcal{K}_R$ switched on should 
  agree with bounds on the $S_2$ or $\tilde{V}_2$ LQ models of Ref.~\cite{Bause:2020xzj}. Therein, however, a 
  stronger estimate was missed for the bounds with only  $\mathcal{K}_{L}$ or  $\mathcal{K}_{R}$, and actual bounds are a factor $2$ stronger. 
  The same applies for all other bounds shown in Table VI of Ref.~\cite{Bause:2020xzj}. Our bounds for $\mathcal{K}_L\,\&\,\mathcal{K}_R$ agree with the ones in Table III of Ref.~\cite{Bause:2020xzj}. We thank Hector Gisbert for confirmation.}

\subsubsection{\texorpdfstring{$c \to u \ell^+ \ell^-$ analysis}{c -> u l+ l-}}  \label{sec:ellell}

For $\ell = \mu$ the most stringent 
upper limits on the Wilson coefficients in Eq.~\eqref{eq:WET} are obtained in Ref.~\cite{Gisbert:2024kob} from $D^0\to\mu^+\mu^-$, $D^+\to\pi^+\mu^+\mu^-$ and $\Lambda^+_c\to p\mu^+\mu^-$ decays.
While the baryonic decay gives the strongest bound for dipole operators $\mathcal{O}_7^{\prime}$
\begin{equation}
 \left|\mathcal{K}_{7}^{\phantom{\prime}}+\mathcal{K}_{7}^{\prime}\right| ,\:  \left|\mathcal{K}_{7}^{\phantom{\prime}}-\mathcal{K}_{7}^{\prime}\right| \lesssim 0.24 \:,
\end{equation}
constraints on the vector and axial operators $\mathcal{O}_{9,10}^{(\prime)}$ are currently dominated by $D$-meson decays. For $\mathcal{O}_{10}^{(\prime)}$ the constraints are a combination of   semileptonic and leptonic $D$ meson decay channels, while $\mathcal{O}_{9}^{(\prime)}$ bounds include all channels, including the $\Lambda^+_c$ to eliminate some otherwise flat directions. The current upper limits
for muons read  
\begin{align}
     \left|\mathcal{K}_{9}^{\mu\mu}+\mathcal{K}_{9}^{\prime\,\mu\mu}\right| \lesssim 0.85\:, \quad \left|\mathcal{K}_{9}^{\mu\mu}-\mathcal{K}_{9}^{\prime\,\mu\mu}\right| \lesssim 1.3\phantom{0} \:, \quad \left|\mathcal{K}_{9}^{(\prime)\,\mu\mu}\right| \lesssim 1.1\phantom{0}\:,\\
    \left|\mathcal{K}_{10}^{\mu\mu}+\mathcal{K}_{10}^{\prime\,\mu\mu}\right| \lesssim 0.85\:, \quad \left|\mathcal{K}_{10}^{\mu\mu}-\mathcal{K}_{10}^{\prime\,\mu\mu}\right| \lesssim 0.52\:, \quad \left|\mathcal{K}_{10}^{(\prime)\,\mu\mu}\right| \lesssim 0.7\phantom{0}  \:.  \label{eq:K10limit}
\end{align}

For $\ell=e$ the bounds from rare decays are significantly weaker with orders of magnitude weaker experimental limits on the branching fractions and higher lepton mass suppression, especially relevant for $D^0\to e^+e^-$. Ref.~\cite{deBoer:2015boa} gives bounds\footnote{In Eq.~$(32)$ of Ref.~\cite{deBoer:2015boa}  it should read $\left|C_{9,10}^{(e)}+C_{9,10}^{(e)\prime}\right|\lesssim 4$ and $\left|C_7\left(C_{9}^{(e)}+C_{9}^{(e)\prime}\right)^\ast\right|\lesssim 2$, with relative plus signs as both bounds stem from $D^+\to \pi^+ e^+e^-$ decays.} from $D^+\to \pi^+ e^+ e^-$ decays 
\begin{equation} \label{eq:Keelimit}
    \left|\mathcal{K}_{9,10}^{ee}+\mathcal{K}_{9,10}^{\prime\,ee}\right| \lesssim 3.5 \:,
\end{equation}
which are a factor $4$ smaller than the ones for muons. For other combinations of Wilson coefficients we obtain 
\begin{align}
    \left|\mathcal{K}_{9,10}^{ee}-\mathcal{K}_{9,10}^{\prime\,ee}\right| \lesssim 13 \:,
\end{align}
from $\mathcal{B}\left(\Lambda^+_c\to p e^+ e^- \right) < 5.5\cdot 10^{-6}$ at $90\%$ CL. \cite{BaBar:2011ouc}. Constraints from $D^0\to e^+ e^-$ are very poor with $\left|\mathcal{K}_{10}^{ee}-\mathcal{K}_{10}^{\prime\,ee}\right| \lesssim 6\cdot 10^2$ using $\mathcal{B}(D^0\to e^+e^-) < 7.9\cdot 10^{-8}$ at 90\% CL.~\cite{Belle:2010ouj}.
Combining limits on the sum and the difference yields a rather weak limit
\begin{equation}
\left|\mathcal{K}_{9,10}^{(\prime)\,ee} \right|\lesssim 9.5\, . 
\end{equation}

Additionally bounds for all lepton flavors exist from Drell-Yan data~\cite{Bause:2020xzj,Angelescu:2020uug,Fuentes-Martin:2020lea}. For $|\Delta c| = |\Delta u|=1$ with muons they are weaker than the ones from  rare decays, while for electrons they give the leading constraint~\cite{Bause:2020xzj,Angelescu:2020uug,Fuentes-Martin:2020lea} 
\begin{equation}
    \sqrt{\left|\mathcal{K}_{L}^{\mu\mu(ee)}\right|^2 + \left|\mathcal{K}_{R}^{\mu\mu(ee)}\right|^2 + \left|\mathcal{K}_{9}^{\mu\mu(ee)}+\mathcal{K}_{10}^{\mu\mu(ee)}\right|^2 + \left|\mathcal{K}_{9}^{\prime\mu\mu(ee)}+\mathcal{K}_{10}^{\prime\mu\mu(ee)}\right|^2} \lesssim 1.6(2.9)\:. \label{eq:bound_KLR}
\end{equation}
where we used (\ref{eq:KL910}) and made the lepton flavor explicit.

Barring large cancellations between SMEFT Wilson coefficients in the  $|\Delta s| = |\Delta d|=1$  decay amplitude, two orders of magnitude stronger constraints exist from  $SU(2)_L$ (\ref{eq:WET_correlations}) and  $K \to \pi \nu \bar \nu$ decays~\cite{Bause:2020auq}
\begin{equation}
    -1.9\cdot 10^{-2} \lesssim \left|\mathcal{K}_{L}^{\ell\ell}\right|_{\text{kaon}} \lesssim 0.7\cdot 10^{-2} \:.
\end{equation}
This constraint holds for operators with doublet leptons and doublet quarks, but, for instance, not for $\mathcal{Q}_{qe}$ or $\mathcal{Q}_{\ell u}$.

\begin{table}    
\caption{Null-test observables $\langle A_{i}^{\ell} \rangle$ in 
$\Lambda^+_c\to p\mu^+\mu^-$ decays (top) and $\Lambda^+_c\to pe^+e^-$  (bottom)  for NP scenarios 
compatible with current experimental bounds from low and high $p_T$,  Eqs.~(\ref{eq:K10limit}-\ref{eq:bound_KLR}).
The main source of uncertainty stems from hadronic parameters $a_\mathrm{R}$ and strong phases $\delta_{\mathrm{R}}$. }\label{tab:null_tests_NP}

\renewcommand{\arraystretch}{1.5}
\begin{adjustbox}{max width=\textwidth}
\begin{tabular}{llllllll}
    \hline
     Observable                                                                   & bin                & $\mathcal{K}_{10}^{\mu\mu} = 0.7$   & $\mathcal{K}_{10}^{\prime\,\mu\mu}  = 0.7$   & $\mathcal{K}_{10}^{\mu\mu} = \mathcal{K}_{10}^{\prime\,\mu\mu} = 0.425$   & $\mathcal{K}_{10}^{\mu\mu} = - \mathcal{K}_{10}^{\prime\,\mu\mu} = 0.26$   & $\mathcal{K}_{9}^{\prime\,\mu\mu} = -\mathcal{K}_{10}^{\prime\,\mu\mu} = 0.8$   & $\mathcal{K}_{9}^{\mu\mu} = -\mathcal{K}_{10}^{\mu\mu} = 0.8$   \\
                                                                                  &                    & in \%                               & in \%                                        & in \%                                                                                  & in \%                                                                        & in \%                                                                             & in \% \\
    \hline
     $\langle A_{\text{FB}}^{\mu}\rangle$                                        & $\text{low } \phi$  & $[-3.8,+3.9]$           & $0$                              & $[-2.4,+2.3]$                                                    & $[-1.4,+1.4]$                                          & $[+0.57,+0.95]$                                               & $[-4.4,+4.4]$                                \\
     $\langle A_{\text{FB}}^{\mu}\rangle$                                        & $\text{high } \phi$ & $[-2.8,+2.8]$           & $0$                              & $[-1.7,+1.7]$                                                    & $[-1.0,+1.0]$                                          & $[+0.18,+0.26]$                                               & $[-3.1,+3.2]$                                \\
     $\langle A_{\text{FB}}^{\mu}\rangle$                                        & $\phi$              & $[-2.3,+2.3]$           & $0$                              & $[-1.4,+1.4]$                                                    & $[-0.87,+0.87]$                                          & $[+0.43,+0.56]$                                               & $[-2.8,+2.4]$                                \\
    \hline
     $\langle A_{K_{13}}^{\mu}\rangle\:/\:P_{\Lambda_c}$                         & $\text{low } \phi$  & $[-2.1,+2.1]$           & $[-0.94,+0.94]$                   & $[-0.73,+0.73]$                                                    & $[-1.2,+1.2]$                                          & $[-1.4,+0.75]$                                              & $[-2.4,+2.4]$                                \\
     $\langle A_{K_{13}}^{\mu}\rangle\:/\:P_{\Lambda_c}$                         & $\text{high } \phi$ & $[-1.5,+1.5]$           & $[-0.69,+0.69]$                   & $[-0.52,+0.52]$                                                    & $[-0.83,+0.83]$                                          & $[-0.90,+0.66]$                                              & $[-1.7,+1.8]$                                \\
     $\langle A_{K_{13}}^{\mu}\rangle\:/\:P_{\Lambda_c}$                         & $\phi$              & $[-1.3,+1.3]$           & $[-0.57,+0.57]$                   & $[-0.44,+0.44]$                                                    & $[-0.70,+0.69]$                                          & $[-0.87,+0.41]$                                              & $[-1.6,+1.3]$                                \\
    \hline
     $\langle A_{K_{22}}^{\mu} \rangle\:/\:P_{\Lambda_c}$                        & $\text{low } \phi$  & $0$                      & $[-0.76,+0.76]$                   & $[-0.46,+0.46]$                                                    & $[-0.28,+0.28]$                                          & $[-0.86,+0.86]$                                              & $0$                                               \\
     $\langle A_{K_{22}}^{\mu} \rangle\:/\:P_{\Lambda_c}$                        & $\text{high } \phi$ & $0$                      & $[-0.54,+0.54]$                   & $[-0.33,+0.33]$                                                    & $[-0.20,+0.20]$                                          & $[-0.62,+0.62]$                                              & $0$                                               \\
     $\langle A_{K_{22}}^{\mu} \rangle\:/\:P_{\Lambda_c}$                        & $\phi$              & $0$                      & $[-0.46,+0.45]$                   & $[-0.28,+0.28]$                                                    & $[-0.17,+0.17]$                                          & $[-0.52,+0.52]$                                              & $0$                                               \\
    \hline
     $\langle A_{K_{24}}^{\mu} \rangle\:/\:P_{\Lambda_c}$                        & $\text{low } \phi$  & $[-5.0,+5.0]$           & $[-1.5,+1.5]$                   & $[-2.1,+2.1]$                                                    & $[-2.4,+2.4]$                                          & $[-2.5,+0.93]$                                              & $[-5.6,+5.6]$                                \\
     $\langle A_{K_{24}}^{\mu} \rangle\:/\:P_{\Lambda_c}$                        & $\text{high } \phi$ & $[-3.6,+3.6]$           & $[-1.1,+1.1]$                   & $[-1.5,+1.5]$                                                    & $[-1.7,+1.7]$                                          & $[-1.5,+0.96]$                                              & $[-4.0,+4.1]$                                \\
     $\langle A_{K_{24}}^{\mu} \rangle\:/\:P_{\Lambda_c}$                        & $\phi$              & $[-3.0,+3.0]$           & $[-0.91,+0.91]$                   & $[-1.3,+1.3]$                                                    & $[-1.5,+1.4]$                                          & $[-1.6,+0.47]$                                              & $[-3.7,+3.1]$                                \\
    \hline
    \\
    \hline
   Observable                                          & bin                & $\mathcal{K}_{10}^{ee} = 2.05$   & $\mathcal{K}_{10}^{\prime\,ee} = 2.05$   & $\mathcal{K}_{10}^{ee} = \mathcal{K}_{10}^{\prime\,ee} = 1.45$   & $\mathcal{K}_{10}^{ee} = - \mathcal{K}_{10}^{\prime\,ee} = 1.45$   & $\mathcal{K}_{9}^{\prime\,ee} = -\mathcal{K}_{10}^{\prime\,ee} = 1.45$   & $\mathcal{K}_{9}^{ee} = -\mathcal{K}_{10}^{ee} = 1.45$   \\
                                                        &                    & in \%                           & in \%                                   & in \%                                                            & in \%                                                              & in \%                                                                    & in \%                                                                                                                       \\
 \hline
$\langle A_{\text{FB}}^{e}\rangle$                 & $\text{low } \phi$  & $[-11.2,+11.2]$               & $0$                        & $[-8.0,+8.0]$                                                 & $[-7.8,+7.8]$                                                   & $[+1.8,+3.2]$                                                          & $[-8.1,+7.8]$                                         \\
 $\langle A_{\text{FB}}^{e}\rangle$                 & $\text{high } \phi$ & $[-8.2,+8.2]$                 & $0$                        & $[-5.9,+5.9]$                                                 & $[-5.8,+5.8]$                                                   & $[+0.61,+0.91]$                                                          & $[-5.7,+6.0]$                                         \\
 $\langle A_{\text{FB}}^{e}\rangle$                 & $\phi$             & $[-6.8,+6.9]$                 & $0$                        & $[-4.9,+4.9]$                                                 & $[-4.8,+4.8]$                                                   & $[+1.4,+1.9]$                                                          & $[-5.5,+3.9]$                                         \\

\hline
$\langle A_{K_{13}}^{e} \rangle\:/\:P_{\Lambda_c}$ & $\text{low } \phi$  & $[-6.2,+6.2]$                 & $[-2.7,+2.7]$                         & $[-2.5,+2.5]$                                                 & $[-6.2,+6.2]$                                                   & $[-3.0,+0.89]$                                                         & $[-4.5,+4.4]$                                         \\
 $\langle A_{K_{13}}^{e} \rangle\:/\:P_{\Lambda_c}$ & $\text{high } \phi$ & $[-4.6,+4.6]$                 & $[-2.1,+2.1]$                         & $[-1.8,+1.8]$                                                 & $[-4.7,+4.7]$                                                   & $[-1.8,+1.1]$                                                         & $[-3.2,+3.4]$                                         \\
 $\langle A_{K_{13}}^{e} \rangle\:/\:P_{\Lambda_c}$ & $\phi$             & $[-3.8,+3.8]$                 & $[-1.7,+1.7]$                         & $[-1.5,+1.5]$                                                 & $[-3.8,+3.8]$                                                   & $[-1.9,+0.39]$                                                         & $[-3.1,+2.2]$                                         \\

 \hline
$\langle A_{K_{22}}^{e} \rangle\:/\:P_{\Lambda_c}$ & $\text{low } \phi$  & $0$                              & $[-2.2,+2.2]$                         & $[-1.6,+1.6]$                                                 & $[-1.5,+1.5]$                                                   & $[-1.6,+1.6]$                                                         & $0$                                                      \\
 $\langle A_{K_{22}}^{e} \rangle\:/\:P_{\Lambda_c}$ & $\text{high } \phi$ & $0$                              & $[-1.6,+1.6]$                         & $[-1.2,+1.2]$                                                 & $[-1.1,+1.1]$                                                   & $[-1.2,+1.2]$                                                         & $0$                                                      \\
 $\langle A_{K_{22}}^{e} \rangle\:/\:P_{\Lambda_c}$ & $\phi$             & $0$                              & $[-1.4,+1.4]$                         & $[-0.97,+0.97]$                                                 & $[-0.94,+0.94]$                                                   & $[-0.96,+0.96]$                                                         & $0$                                                      \\

 \hline
$\langle A_{K_{24}}^{e} \rangle\:/\:P_{\Lambda_c}$ & $\text{low } \phi$  & $[-14.3,+14.3]$               & $[-4.3,+4.3]$                         & $[-7.2,+7.2]$                                                 & $[-12.9,+13.0]$                                                 & $[-6.0,+0.57]$                                                         & $[-10.4,+10.1]$                                       \\
 $\langle A_{K_{24}}^{e} \rangle\:/\:P_{\Lambda_c}$ & $\text{high } \phi$ & $[-10.6,+10.6]$               & $[-3.3,+3.3]$                         & $[-5.2,+5.2]$                                                 & $[-9.8,+9.8]$                                                   & $[-3.3,+1.4]$                                                         & $[-7.3,+7.8]$                                         \\
 $\langle A_{K_{24}}^{e} \rangle\:/\:P_{\Lambda_c}$ & $\phi$             & $[-8.8,+8.8]$                 & $[-2.7,+2.7]$                         & $[-4.4,+4.4]$                                                 & $[-8.0,+8.0]$                                                   & $[-3.6,+0.003]$                                                         & $[-7.1,+5.1]$                                         \\

 \hline
\end{tabular}
\end{adjustbox}
\end{table}

To illustrate the reach of a future measurement of the null tests for polarized $\Lambda^+_c \to p\ell^+\ell^-$ decays we evaluate the null tests for fixed Wilson coefficients that fulfill the available constraints and vary the strong phases $\delta_\mathrm{R}$ and hadronic parameters $a_{\mathrm{R}}$ as the main source of uncertainty. The results are shown in Table~\ref{tab:null_tests_NP} for  
$\ell=e$ (top) and $\ell=\mu$ (bottom) and 
three $q^2$ bins around the $\phi$ resonance following Ref.~\cite{LHCb:2025bfy} as we expect here the best sensitivity for a first measurement. 
The $\sqrt{q^2}$ regions low $\phi$, high $\phi$ and $\phi$ are defined as $[979.46,1019.46]\mev$, $[1019.46,1059.46]\mev$ and $[979.46,1059.46]\mev$, respectively.  Differences in the predictions between dielectron and dimuon modes arise solely from the allowed ranges of the Wilson coefficients, which are larger for electrons than for the muons.
We factor out the $\Lambda^+_c$-polarization in  the null tests that are only available in the polarized case. 
We find the largest values for the polarization-specific null tests for $\ell=\mu$ as $\langle A_{K_{24}}^{\mu}\rangle_{\text{low }\phi}$ within $\left[-5.6,5.6\right]\cdot P_{\Lambda_c}\,\%$ in the NP scenario $\mathcal{K}^{\mu\mu}_{9}=-\mathcal{K}^{\mu\mu}_{10}=0.8$.
For $\ell=e$ maximum values are obtained in the   scenario $\mathcal{K}^{ee}_{10}=2.05$ with
$\langle A_{K_{24}}^{e}\rangle_{\text{low }\phi}$ within $\left[-14.3,14.3\right]\cdot P_{\Lambda_c}\,\%$,
larger than the reach for the muons due to weaker bounds in the electron sector (\ref{eq:bound_KLR}).

Additionally, $\langle A_{K_{21}}^\ell\rangle$ is a test of NP in dipole operators. 
We obtain the largest values for $\mathcal{K}_7=0.24$ with $\langle A_{K_{21}}^e\rangle_{\text{low }\phi} = \left[-0.82,0.82\right]\cdot P_{\Lambda_c}\,\%$. 
As already anticipated, this is smaller than some of the other, GIM-protected  asymmetries due to stronger constraints on the dipole coefficients from $D \to \rho \gamma$ and $\Lambda^+_c \to p \ell^+ \ell^-$ \cite{Gisbert:2024kob}. 
An additional suppression arises because the angular observable $K_{21}$ vanishes in the low-recoil limit exploiting endpoint relations (\ref{eq:EPrelations}) \cite{Blake:2017une}.

\subsection{New Physics Correlations \label{sec:NP}}

\begin{figure}
    \centering
    \includegraphics[width=0.7\linewidth]{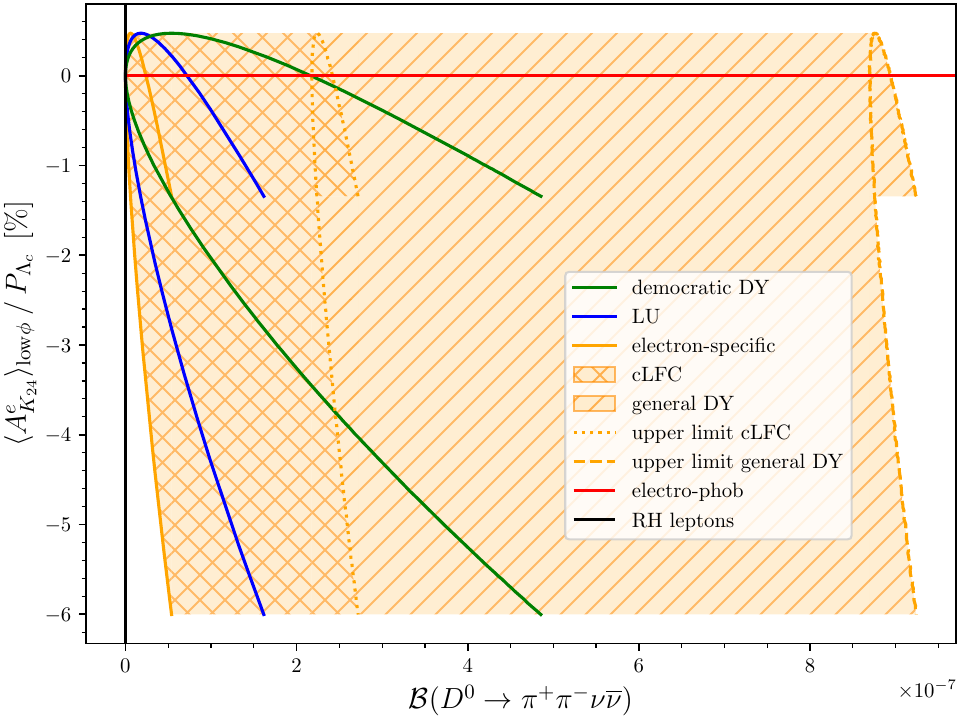}
    \caption{Correlation between $\mathcal{B}(D^0\to \pi^+\pi^-\nu\overline{\nu})$ and the angular observable $\langle A^e_{K_{24}}\rangle_{\text{low }\phi} \:/\: P_{\Lambda_c}$ in $\Lambda^+_c \to p e^+ e^-$ decays for fixed strong phases $(\delta_\rho = 0.0,\,\delta_{\omega-\rho}=6.0,\,\delta_{\phi-\rho}=5.7)$ and real Wilson coefficients within allowed constraints. We use  non-zero $\mathcal{K}_R^U$ and $\mathcal{C}_R^U$ to highlight correlations via Eq.~\eqref{eq:WET_correlations}. 
    The correlation is one-to-one for electron-specific (solid orange), LU (blue) and democratic (green) flavor scenarios, and gives a band for cLFC and general flavor structure with upper limits in dotted and dashed, respectively. 
    The lower limit is the one of the electron-specific scenario. An electro-phob scenario is shown in red (horizontal line), and one which couples only to RH leptons  in black (vertical line) to illustrate complementarity in NP reach. }
    \label{fig:correlations_A_K24ee_low_phi}
\end{figure}

In NP scenarios featuring predominantly operators with  a left-handed lepton  and right-handed quark current, that is, with  $\mathcal{K}_9^{\prime\,\ell\ell} = - \mathcal{K}_{10}^{\prime\,\ell\ell}$,  or equivalently $\mathcal{K}_R$, correlations exist between dineutrino modes and charged lepton modes of the same quark flavors via Eq.~\eqref{eq:WET_correlations}. To demonstrate this we show in Fig.~\ref{fig:correlations_A_K24ee_low_phi} $\langle A_{K_{24}}^{e}\rangle_{\text{low }\phi}$ in $\Lambda^+_c \to p e^+ e^-$ decays against $\mathcal{B}\left(D^0\to\pi^+\pi^-\nu\overline{\nu}\right)$.
We use the data-driven form factors for the latter and fixed values for the hadronic inputs.
Specifically,  we use central values for form factors and  fixed values of the other main source of theoretical uncertainty,
the strong phases in $\Lambda^+_c\to p\ell^+\ell^-$ decays. We choose values of the phases that maximize the NP contribution to the angular observable.
We further assume real Wilson coefficients.

As detailed in Sec.~\ref{sec:EFT} we consider different lepton flavor scenarios to compute 
the contributions to the dineutrino modes. For lepto-specific (orange), LU (blue) and democratic (green) a given dineutrino branching fraction 
predicts  the angular polarization asymmetry, and vice versa.
This is different for cLFC (cross pattern) and general (slash pattern) flavor scenarios, which allow for independent contributions to other lepton flavors, which affect only the dineutrino mode.
For these flavor scenarios  a fixed value of $\langle A^e_{K_{24}}\rangle_{\text{low }\phi}$ implies a range for the branching fraction $\mathcal{B}\left(D^0\to\pi^+\pi^-\nu\overline{\nu}\right)$.
The lower limit of these ranges is identical in both cases to the correlation of the lepto-specific case and the upper limit
is indicated by the dotted and dashed line, respectively.
Additionally, we show some rather trivial scenarios to highlight the complementarity in the NP reach between the two modes.
First, an electro-phob scenario (red), featuring only a signal in $D^0\to\pi^+\pi^-\nu\overline{\nu}$, secondly a scenario with NP that only involve right handed lepton (black) via $\mathcal{K}_9^{ee}=\mathcal{K}_{10}^{ee}$, thus only giving a signal for $\Lambda^+_c\to pe^+e^-$.  

\begin{figure}
    \centering
    \includegraphics[width=0.7\linewidth]{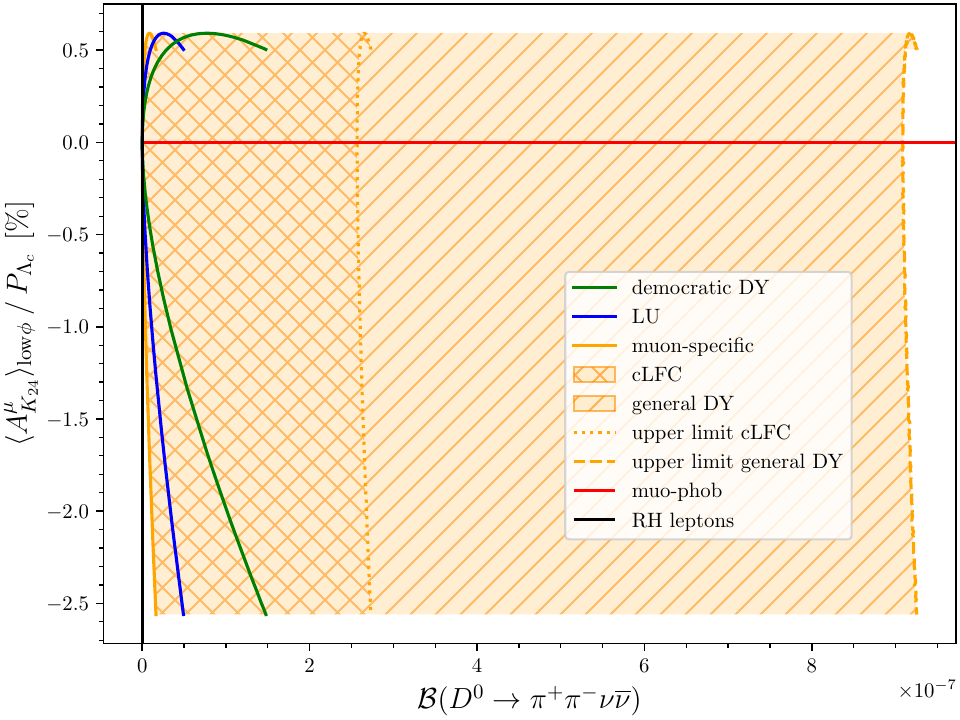}
    \caption{$\mathcal{B}(D^0\to \pi^+\pi^-\nu\overline{\nu})$ against $\langle A^\mu_{K_{24}}\rangle_{\text{low }\phi} \:/\: P_{\Lambda_c}$
    in  $\Lambda_c^+ \to p \mu^+ \mu^-$ decays, see Fig.~\ref{fig:correlations_A_K24ee_low_phi}. Here the angular observable is maximized by the strong phases $\delta_\rho=5.6$, $\delta_{\omega-\rho}=2.5$ and $\delta_{\phi-\rho}=6.3$.}
        \label{fig:correlations_A_K24mumu_low_phi}
\end{figure}

In Fig.~\ref{fig:correlations_A_K24mumu_low_phi} we show analogous correlations and complementarity between  dineutrino and dimuon modes.
The asymmetry in $\Lambda_c^+$-decays cannot be as large as the one for dieleectrons, as  NP with muons is presently subject to stronger experimental constraints.

\section{Experimental sensitivity\label{sec:exp}}
In the following, we estimate the sensitivity to search for the decay \Dtoppnn and to measure null-test asymmetries in polarized \Lctopmm decays at future $e^+e^-$ colliders operating at center-of-mass energies close to the $Z$-pole mass. We use FCC-ee with $6 \times 10^{12}$ produced $Z$ bosons as a reference, although the obtained results could be extended to similar facilities, such as CEPC. Further technical information about the analysis is available in Ref.~\cite{hacheney_2025_e3nw3-fx653}. 

\subsection{Simulated data and detector model} \label{sec:experiment}
The event generator \texttt{PYTHIA8}~\cite{Pythia8_reference} simulates the $e^+e^-$ collisions at a center-of-mass energy $\sqrt{s}=91\gev$ and handles the hadronization of outgoing quarks. For background studies, we simulate inclusive $Z\to q\bar{q}$ decays ($q = u,d,s,c,b$) and $Z\to \tau^+\tau^-$ decays. The latter are relevant for the \Dtoppnn search because neutrinos from $\tau$ decays can mimic the missing momentum signature of the signal decays. Unstable particles are decayed using \texttt{PYTHIA8} or \texttt{EvtGen}~\cite{EvtGen_reference} with branching fractions taken from available experimental measurements or theory predictions~\cite{Belle2DecayFile}. Final-state radiation is generated using \texttt{PHOTOS}~\cite{Davidson:2010ew}. The signal decays \Dtoppnn and \Lctopmm are simulated in $Z\to c\bar{c}$ events with \texttt{EvtGen}, assuming a flat distribution of final-state particles in phase space (PHSP). We generate $1\times 10^{8}$ events for each background process, and $1\times 10^{6}$ for each of the two signal processes. The background processes are then merged according to their expected rates~\cite{PDG2024}.

\begin{figure}[tb]
\centering
\subfloat{{\includegraphics[width=0.5\textwidth]{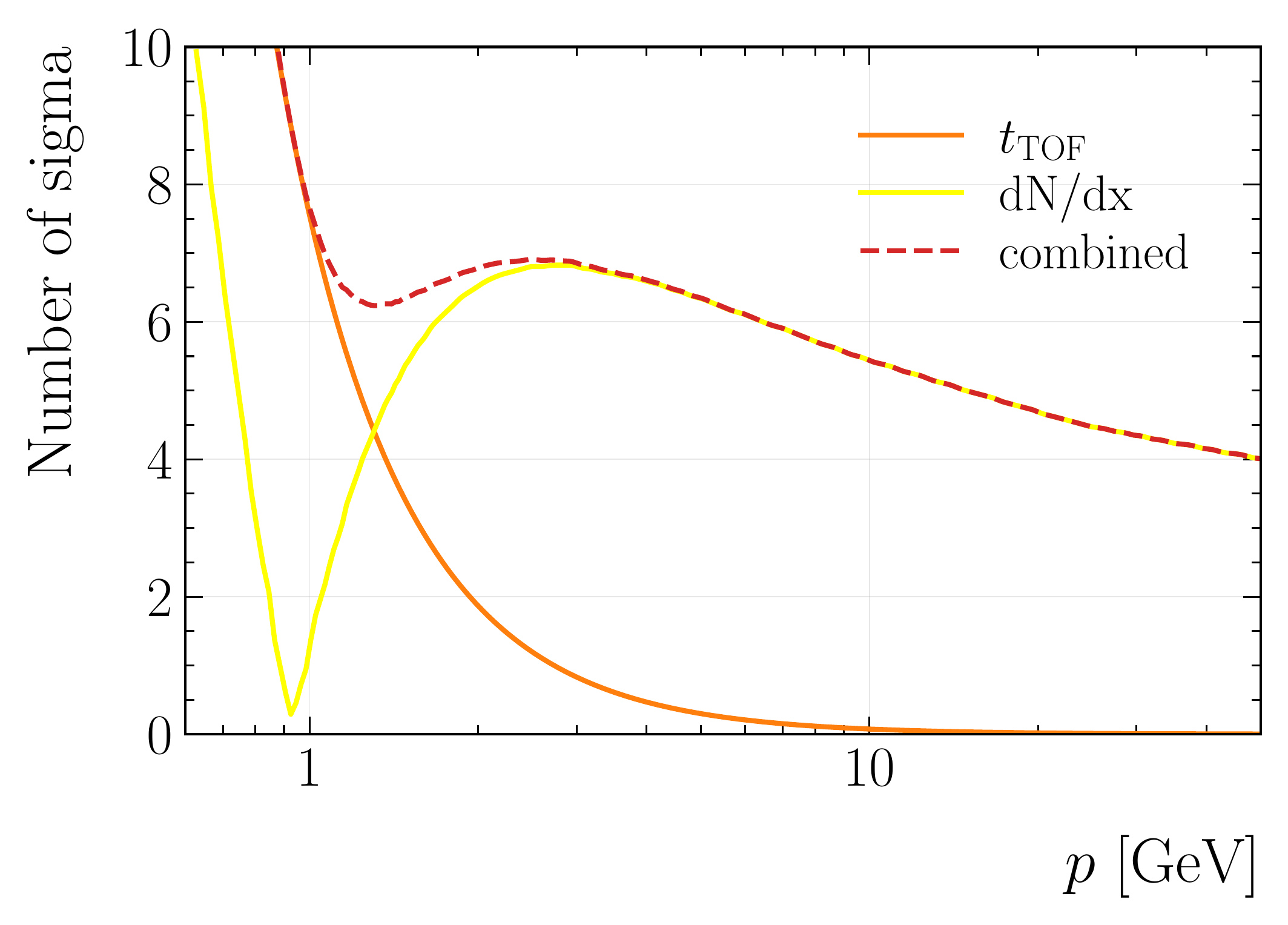} }}%
\subfloat{{\includegraphics[width=0.5\textwidth]{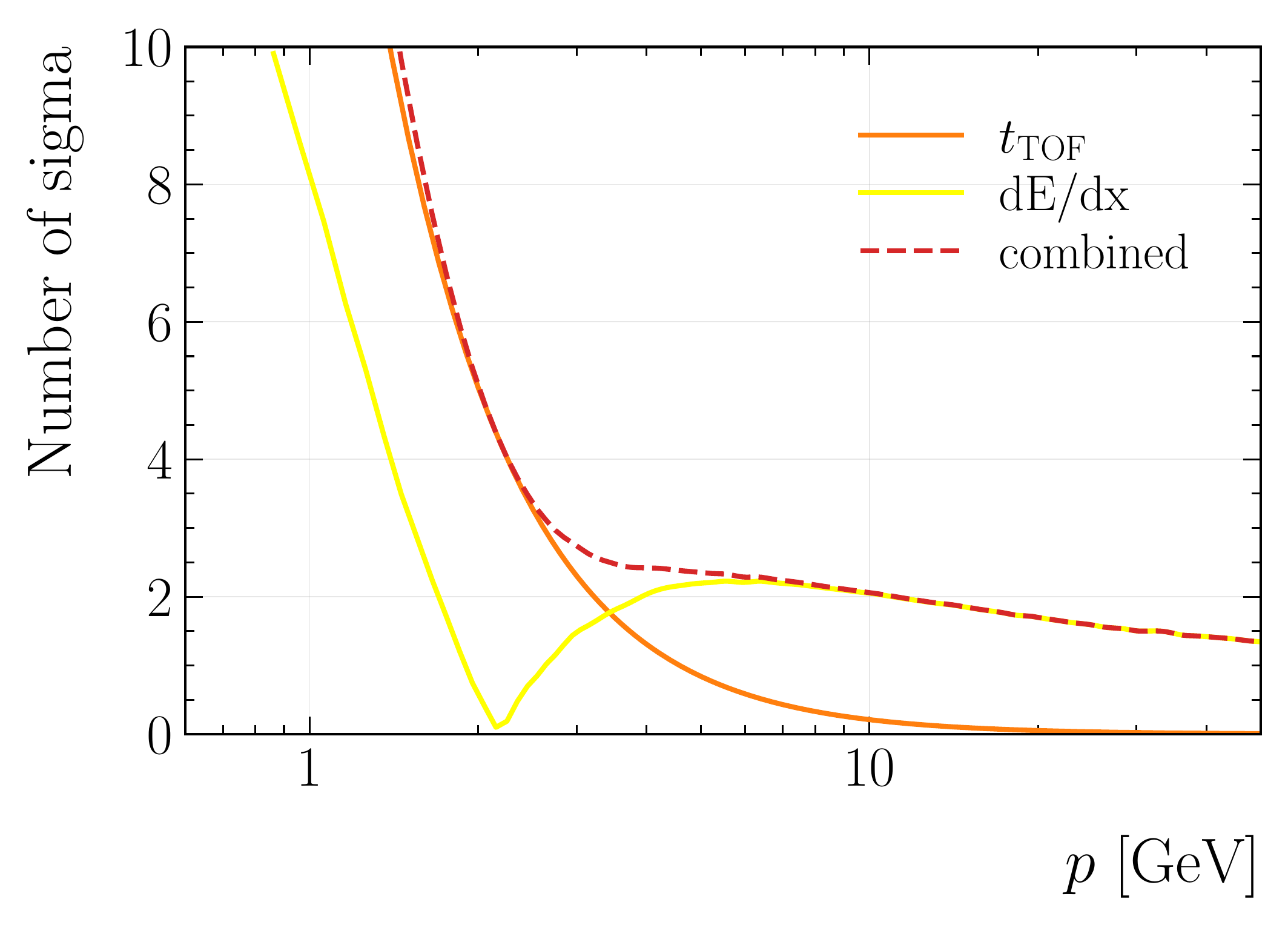} }}%
\caption{Assumed separation power (in units of Gaussian standard deviations) between (left) pions and kaons and (right) kaons and protons using time-of-flight ($t_{\rm TOF}$), drift-chamber cluster-counting ($dN/dx$), and energy-loss ($dE/dx$) measurements within the IDEA detector concept.}\label{fig:PID-performance-FCC}
\end{figure}

A parametric model of the detector, based on the \texttt{DELPHES} software package~\cite{Delphes_reference} and the Innovative Detector for Electron-positron Accelerators (IDEA) concept~\cite{IDEA_detector,FCC:2025lpp}, is used in the study. IDEA features a cylindrical geometry with silicon vertex detectors close to the beam pipe, a drift chamber, and a layer of silicon micro-strip detectors placed into a \SI{2}{\tesla} solenoid, for quasi-continuous tracking of charged particles. The tracking system is surrounded by a  dual-readout calorimeter and by muon chambers. Identification of charged hadrons is performed by combining information from drift-chamber cluster-counting and energy-loss measurements with time-of-flight information from the silicon micro-strip detector. The assumed particle-identification performance relies on estimates from Ref.~\cite{FCC_ee_the_lepton_collider}, with kaon-to-pion and kaon-to-proton separations as reported in Fig.~\ref{fig:PID-performance-FCC}. Identification of muons relies also on information from the muon chambers. In line with previous FCC-ee studies of semileptonic rare $B$ decays~\cite{FCCee_ESPPU}, we assume an estimated pion-to-muon misidentification probability of 0.4\%. 

\subsection{Search for \texorpdfstring{\Dtoppnn}{D0 -> pi pi nu nubar} decays\label{sec:analysis_D2hhnunu}}
In this section, we estimate the sensitivity to search for \Dtoppnn decays at FCC-ee. The event selection performance, and hence the sensitivity, is expected to depend on final-state kinematics and thus on the unknown decay model. To account for this, we weight the simulated signal candidates as a function of $p^2$ and $q^2$ to reproduce each of the three decay models described in Sec.~\ref{sec:pipinunu_theo}. Results are presented separately for each model and without any weighting, \textit{i.e.}, using the PHSP model employed in generation.

\subsubsection{Candidate reconstruction and selection}
Signal candidates are reconstructed from pairs of oppositely charged pions that are significantly displaced from the $e^+e^-$ interaction vertex (PV). The PV is determined from an iterative fit to the reconstructed charged particles in the event. The fit constrains the PV position to be within the $e^+e^-$ luminous region. At each iteration, charged particles that increase the PV fit $\chi^2$ by more than 25 are treated as inconsistent with originating from the primary interaction and removed from the next iteration until no charged particles are removed. Charged particles that are removed from the PV fit are considered as candidate pions to form \Dtoppnn decays. These pion candidates include both correctly identified true pions and kaons that have been misidentified as pions with misidentification rates derived from Fig.~\ref{fig:PID-performance-FCC}. Misidentification of other particle species into pions is assumed to be negligible. 

Pairs of pion candidates are formed by combining oppositely charged pions each having a distance of closest approach (impact parameter) smaller than 1.0\mm or 2.0\mm. The pion pairs are fitted to a common vertex. Vertices are required to satisfy fit quality requirements and are rejected if they form, in combination with other charged particles in the event, alternative secondary vertices meeting the same quality criterion. This procedure removes background from multibody decays involving three or more charged particles. The remaining two-pion vertices are required to have masses smaller than 1.8\gev, opening angles smaller than 0.9\rad, energies greater than 1.0\gev, impact parameters smaller than 1.0\mm, and flight distances smaller than 1.0\cm. 

The presence of two neutrinos leads to a large non-reconstructed (missing) momentum in the event. The missing momentum $\vec{p}_\mathrm{miss}$ is computed from the momenta, $\vec{p}$, of the reconstructed particles as
\begin{equation}
    \vec{p}_\mathrm{miss} = -\!\!\!\sum_{i = 1}^{n_\text{particles}}\!\!\vec{p}_i\,,
\end{equation}
where $i$ runs over all reconstructed charged and neutral particles of the event. Candidate \Dtoppnn decays are formed by combining the $\pi^+\pi^-$ candidates and the missing momentum. Only $D^0$ candidates with missing momenta larger than 1.2\gev and with masses smaller than 3.0\gev are considered. The difference between the impact parameters of the $\pi^+\pi^-$ system and of the $D^0$ candidate is required to be larger than $-0.5\mm$. The direction angle is defined as the angle between the momentum and the vector connecting the primary and secondary vertices. The difference between the direction angle of the $\pi^+\pi^-$ system and of the $D^0$ candidate must be larger than $-0.5$\rad. The corrected mass,
\begin{equation}
    m_{\mathrm{corr}} = \sqrt{m^2(\pi^+\pi^-) + p^2_{\perp}(\pi^+\pi^-)} + p_{\perp}(\pi^+\pi^-),
\end{equation}
with $p_{\perp}(\pi^+\pi^-)$ being the component of the dipion momentum perpendicular to the $D^0$ flight direction, partially accounts for the missing neutrinos and is required to be larger than 0.6\gev to efficiently suppress backgrounds from fully reconstructed decays of neutral kaons.

To suppress background from $Z\to\tau^+\tau^-$ decays, the number of reconstructed particles in the event must be larger than 15, and the number of reconstructed charged particles smaller than 12. In the $e^+e^-$ center-of-mass frame, quarks and leptons from $Z$ decays are produced in opposite directions, with the decay products of one quark or lepton isolated from those of the other and contained in opposite hemispheres. The boundary between the hemispheres is experimentally defined by the plane perpendicular to the thrust vector $\hat{t}$ that maximizes the value
\begin{equation} \label{eq:thrust}
     \frac{\sum_i |\vec{p}_i \cdot \hat{t}|}{\sum_i |\vec{p}_i |}\,,
\end{equation}
where the index $i$ runs over all reconstructed charged and neutral particles of the event. To suppress semileptonic charm- and bottom-hadron decays, no muons or electrons must be reconstructed in the signal hemisphere. Requirements on the outputs of a jet-flavor tagger~\cite{Bedeschi_2022}, which provides probabilities for jets resulting from $Z\to q\bar{q}$ decays to originate from a given flavor $q=u/d,s,c,b$, are used to suppress events inconsistent with $Z\to c\bar{c}$ decays.

At this stage of the selection, signal \Dtoppnn decays are selected with approximately $46.0\%$ efficiency, with loss of efficiency mainly due to requiring the $\pi^+\pi^-$ vertex to be well separated from the primary vertex. Background events originating from $Z \to q\bar{q}$ with $q\neq c$ and $Z\to\tau^+\tau^-$ decays are suppressed by more than $99.8\%$. To further suppress background events from $Z\to c\bar{c}$ decays, which at this stage is selected with $19.0\%$ efficiency, we use a classifier based on a boosted decision tree (BDT) with adaptive boosting~\cite{Breiman,Roe,FREUND1997119}. To maximize signal-to-background separation, we input to the BDT a set of variables that describe the event topology and kinematics such as the number of charged particles used in the PV fit and their mass; the number of reconstructed particles, charged particles, muons and electrons in the events; the number of 
secondary vertices in the event; the number of charged particles reconstructed within a 0.5-radius cone around the $\pi^+\pi^-$-momentum direction; the number of $\pi^0\to\gamma\gamma$ candidates reconstructed within a 0.4-radius cone around the $\pi^+\pi^-$-momentum direction; the $\pi^+\pi^-$ energy; the $D^0$ flight distance; the difference between the impact parameters of the $\pi^+\pi^-$ system and of the $D^0$ candidate; the total energy of the event and the total energy from the reconstructed charged particles; the missing momentum; the difference between the masses of the $\pi^+\pi^-$ system and of the $D^0$ candidate; the corrected mass; the asymmetry in the energies of the hemispheres, computed using all reconstructed particles; the cosine of the angle between the $\pi^+\pi^-$ momentum and the thrust axis; and combinations of the jet-flavor probabilities. To suppress background from exclusive $D^0 \to K^0_{S,L}\pi^+\pi^-$ and other multibody charmed-hadron decays, we input to the BDT also the imbalance between the $\pi^+$ and $\pi^-$ momenta, computed as $[p(\pi^+)-p(\pi^-)]/[p(\pi^+)+p(\pi^-)]$, and the mass of $K^0_{S,L}\pi^+\pi^-$ candidates reconstructed by combining the dipion system with $K^0_{S,L}$ candidates found in its vicinity. For validation purposes, the simulated data are split into training (two-thirds) and testing (one-third) subsets and agreement between the samples is found.

\begin{figure}[tb]
    \centering
    \includegraphics[width=0.55\textwidth]{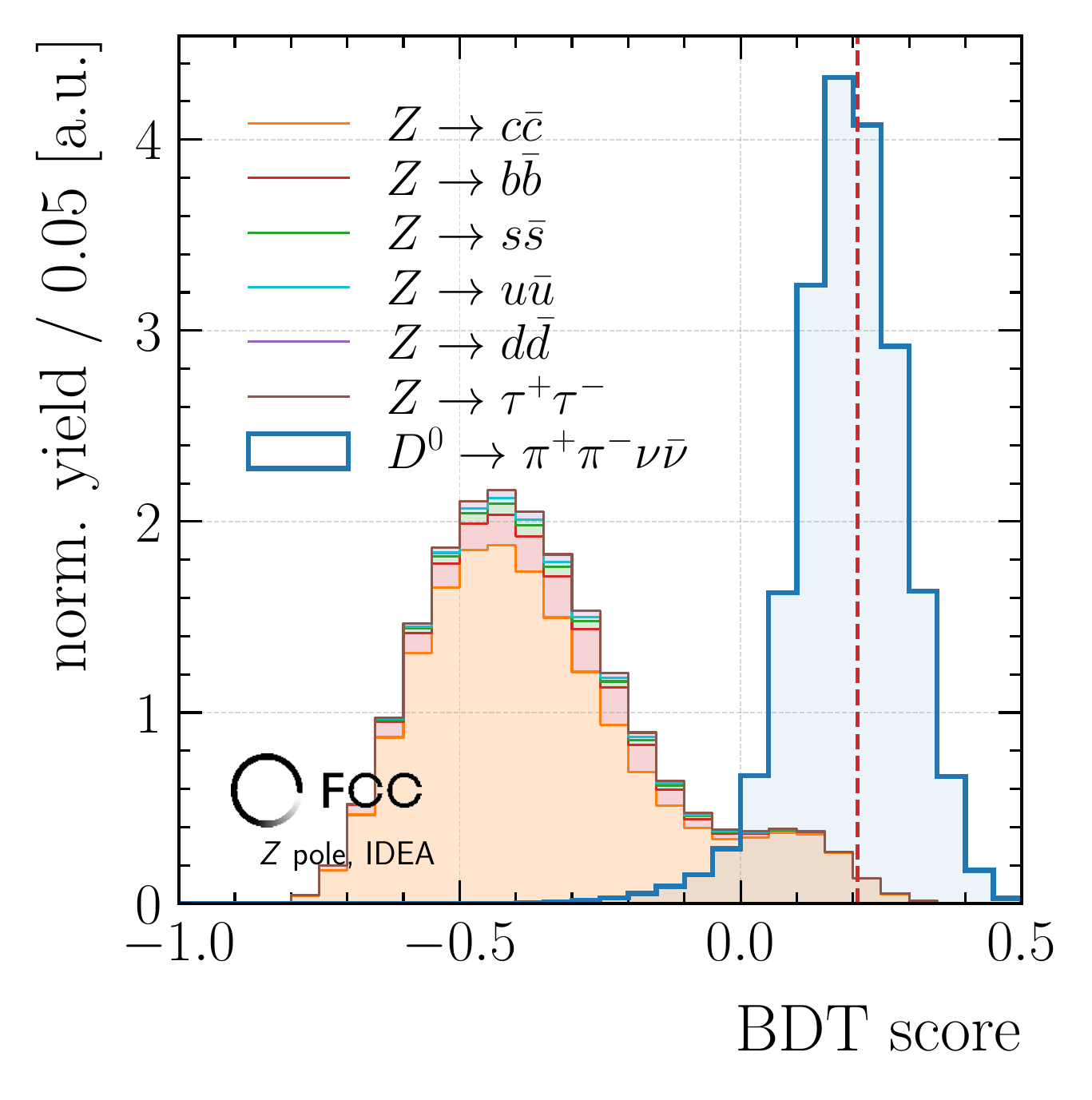}
    \caption{Distributions of the BDT output for signal generated with a PHSP decay model, and for the sum of the background processes. The red dotted line represents the optimal lower threshold used in the analysis.}
    \label{fig:BDT1_BDT2_score}
\end{figure}

The BDT-output distribution is shown in Fig.~\ref{fig:BDT1_BDT2_score}. A lower threshold on the BDT output is chosen by maximizing the figure of merit $\epsilon_{\mathrm{BDT}}/(1.5 + \sqrt{B})$, where $\epsilon_{\mathrm{BDT}}$ and $B$ are the signal efficiency and the expected background yield for a given BDT requirement, respectively. In comparison to other frequently used figures of merit, this metric has the advantage of being independent of the signal rate~\cite{Punzi:2003bu}. The BDT requirement that maximizes the figure-of-merit depends on the assumed signal-decay model and has signal efficiencies in the range 38--55\% for corresponding background rejections of 98--99\%. After the BDT requirement, the remaining background originates predominantly from events containing $D^0\to K^0_{S,L}\pi^+\pi^-$ decays where the $K^0_{S,L}$ meson is not reconstructed, other partially reconstructed multibody decays of charmed hadrons, and three-prong decays of $\tau$ leptons.

\subsubsection{Sensitivity estimation\label{sec:dtopnn-sensitivity}}
The measurement sensitivity (in units of Gaussian standard deviations $\sigma$) is estimated from the expected signal ($S$) and background ($B$) yields as $S/\sqrt{S+B}$, without accounting for systematic uncertainties or for the assumed background models used in generation.\footnote{The impact of these effects is beyond the scope of this study, and is not expected to qualitatively change its conclusions.}
The signal yield is computed as a function of the signal branching fraction $\mathcal{B}(\Dtoppnn)$ as
\begin{equation}
 S = 2 \ N_Z \ \mathcal{B}(Z \to c\bar{c}) \ f_{D^0} \ \mathcal{B}(\Dtoppnn)\ \epsilon,
\end{equation} \label{eq:signal_estimation}
where the factor 2 accounts for the production of both a $c$ and $\bar{c}$ quark in the $Z$ boson decay, $N_Z=6\times10^{12}$ is the assumed number of produced $Z$ bosons, $\mathcal{B}$ indicates the branching fraction for the given process~\cite{PDG2024}, $f_{D^0} = 0.59 $ is the charm fragmentation fraction into a $ D^0 $ meson~\cite{Lisovyi_2016}, and $\epsilon$ is the signal efficiency. The background yield is estimated by summing over all considered background processes as
\begin{equation}
B = N_Z \sum_f \mathcal{B}(Z \to f\bar{f}) \ \epsilon_{f\bar{f}}\,,
\end{equation} \label{eq:bkg_estimation}
where $f=u,d,s,c,b,\tau$.

\begin{figure}[tb]
\centering
\includegraphics[width=0.8\textwidth]{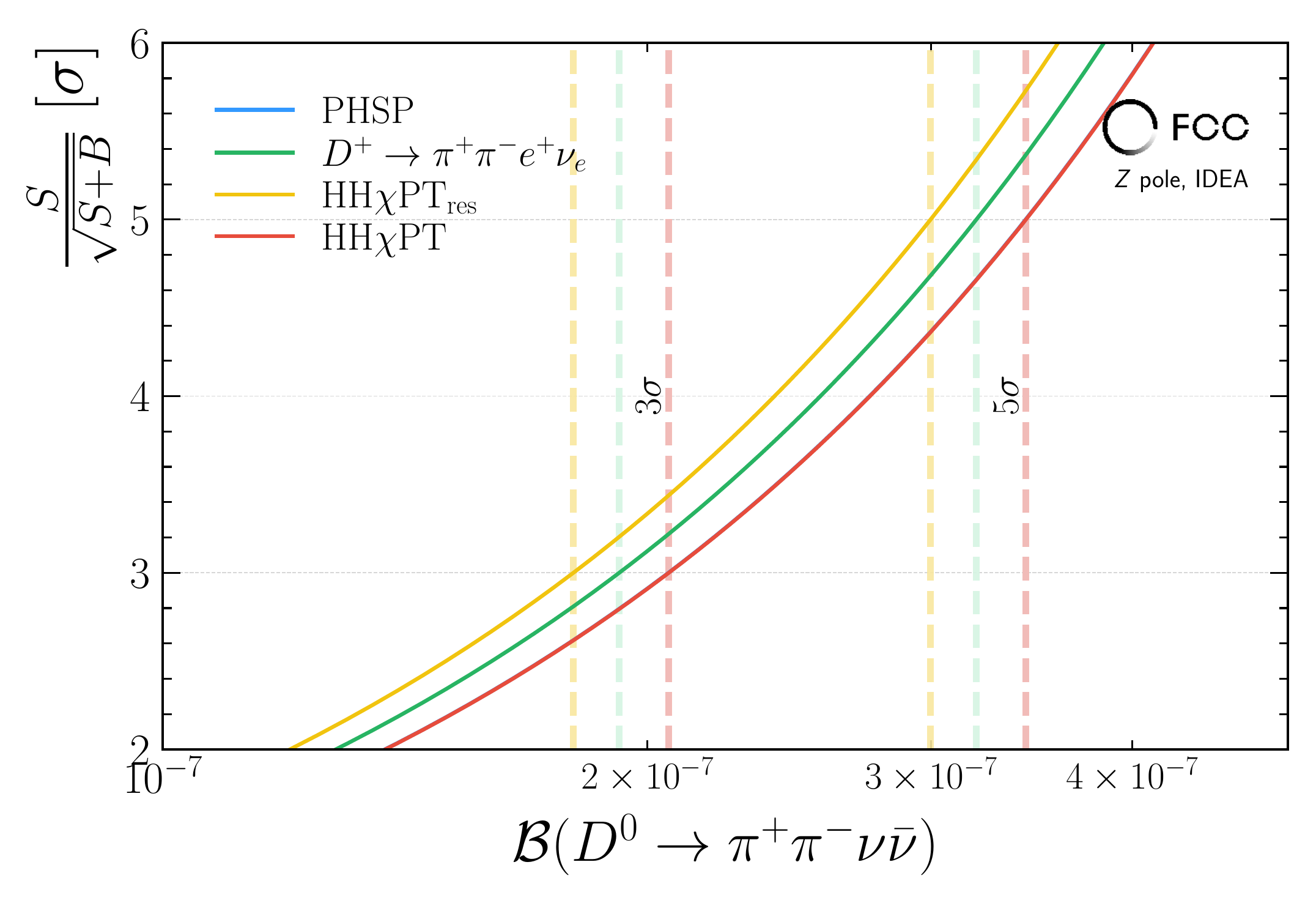}\\
\caption{Sensitivity as a function of the \Dtoppnn branching fraction for different decay models, with $3\sigma$ and $5\sigma$ thresholds. Notice that the curves corresponding to the PHSP and HH$\chi$PT models overlap almost exactly.\label{fig:sensitivity}}
\end{figure}

The sensitivity as a function of the assumed signal branching fraction $\mathcal{B}(\Dtoppnn)$, and for each decay–model hypothesis, is shown in Fig.~\ref{fig:sensitivity}. Significances of 3 and $5\sigma$ can be reached for branching fractions as small as $1.8\text{--}2.1\times 10^{-7}$ and $2.9\text{--}3.4\times 10^{-7}$, respectively, depending on the assumed decay model. These results would represent an improvement of roughly three orders of magnitude over the current experimental limits on $c\to u\nu\bar{\nu}$ transitions~\cite{BESIII:2021slf,Belle:2016qek}. They would probe the parameter space allowed by NP models and constrain their lepton flavor structure, as discussed in Sec.~\ref{sec:pipinunu_theo}. 
While experiments at the (HL-)LHC are not well suited to search for dineutrino final states, current flavor experiments at $e^+e^-$ colliders and their future upgrades (Belle II, BESIII, STCF) are expected to improve the current limits before FCC-ee starts taking data. These experiments, however, will access much smaller samples of $c\bar{c}$ events and are not expected to achieve similar sensitivities as FCC-ee.

\subsection{Angular asymmetries in polarized \texorpdfstring{\Lctopmm}{Lambdac -> p mu mu} decays\label{sec:analysis_Lc2pmumu}}
As discussed in Sec.~\ref{sec:l2pmumu_theo}, a future $e^+e^-$ collider would enable access to a range of angular observables in semileptonic rare $\Lc \to  p \ell^+ \ell^-$ decays originating from polarized \Lc baryons, out of which many are null tests of the SM that can also be correlated with the \Dtoppnn branching fraction. In this section, we estimate the sensitivity to a generic null-test asymmetry in \Lctopmm decays, taken as a proxy for future measurements of angular observables. We briefly outline a candidate reconstruction and selection to determine expected signal and background yields.

\subsubsection{Candidate reconstruction and selection}
Candidate \Lctopmm decays are formed by combining candidate protons with transverse momenta larger than 4.0\gev and IP larger than 2.5\mum, with pairs of oppositely charged candidate muons each having transverse momentum larger than 300\mev. Candidate protons include correctly identified true protons and kaons that have been misidentified with rates derived from Fig.~\ref{fig:PID-performance-FCC}. Candidate muons include correctly identified true muons and pions that have been misidentified as muons with a rate of 0.4\%. Misidentification of other particles into either protons or muons is assumed to be negligible. The three final-state particles must form a good-quality vertex with an invariant mass, $m(p\mu^+\mu^-)$, in the range $[2.275,2.300]$\gev and a reconstructed energy above 20.0\gev. The \Lc candidate must have an IP smaller than 0.26\mm, a flight distance smaller than 2.8\mm, and a direction angle smaller than 2.8\rad. Background originating from $Z\to q\bar{q}$ decays with $q\neq c$ is suppressed using the output of the jet-flavor tagger and requirements on the number of reconstructed neutral and charged particles in the event and around the signal candidate. 
The selection efficiency is about $43.4\%$ for the signal decays and $\mathcal{O}(10^{-9})$ for background processes. After selection, the background consists mainly of randomly associated charged particles that accidentally fulfill the selection requirements, with one or two particles misidentified. The background appears to be smoothly distributed in $m(p\mu^+\mu^-)$, while the signal peaks at the known value of the \Lc mass with a resolution of about 2\mev.

\subsubsection{Sensitivity estimation}
The expected signal and background yields are evaluated following the same approach as in Sec.~\ref{sec:dtopnn-sensitivity}. The fragmentation of charm quarks into \Lc baryons is taken to be $0.06$~\cite{Lisovyi_2016}, and the \Lctopmm branching fraction is approximated as the sum of the branching fractions of the dominant intermediate hadronic resonances $R=\eta,\rho,\omega,\phi$ with subsequent decay of $R\to\mu^+\mu^-$, giving $\mathcal{B}(\Lctopmm) = 4.15\times 10^{-7}$~\cite{PDG2024}. This translates into an expected signal yield of approximately $1.5 \times 10^4$ for a sample of $6\times10^{12}$ $Z$ bosons. The expected background yield is $7 \times 10^3$.

\begin{figure}[tb]
\centering
\includegraphics[width=0.65\textwidth]{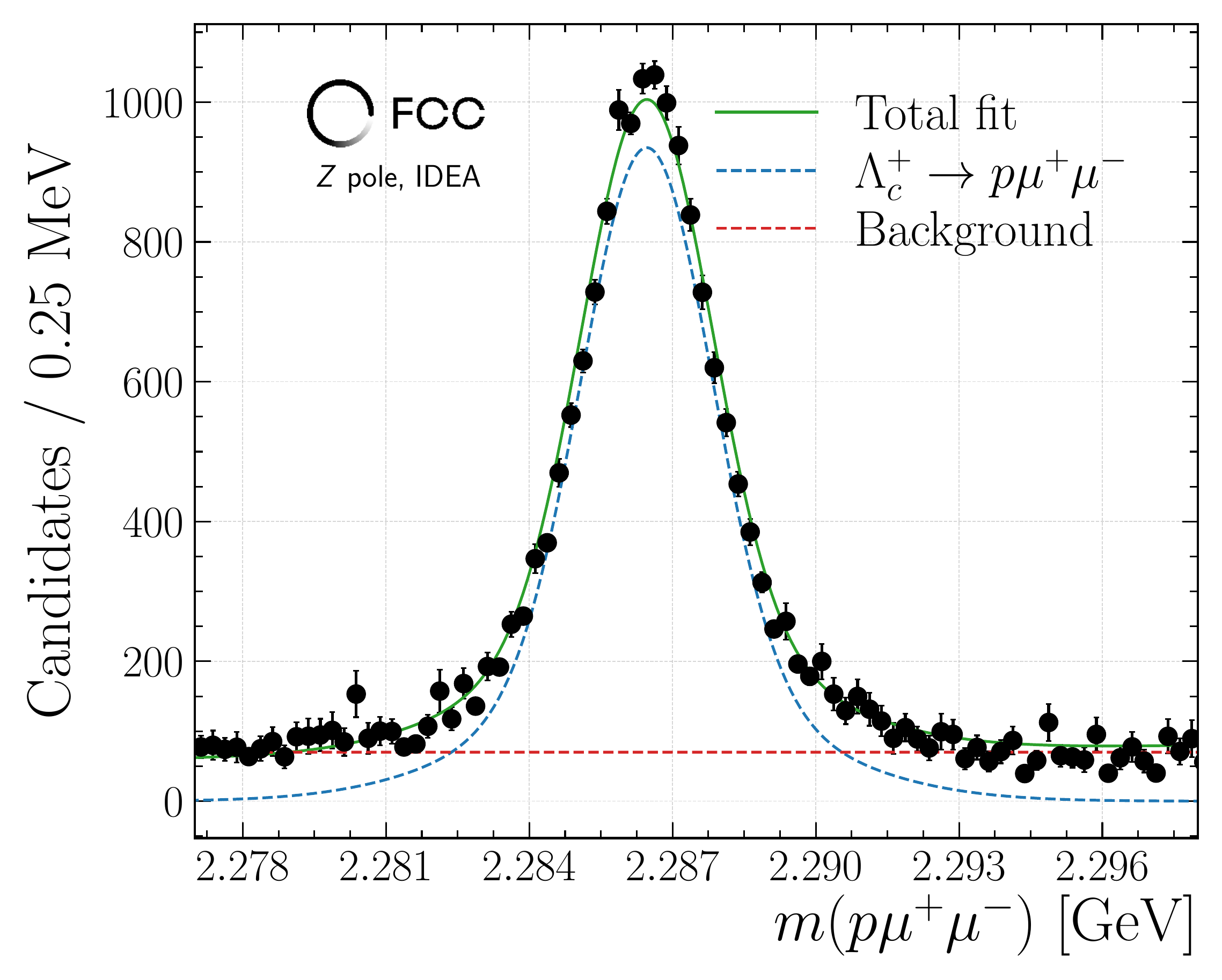} 
\caption{Mass distribution of \Lctopmm candidates for a simulated sample of $6\times10^{12}$ $Z$ bosons, with fit projections overlaid.}
\label{fig:Lc_fits}
\end{figure}

The target null tests can be determined experimentally as yield asymmetries between subsamples defined by angular observables (\textit{e.g.}, see Ref.~\cite{LHCb:2025bfy} for the measurement of the forward-backward asymmetry in unpolarized \Lctopmm decays). The sensitivity is evaluated using simulated data generated with the expected signal and background yields, and randomly divided into two halves to emulate a null-test asymmetry. An extended maximum-likelihood fit to $m(p\mu^+\mu^-)$ distributions of the two subsamples is performed with the asymmetry as a shared parameter (Fig.~\ref{fig:Lc_fits}). In the fit, the $m(p\mu^+\mu^-)$ distribution of the signal component is modeled as a double Gaussian function. The distribution of the background is modeled as an exponential function. The shape parameters are determined by the fit together with the signal and background yields and asymmetries. The estimated statistical uncertainty in the null-test asymmetry, when considering the full dimuon-mass range, is 0.9\%. It is important to note that the statistical uncertainty in an asymmetry measurement depends solely on the signal-to-background ratio and the value of the asymmetry itself, which is assumed to be zero. Consequently, the estimated sensitivity is representative of all null-test observables defined in Sec.~\ref{sec:l2pmumu_theo}.

Since the sensitivity to NP can be enhanced by measuring asymmetries as a function of dimuon mass, the study is repeated in the three dimuon-mass intervals, centered around the $\phi$-meson mass peak, defined in Sec.~\ref{sec:nulltest_pred} ($\text{low }\phi$, $\text{high }\phi$, and $\phi$). The expected yields in these intervals are estimated using the known branching fraction of $\Lc \to p \phi(\to \mu^+\mu^-)$ and the ratio of yields reported in Ref.~\cite{LHCb:2025bfy}. The resulting asymmetry measurements reach uncertainties of 1.5\% in the $\text{low } \phi$ region, and 1.3\% in the $\text{high } \phi$ region, while the combined $\phi$ region achieves a precision of 1.0\%. Systematic uncertainties are not evaluated in this study but are expected to be negligible compared to the statistical uncertainties for asymmetries of this type. The sensitivity can be compared to possible NP effects discussed in Table~\ref{tab:null_tests_NP}, which can be as large as $\sim 5 \% \cdot P_{\Lambda_c}$.

Dedicated studies of electron reconstruction and identification will be necessary to provide robust sensitivity estimates for $\Lc \to p e^+e^-$ decays and are left for future work. Nevertheless, preliminary studies with the IDEA detector concept indicate that the performance for dielectron final states will be comparable to that for dimuon modes~\cite{FCCee_ESPPU}. This is particularly promising given the weaker experimental bounds on Wilson coefficients in the dielectron channel, such that asymmetries can reach values exceeding 14\% (see Table~\ref{tab:null_tests_NP}).

These sensitivities can be directly compared to those expected at current flavor experiments and their future upgrades. After its second upgrade, LHCb is expected to achieve unprecedented precision in measurements of \Lctopmm decays with unpolarized \Lc baryons~\cite{LHCb:2018roe}. In contrast, FCC-ee offers complementary advantages for observables in \Lctopmm that rely on polarization, as inferred from signal yields in existing measurements of hadronic polarized \Lc decays from $b$ hadrons at LHCb~\cite{LHCb:2017hwf,LHCb:2022sck} that allow for naive scaling. Prospects are particularly strong in the dielectron channel. Other experiments (Belle II, BESIII, STCF) are not expected to make significant contributions to these measurements due to their limited luminosity or small polarization of produced \Lc baryons.

\section{Conclusions \label{sec:con}}

We explore the potential to probe BSM effects through rare charm decays at future $e^+e^-$ colliders operating at the $Z$ pole. 
We consider two decay modes that invite studies in null-test observables in a clean environment with high $c \bar c$ production rates. 

First, $D^0 \to \pi^+ \pi^- \nu \bar \nu$, resulting in a missing-energy signal,
with branching fractions induced by heavy NP as large as a few $\times 10^{-6}$ \cite{Bause:2020xzj}.
The theoretical description is rather simple as the neutrinos factorize in the decay amplitudes and only hadronic transition form factors matter.
Using different models for the hadronic matrix elements, including one obtained from recent $D^+ \to \pi^+ \pi^- e^+ \nu$ measurements, we find a spread in BSM predictions, see Table~\ref{tab:BR_upper_limits}. However, 
the order of magnitude remains intact, allowing to probe lepton-flavor patterns~\cite{Bause:2020auq}. As the $c \to u \nu \bar \nu$ transitions are SM null tests, an observation would provide a clear signal of NP. Improved knowledge of the form factors from lattice QCD or other means would allow more precise interpretation of experimental results in terms of BSM couplings. 

Second, we study the decay $\Lambda^+_c \to p \ell^+ \ell^-$, with $\ell=e,\mu$, taking advantage of the initial state polarization in $Z \to c \bar c$ production~\cite{deBoer:2017que}. We work out the phenomenology of angular asymmetries that become accessible with finite $\Lambda^+_c$ polarization, see Eqs.~(\ref{eq:afbs}). Focusing on dilepton mass regions near the $\phi$ peak, we find asymmetries catalyzed by resonances \cite{Fajfer:2012nr,Golz:2021imq} up to $\sim 5\%\cdot P_{\Lambda_c}$ ($\sim 14\%\cdot P_{\Lambda_c}$) for dimuon (dielectron) modes, see Table~\ref{tab:null_tests_NP}.
These observables are GIM-protected null tests of the SM \cite{Golz:2022alh} and their observation heralds NP.
Local $q^2$-dependent asymmetries can be even larger \cite{Golz:2021imq,Golz:2022alh}, but require matching experimental strategies.

The SMEFT framework allows to study correlations between the dilepton and dineutrino modes, illustrated in Fig.~\ref{fig:correlations_A_K24ee_low_phi}.
Synergies arise as some NP scenarios contribute to one mode but not the other. 
For instance, NP purely in LFV or $\tau$-couplings is not relevant for $\Lambda^+_c \to p \ell^+ \ell^-$ decays, whereas
NP in right-handed charged leptons does not induce dineutrino modes. Scenarios with left-handed electrons or muons show
correlations between the modes that provide information on additional flavor couplings.

Using the IDEA detector concept at FCC-ee with $6 \times 10^{12}$ produced $Z$ bosons as an example, we estimated sensitivities for $D^0 \to \pi^+\pi^-\nu\bar{\nu}$ and angular asymmetries in polarized $\Lambda^+_c \to p \mu^+\mu^-$. For the dineutrino study, branching fractions as low as $\sim 2 \times 10^{-7}$ can be evidenced at $3\sigma$, with $\sim 3 \times 10^{-7}$ potentially observable (Fig.~\ref{fig:sensitivity}), well within the range predicted by NP models. This enables the probe of 
lepton-flavor predictions, including cLFC, see Table \ref{tab:BR_upper_limits}. 
The sensitivity shows only mild dependence on the hadronic transversity form-factor model ($\sim 10\%$). 
For angular asymmetries in $\Lambda^+_c \to p \mu^+\mu^-$, we find that measurable asymmetries can be detected at the 1\% level in the dimuon mode. 
Dedicated studies of dielectron channels are left for future work, however, similar sensitivities are expected. Thus, possible NP effects in angular asymmetries are accessible to FCC-ee searches.

We further suggest performing an auxiliary measurement at the $Z$ pole of depolarization effects in $c \to \Lambda^+_c$ hadronization 
\cite{Falk:1993rf}, to validate estimates of 
$P_{\Lambda_c}$ \cite{deBoer:2017que}. This could be achieved using SM-dominated $\Lambda^+_c$ decays with minimal NP pollution, such as $\Lambda^+_c \to \Lambda \ell^+ \nu$, thereby supporting precision null-test analyses. Additional methods to measure the $\Lambda^+_c$  polarization at current LHC experiments can be found in Refs.~\cite{LHCb:2022sck,LHCb:2023crj,Galanti:2015pqa,Kats:2015zth}. These methods could also be used at future facilities.

To summarize, our results demonstrate strong potential for next-generation $e^+e^-$ colliders to lead searches for rare charm decays with invisible particles and precision measurements requiring polarization, significantly extending their physics reach.

\section*{Acknowledgments}
We thank Hector Gisbert and Tom Magorsch for useful exchanges on rare charm decays, and Felix Erben on lattice computations.  We thank the FCC-ee Physics Performance Group for insightful discussions and useful feedback regarding the analysis procedure, in particular Emmanuel Perez, Guy Wilkinson and Andrea Sciandra. D.M. and L.R. appreciate support by the DFG within the Emmy
Noether Program under grant number MI 2869/1-1.


\clearpage
\appendix

\section{\texorpdfstring{$\Lambda_c\to p$}{Lambdac to p} form factors}
\label{app:formfactors}

The $\Lambda_c\to p$ transitions are described by ten independent form factors.
We use the helicity-based definition \cite{Feldmann:2011xf} with the form factors $f_0,g_0,f_+,g+,f_{\bot},g_{\bot},h_+,h_{\bot},\tilde{h}_+$ 
and $\tilde{h}_{\bot}$ defined as
  \begin{align}
    \mel{p^+(p_p,\lambda_p)}{\:\overline{u}\gamma^\mu c\:}{\Lambda_c(p_{\Lambda_c},\lambda_{\Lambda_c})} 
    &= \overline{u}_p(p_p,\lambda_p)  \left[ f_0(q^2)(m_{\Lambda_c} - m_p ) \frac{q^\mu}{q^2}  \right. \\
            &\qquad\quad+ f_+(q^2)\frac{m_{\Lambda_c}+m_p}{s_+}\left( p_{\Lambda_c}^\mu + p_p^\mu - (m_{\Lambda_c}^2-m_p^2) \frac{q^\mu}{q^2}  \right) \nonumber\\
            &\qquad\quad+ \left. f_{\bot}(q^2) \left( \gamma^\mu - \frac{2m_p}{s_+}p_{\Lambda_c}^\mu - \frac{2m_{\Lambda_c}}{s_+}p_p^\mu \right) \right]  u_{\Lambda_c}(p_{\Lambda_c},\lambda_{\Lambda_c}) \:,\nonumber\\
    \mel{p^+(p_p,\lambda_p)}{\:\overline{u}\gamma^\mu\gamma_5 c\:}{\Lambda_c(p_{\Lambda_c},\lambda_{\Lambda_c})} 
    &= - \overline{u}_p(p_p,\lambda_p) \gamma_5 \left[ g_0(q^2)(m_{\Lambda_c} + m_p ) \frac{q^\mu}{q^2}  \right. \\
            &\qquad\quad+ g_+(q^2)\frac{m_{\Lambda_c}-m_p}{s_-}\left( p_{\Lambda_c}^\mu + p_p^\mu - (m_{\Lambda_c}^2-m_p^2) \frac{q^\mu}{q^2}  \right) \nonumber\\
            &\qquad\quad+ \left. g_{\bot}(q^2) \left( \gamma^\mu + \frac{2m_p}{s_-}p_{\Lambda_c}^\mu - \frac{2m_{\Lambda_c}}{s_-}p_p^\mu \right) \right]  u_{\Lambda_c}(p_{\Lambda_c},\lambda_{\Lambda_c}) \:,\nonumber \\
    \mel{p^+(p_p,\lambda_p)}{\overline{u}i\sigma^{\mu\nu}q_\nu c}{\Lambda_c(p_{\Lambda_c},\lambda_{\Lambda_c})}  &= -\overline{u}_p(p_p,\lambda_p) \Bigg[
    h_+(q^2) \frac{q^2}{s_+} \left( p_{\Lambda_c}^\mu + p_p^\mu - \left(m_{\Lambda_c}^2 -m_p^2\right)\frac{q^\mu}{q^2}\right) \nonumber \\
    &+h_\perp(q^2)\left(m_{\Lambda_c} +m_p\right)\bigg(\gamma^\mu-\frac{2m_p}{s_+} p_{\Lambda_c}^\mu 
    - \frac{2m_{\Lambda_c}}{s_+}p_p^\mu\bigg)\Bigg] u_{\Lambda_c}(p_{\Lambda_c},\lambda_{\Lambda_c})\:,\\
    \mel{p^+(p_p,\lambda_p)}{\overline{u}i\sigma^{\mu\nu}q_\nu \gamma_5 c}{\Lambda_c(p_{\Lambda_c},\lambda_{\Lambda_c})}  &= -\overline{u}_p(p_p,\lambda_p) \gamma_5\Bigg[
    \tilde{h}_+(q^2) \frac{q^2}{s_-} \left( p_{\Lambda_c}^\mu + p_p^\mu - \left(m_{\Lambda_c}^2 -m_p^2\right)\frac{q^\mu}{q^2}\right) \nonumber\\
    &+\tilde{h}_\perp(q^2)\left(m_{\Lambda_c} +m_p\right)\bigg(\gamma^\mu-\frac{2m_p}{s_-} p_{\Lambda_c}^\mu 
    - \frac{2m_{\Lambda_c}}{s_-}p_p^\mu\bigg)\Bigg] u_{\Lambda_c}(p_{\Lambda_c},\lambda_{\Lambda_c})\:,
\end{align}
where $s_\pm=(m_{\Lambda_c}\pm m_p)^2 - q^2$. At the zero hadronic recoil endpoint is $q^2$ maximal,
$q^2_{\text{max}} = (m_{\Lambda_c}-m_p)^2$  and $s_-=0$.

The following endpoint relations hold \cite{Golz:2021imq}
\begin{equation}
  \begin{aligned}
    f_0(0) &= f_+(0)\,, \quad \quad &g_\perp(q^2_{\text{max}}) &= g_+(q^2_{\text{max}})\,, \\
    g_0(0) &= g_+(0)\,, \quad \quad &\tilde{h}_\perp(q^2_{\text{max}}) &= \tilde{h}_+(q^2_{\text{max}})\,, \\
    h_\perp(0) &= \tilde{h}_\perp(0)\:. \quad\quad &
  \end{aligned} \label{eq:EPrelations}
\end{equation}
The form factors are available from lattice QCD \cite{Meinel:2017ggx}. 
While  $h_\perp(0) =\tilde{h}_\perp(0)$ was not imposed in the fit of Ref.~\cite{Meinel:2017ggx}, it
is numerically satisfied within uncertainties \cite{Golz:2021imq}.
Additionally, the measurement of the branching fraction of $\Lambda_c\to n e^+\nu_e$ decays by BESIII is in agreement with the form factors  \cite{Meinel:2017ggx} within one standard deviation~\cite{BESIII:2024mgg}.

\section{Helicity amplitudes}
\label{app:helicity_amps}
Following \cite{Gutsche:2013pp,Golz:2022alh} we define the contributions to the angular coefficients of eq.~\eqref{eq:K_i} as 
\begin{equation}
\begin{aligned}
    S^{m m^\prime} &= N^2\,\mathrm{Re}\left[ \mathcal{H}_{\frac{1}{2},0}^{m,0} \mathcal{H}_{\frac{1}{2},0}^{m^\prime,0,\,\ast} + \mathcal{H}_{-\frac{1}{2},0}^{m,0} \mathcal{H}_{-\frac{1}{2},0}^{m^\prime,0,\,\ast}  \right] \:,\\
    S_P^{m m^\prime} &= N^2\,\mathrm{Re}\left[ \mathcal{H}_{\frac{1}{2},0}^{m,0} \mathcal{H}_{\frac{1}{2},0}^{m^\prime,0,\,\ast} - \mathcal{H}_{-\frac{1}{2},0}^{m,0} \mathcal{H}_{-\frac{1}{2},0}^{m^\prime,0,\,\ast}  \right] \:,\\
    U^{m m^\prime} &= N^2\,\mathrm{Re}\left[ \mathcal{H}_{\frac{1}{2},1}^{m,1} \mathcal{H}_{\frac{1}{2},1}^{m^\prime,1,\,\ast} + \mathcal{H}_{-\frac{1}{2},-1}^{m,1} \mathcal{H}_{-\frac{1}{2},-1}^{m^\prime,1,\,\ast}  \right] \:,\\
    P^{m m^\prime} &= N^2\,\mathrm{Re}\left[ \mathcal{H}_{\frac{1}{2},1}^{m,1} \mathcal{H}_{\frac{1}{2},1}^{m^\prime,1,\,\ast} - \mathcal{H}_{-\frac{1}{2},-1}^{m,1} \mathcal{H}_{-\frac{1}{2},-1}^{m^\prime,1,\,\ast}  \right] \:,\\
    L^{m m^\prime} &= N^2\,\mathrm{Re}\left[ \mathcal{H}_{\frac{1}{2},0}^{m,1} \mathcal{H}_{\frac{1}{2},0}^{m^\prime,1,\,\ast} + \mathcal{H}_{-\frac{1}{2},0}^{m,1} \mathcal{H}_{-\frac{1}{2},0}^{m^\prime,1,\,\ast}  \right] \:,\\
    L_P^{m m^\prime} &= N^2\,\mathrm{Re}\left[ \mathcal{H}_{\frac{1}{2},0}^{m,1} \mathcal{H}_{\frac{1}{2},0}^{m^\prime,1,\,\ast} - \mathcal{H}_{-\frac{1}{2},0}^{m,1} \mathcal{H}_{-\frac{1}{2},0}^{m^\prime,1,\,\ast}  \right] \:,\\
    J_{1P}^{m m^\prime} &= N^2\,\mathrm{Re}\left[ \mathcal{H}_{\frac{1}{2},1}^{m,1} \mathcal{H}_{\frac{1}{2},0}^{m^\prime,1,\,\ast} + \mathcal{H}_{-\frac{1}{2},0}^{m,1} \mathcal{H}_{-\frac{1}{2},-1}^{m^\prime,1,\,\ast} + \mathcal{H}_{\frac{1}{2},0}^{m,1} \mathcal{H}_{\frac{1}{2},1}^{m^\prime,1,\,\ast} + \mathcal{H}_{-\frac{1}{2},-1}^{m,1} \mathcal{H}_{-\frac{1}{2},0}^{m^\prime,1,\,\ast}  \right] \:,\\
    J_{2P}^{m m^\prime} &= N^2\,\mathrm{Re}\left[ \mathcal{H}_{\frac{1}{2},0}^{m,1} \mathcal{H}_{\frac{1}{2},1}^{m^\prime,1,\,\ast} - \mathcal{H}_{-\frac{1}{2},-1}^{m,1} \mathcal{H}_{-\frac{1}{2},0}^{m^\prime,1,\,\ast}  \right] \:,\\
    J_{3P}^{m m^\prime} &= N^2\,\mathrm{Im}\left[ \mathcal{H}_{\frac{1}{2},1}^{m,1} \mathcal{H}_{\frac{1}{2},0}^{m^\prime,1,\,\ast} + \mathcal{H}_{-\frac{1}{2},0}^{m,1} \mathcal{H}_{-\frac{1}{2},-1}^{m^\prime,1,\,\ast} - \mathcal{H}_{\frac{1}{2},0}^{m,1} \mathcal{H}_{\frac{1}{2},1}^{m^\prime,1,\,\ast} - \mathcal{H}_{-\frac{1}{2},-1}^{m,1} \mathcal{H}_{-\frac{1}{2},0}^{m^\prime,1,\,\ast}  \right] \:,\\
    J_{4P}^{m m^\prime} &= N^2\,\mathrm{Im}\left[ \mathcal{H}_{\frac{1}{2},0}^{m,1} \mathcal{H}_{\frac{1}{2},1}^{m^\prime,1,\,\ast} - \mathcal{H}_{-\frac{1}{2},-1}^{m,1} \mathcal{H}_{-\frac{1}{2},0}^{m^\prime,1,\,\ast}  \right] \:,\\
\end{aligned} \label{eq:helicity_amplitudes}
\end{equation}
in terms of helicity amplitudes $\mathcal{H}_{\lambda_p,\lambda_\gamma}^{m,J_\gamma}$, where $J_\gamma$ is 
the spin of the dilepton system, which can assume the values $0,+1$, and $\lambda_p$ and $\lambda_\gamma$ are the helicities of 
the proton and the effective current $\gamma^\ast \to \ell^+\ell^-$ respectively. The superscript $m$ distinguishes the contributions 
from leptonic vector currents ($m=1$) including contributions from $\mathcal{K}_7^{(\prime)}$ and $\mathcal{K}_9^{(\prime)}$ 
and axial vector currents ($m=2$) including contributions from $\mathcal{K}_{10}^{(\prime)}$. The helicity amplitudes in 
terms of the Wilson coefficients and the hadronic form factors $f_i^{\Lambda_c\to p}$, defined in appendix~\ref{app:formfactors}, have been calculated in \cite{Golz:2021imq}.
Using the above  helicity amplitudes we obtain for the contributions in eq.~\eqref{eq:helicity_amplitudes}
\begin{equation}
    \begin{aligned}
        J_{1P}^{12} &= 4\sqrt{2}\,N^2\, \bigg(
        \bigg[\mathrm{Re}\left\{ (\mathcal{K}_{7} + \mathcal{K}_{7}^\prime ) (\mathcal{K}_{10} + \mathcal{K}_{10}^\prime)^\ast \right\} m_c \left(f_+ h_\perp \frac{(m_{\Lambda_c}+m_{p})^2}{q^2} +  f_\perp h_+ \right) \\
        &\quad+ \mathrm{Re}\left\{ (\mathcal{K}_{9} + \mathcal{K}_{9}^\prime ) (\mathcal{K}_{10} + \mathcal{K}_{10}^\prime)^\ast \right\} f_\perp f_+ (m_{\Lambda_c}+m_{p})
        \bigg] s_- \\
        &\quad+ \bigg[\mathrm{Re}\left\{ (\mathcal{K}_{7} - \mathcal{K}_{7}^\prime ) (\mathcal{K}_{10} - \mathcal{K}_{10}^\prime)^\ast \right\} m_c \left(g_+ \tilde{h}_\perp \frac{(m_{\Lambda_c}-m_{p})^2}{q^2} +  g_\perp \tilde{h}_+ \right) \\
        &\quad+ \mathrm{Re}\left\{ (\mathcal{K}_{9} - \mathcal{K}_{9}^\prime ) (\mathcal{K}_{10} - \mathcal{K}_{10}^\prime)^\ast \right\} g_\perp g_+ (m_{\Lambda_c}-m_{p})
        \bigg] s_+
    \bigg) \:,\\
        J_{2P}^{11} &= 2\sqrt{2}  N^2\,\bigg(
         \left(\left|\mathcal{K}_{9}\right|^2-\left|\mathcal{K}_{9}^\prime\right|^2 \right) \left( f_\perp g_+ (m_{\Lambda_c}-m_{p}) + f_+ g_\perp (m_{\Lambda_c}+m_{p})\right) \\
      &\quad + \frac{4 m_c^2}{q^2} \left(\left|\mathcal{K}_7\right|^2 - \left|\mathcal{K}_7^\prime\right|^2 \right) \left( h_\perp \tilde{h}_+ (m_{\Lambda_c}+m_{p}) + h_+ \tilde{h}_\perp (m_{\Lambda_c}-m_{p}) \right) \\
      &\quad + 2 m_c \mathrm{Re} \left\{ (\mathcal{K}_{9} - \mathcal{K}_{9}^\prime ) (\mathcal{K}_{7} + \mathcal{K}_{7}^\prime)^\ast \right\} \left( g_+ h_\perp \frac{(m_{\Lambda_c}^2-m_{p}^2)}{q^2}+ g_\perp h_+\right) \\
      &\quad + 2 m_c \mathrm{Re} \left\{ (\mathcal{K}_{9} + \mathcal{K}_{9}^\prime ) (\mathcal{K}_{7} - \mathcal{K}_{7}^\prime)^\ast \right\} \left( f_+ \tilde{h}_\perp \frac{(m_{\Lambda_c}^2-m_{p}^2)}{q^2}+ f_\perp \tilde{h}_+\right) 
    \bigg) \sqrt{s_- s_+} \:,\\
        J_{2P}^{22} &=  2\sqrt{2} N^2\,
         \left(\left|\mathcal{K}_{10}\right|^2-\left|\mathcal{K}_{10}^\prime\right|^2 \right) \big( f_\perp g_+ (m_{\Lambda_c}-m_{p}) + f_+ g_\perp (m_{\Lambda_c}+m_{p})
    \big) \sqrt{s_- s_+} \:,\\
        J_{3P}^{12} &= -4\sqrt{2}\,N^2\, m_c \bigg(
        s_-\,\left[ f_\perp\, h_+\, - f_+ \, h_\perp\,  \frac{\left(m_{\Lambda_c}+m_{p}\right)^2}{q^2}  \right] \mathrm{Im}\left\{ (\mathcal{K}_9+\mathcal{K}_9^\prime)(\mathcal{K}_7^\ast+\mathcal{K}_7^{\prime\,\ast}) \right\}   \\
    &\quad+ s_+\,\left[g_\perp\,\tilde{h}_+ - g_+\, \tilde{h}_\perp\,\frac{\left(m_{\Lambda_c}-m_{p}\right)^2}{q^2} \right]  \mathrm{Im}\left\{ (\mathcal{K}_9-\mathcal{K}_9^\prime)(\mathcal{K}_7^\ast-\mathcal{K}_7^{\prime\,\ast}) \right\}    
        \bigg) \:,\\
        J_{4P}^{11} &= -4\sqrt{2}\,N^2\, m_c \bigg(
        s_-\,\left[ f_\perp\, h_+\, - f_+ \, h_\perp\,  \frac{\left(m_{\Lambda_c}+m_{p}\right)^2}{q^2}  \right] \mathrm{Im}\left\{ (\mathcal{K}_9+\mathcal{K}_9^\prime)(\mathcal{K}_7^\ast+\mathcal{K}_7^{\prime\,\ast}) \right\}   \\
    &\quad+ s_+\,\left[g_\perp\,\tilde{h}_+ - g_+\, \tilde{h}_\perp\,\frac{\left(m_{\Lambda_c}-m_{p}\right)^2}{q^2} \right]  \mathrm{Im}\left\{ (\mathcal{K}_9-\mathcal{K}_9^\prime)(\mathcal{K}_7^\ast-\mathcal{K}_7^{\prime\,\ast}) \right\}    
        \bigg) \:,\\ 
        J_{4P}^{22} &= 0\:,
    \end{aligned}
\end{equation}
with the normalization $N^2 = \frac{G_F^2 \alpha_e^2 \beta_\ell \sqrt{\lambda\left(m_{\Lambda_c}^2,m_{p}^2,q^2\right)} }{3\cdot 2^{11}\pi^5 m_{\Lambda_c}^3}$. 
The $J_{nP}, n=1,2,3,4$ are a new result of this work. Expressions for the other coefficients are given in ref.~\cite{Golz:2021imq,Golz:2022alh}.

\clearpage
\bibliographystyle{jhep}
\bibliography{main} 

@article{LHCb:2023crj,
    author = {Aaij, R. and others},
    collaboration = {LHCb collaboration},
    title = "{$ {\Lambda}_c^{+} $ polarimetry using the dominant hadronic mode}",
    eprint = "2301.07010",
    archivePrefix = "arXiv",
    primaryClass = "hep-ex",
    reportNumber = "LHCb-PAPER-2022-044, CERN-EP-2022-287",
    doi = "10.1007/JHEP07(2023)228",
    journal = "JHEP",
    volume = "07",
    pages = "228",
    year = "2023"
}

@article{Galanti:2015pqa,
    author = "Galanti, Mario and Giammanco, Andrea and Grossman, Yuval and Kats, Yevgeny and Stamou, Emmanuel and Zupan, Jure",
    title = "{Heavy baryons as polarimeters at colliders}",
    eprint = "1505.02771",
    archivePrefix = "arXiv",
    primaryClass = "hep-ph",
    reportNumber = "CP3-15-12",
    doi = "10.1007/JHEP11(2015)067",
    journal = "JHEP",
    volume = "11",
    pages = "067",
    year = "2015"
}

@article{Fajfer:2023nmz,
    author = "Fajfer, Svjetlana and Kamenik, Jernej Fesel and Korajac, Arman and Ko{\v{s}}nik, Nejc",
    title = "{Correlating New Physics Effects in Semileptonic {\ensuremath{\Delta}}C = 1 and {\ensuremath{\Delta}}S = 1 Processes}",
    eprint = "2305.13851",
    archivePrefix = "arXiv",
    primaryClass = "hep-ph",
    doi = "10.1007/JHEP07(2023)029",
    journal = "JHEP",
    volume = "07",
    pages = "029",
    year = "2023"
}

@article{Grunwald:2023nli,
    author = {Grunwald, Cornelius and Hiller, Gudrun and Kr{\"o}ninger, Kevin and Nollen, Lara},
    title = "{More synergies from beauty, top, Z and Drell-Yan measurements in SMEFT}",
    eprint = "2304.12837",
    archivePrefix = "arXiv",
    primaryClass = "hep-ph",
    reportNumber = "DO-TH 23/03",
    doi = "10.1007/JHEP11(2023)110",
    journal = "JHEP",
    volume = "11",
    pages = "110",
    year = "2023"
}

@article{Leskovec:2025gsw,
    author = "Leskovec, Luka and Meinel, Stefan and Petschlies, Marcus and Negele, John and Paul, Srijit and Pochinsky, Andrew",
    title = "{B{\textrightarrow}{\ensuremath{\rho}}{\ensuremath{\ell}}{\ensuremath{\nu}}{\textasciimacron} Resonance Form Factors from B{\textrightarrow}{\ensuremath{\pi}}{\ensuremath{\pi}}{\ensuremath{\ell}}{\ensuremath{\nu}}{\textasciimacron} in Lattice QCD}",
    eprint = "2501.00903",
    archivePrefix = "arXiv",
    primaryClass = "hep-lat",
    doi = "10.1103/PhysRevLett.134.161901",
    journal = "Phys. Rev. Lett.",
    volume = "134",
    number = "16",
    pages = "161901",
    year = "2025"
}

@article{Kats:2015zth,
    author = "Kats, Yevgeny",
    title = "{Measuring c-quark polarization in W+c samples at ATLAS and CMS}",
    eprint = "1512.00438",
    archivePrefix = "arXiv",
    primaryClass = "hep-ph",
    doi = "10.1007/JHEP11(2016)011",
    journal = "JHEP",
    volume = "11",
    pages = "011",
    year = "2016"
}

@article{Achasov:2023gey,
    author = "Achasov, M. and others",
    title = "{STCF conceptual design report (Volume 1): Physics {\&} detector}",
    eprint = "2303.15790",
    archivePrefix = "arXiv",
    primaryClass = "hep-ex",
    doi = "10.1007/s11467-023-1333-z",
    journal = "Front. Phys. (Beijing)",
    volume = "19",
    number = "1",
    pages = "14701",
    year = "2024"
}

@article{BESIII:2009fln,
    author = "Ablikim, M. and others",
    collaboration = "BESIII collaboration",
    title = "{Design and Construction of the BESIII Detector}",
    eprint = "0911.4960",
    archivePrefix = "arXiv",
    primaryClass = "physics.ins-det",
    doi = "10.1016/j.nima.2009.12.050",
    journal = "Nucl. Instrum. Meth. A",
    volume = "614",
    pages = "345--399",
    year = "2010"
}

@article{Briceno:2015tza,
    author = "Brice{\~n}o, Ra{\'u}l A. and Hansen, Maxwell T.",
    title = "{Relativistic, model-independent, multichannel $2\to 2$ transition amplitudes in a finite volume}",
    eprint = "1509.08507",
    archivePrefix = "arXiv",
    primaryClass = "hep-lat",
    reportNumber = "JLAB-THY-15-2140",
    doi = "10.1103/PhysRevD.94.013008",
    journal = "Phys. Rev. D",
    volume = "94",
    number = "1",
    pages = "013008",
    year = "2016"
}

@article{Hiller:2013cza,
    author = "Hiller, Gudrun and Zwicky, Roman",
    title = "{(A)symmetries of weak decays at and near the kinematic endpoint}",
    eprint = "1312.1923",
    archivePrefix = "arXiv",
    primaryClass = "hep-ph",
    reportNumber = "DO-TH-13-23, QFET-2013-08, EDINBURGH--13-22, CP3-ORIGINS-2013-037 -, DIAS-2013-37",
    doi = "10.1007/JHEP03(2014)042",
    journal = "JHEP",
    volume = "03",
    pages = "042",
    year = "2014"
}

@article{Fajfer:1998dv,
    author = "Fajfer, S. and Prelovsek, Sasa and Singer, P.",
    title = "{Long distance contributions in $D\to V\gamma$ decays}",
    eprint = "hep-ph/9801279",
    archivePrefix = "arXiv",
    reportNumber = "IJS-TP-98-01, TECHNION-PH-98-01",
    doi = "10.1007/s100520050356",
    journal = "Eur. Phys. J. C",
    volume = "6",
    pages = "471--476",
    year = "1999"
}

@article{Efrati:2015eaa,
    author = "Efrati, Aielet and Falkowski, Adam and Soreq, Yotam",
    title = "{Electroweak constraints on flavorful effective theories}",
    eprint = "1503.07872",
    archivePrefix = "arXiv",
    primaryClass = "hep-ph",
    reportNumber = "LPT-ORSAY-15-23",
    doi = "10.1007/JHEP07(2015)018",
    journal = "JHEP",
    volume = "07",
    pages = "018",
    year = "2015"
}

@article{Brivio:2019ius,
    author = "Brivio, Ilaria and Bruggisser, Sebastian and Maltoni, Fabio and Moutafis, Rhea and Plehn, Tilman and Vryonidou, Eleni and Westhoff, Susanne and Zhang, C.",
    title = "{O new physics, where art thou? A global search in the top sector}",
    eprint = "1910.03606",
    archivePrefix = "arXiv",
    primaryClass = "hep-ph",
    reportNumber = "P3H-19-036, CERN-TH-2019-193",
    doi = "10.1007/JHEP02(2020)131",
    journal = "JHEP",
    volume = "02",
    pages = "131",
    year = "2020"
}

@article{Ai:2024nmn,
    author = "Ai, Xiaocong and others",
    title = "{Flavor Physics at the CEPC: a General Perspective}",
    eprint = "2412.19743",
    archivePrefix = "arXiv",
    primaryClass = "hep-ex",
    doi = "10.1088/1674-1137/adf1f0",
    journal = "Chin. Phys.",
    volume = "49",
    number = "10",
    pages = "103003",
    year = "2025"
}

@article{DeBoer:2018pdx,
    author = "De Boer, Stefan and Hiller, Gudrun",
    title = "{Null tests from angular distributions in $D \to P_1 P_2 l^+l^-$, $l=e,\mu$ decays on and off peak}",
    eprint = "1805.08516",
    archivePrefix = "arXiv",
    primaryClass = "hep-ph",
    reportNumber = "DO-TH 18/11, QFET-2018-09, TTP18-018, DO-TH-18-11",
    doi = "10.1103/PhysRevD.98.035041",
    journal = "Phys. Rev. D",
    volume = "98",
    number = "3",
    pages = "035041",
    year = "2018"
}

@article{Greub:1996wn,
    author = "Greub, Christoph and Hurth, Tobias and Misiak, Mikolaj and Wyler, Daniel",
    title = "{The $c \to u \gamma$ contribution to weak radiative charm decay}",
    eprint = "hep-ph/9603417",
    archivePrefix = "arXiv",
    reportNumber = "SLAC-PUB-7120, ZU-TH-6-96",
    doi = "10.1016/0370-2693(96)00694-6",
    journal = "Phys. Lett. B",
    volume = "382",
    pages = "415--420",
    year = "1996"
}

@article{Belle:2016qek,
    author = "Lai, Y. -T. and others",
    collaboration = "Belle collaboration",
    title = "{Search for $D^{0}$ decays to invisible final states at Belle}",
    eprint = "1611.09455",
    archivePrefix = "arXiv",
    primaryClass = "hep-ex",
    reportNumber = "BELLE-PREPRINT-2013-13, KEK-PREPRINT-2013-51, BELLE-PREPRINT-2016-13, KEK-PREPRINT-2016-51",
    doi = "10.1103/PhysRevD.95.011102",
    journal = "Phys. Rev. D",
    volume = "95",
    number = "1",
    pages = "011102",
    year = "2017"
}

@article{Falk:1993rf,
    author = "Falk, Adam F. and Peskin, Michael E.",
    title = "{Production, decay, and polarization of excited heavy hadrons}",
    eprint = "hep-ph/9308241",
    archivePrefix = "arXiv",
    reportNumber = "SLAC-PUB-6311, JHU-TIPAC-930019",
    doi = "10.1103/PhysRevD.49.3320",
    journal = "Phys. Rev. D",
    volume = "49",
    pages = "3320--3332",
    year = "1994"
}

@article{Fajfer:2012nr,
    author = "Fajfer, Svjetlana and Ko{\v{s}}nik, Nejc",
    title = "{Resonance catalyzed CP asymmetries in $D\to P\ell^+\ell^-$}",
    eprint = "1208.0759",
    archivePrefix = "arXiv",
    primaryClass = "hep-ph",
    reportNumber = "LAL-12-275",
    doi = "10.1103/PhysRevD.87.054026",
    journal = "Phys. Rev. D",
    volume = "87",
    number = "5",
    pages = "054026",
    year = "2013"
}

@article{LHCb-DP-2008-001,
      author         = "Alves~Jr., A. A. and others",
      title          = "{The LHCb detector at the LHC}",
      collaboration  = "LHCb collaboration",
      journal        = "JINST",
      volume         = "3",
      pages          = "S08005",
      doi            = "10.1088/1748-0221/3/08/S08005",
      year           = "2008",
      report         = "LHCb-DP-2008-001",
}

@article{Belle-II:2010dht,
    author = "Abe, T. and others",
    collaboration = "Belle II collaboration",
    title = "{Belle II Technical Design Report}",
    eprint = "1011.0352",
    archivePrefix = "arXiv",
    primaryClass = "physics.ins-det",
    reportNumber = "KEK-REPORT-2010-1",
    month = "11",
    year = "2010",
}

@article{Kwok:2025fza,
    author = {Kwok, T. H. and Polonsky, Z. and Lukashenko, V. and Aebischer, J. and Kilminster, B.},
    title = "{Time-Dependent Precision Measurement of $B_s^0\rightarrow \phi \mu^+\mu^-$ Decay at FCC-$ee$}",
    eprint = {2506.08089},
    archivePrefix = {arXiv},
    primaryClass = {hep-ph},
    reportNumber = {CERN-EP-DRAFT-MISC-2025-007, CERN-TH-2025-095},
    year = {2025},
}

@misc{hacheney_2025_e3nw3-fx653,
  author = {Hacheney, T. and Röhrig, L. and Mitzel, D.S. and Di Canto, A. and Monteil, S.},
  title = "{Sensitivity study of the search for the rare decay ${D}^0 \to \pi^+\pi^- \nu \bar{\nu}$ at FCC-ee}",
  year = {2025},
  publisher = {CERN},
  doi       = {\href{https://doi.org/10.17181/mqhwz-45j26}{DOI:10.17181/mqhwz-45j26}},
}

@article{LHCb:2022sck,
    author = {Aaij, R. and others},
    collaboration = {LHCb collaboration},
    title = "{Amplitude analysis of the $\Lambda_c^+ \to pK^-\pi^+$ decay and $\Lambda_c^+$ baryon polarization measurement in semileptonic beauty hadron decays}",
    eprint = {2208.03262},
    archivePrefix = {arXiv},
    primaryClass = {hep-ex},
    reportNumber = {LHCb-PAPER-2022-002, CERN-EP-2022-124},
    doi = {10.1103/PhysRevD.108.012023},
    journal = {Phys. Rev. D},
    volume = {108},
    number = {1},
    pages = {012023},
    year = {2023}
}

@article{LHCb:2017hwf,
    author = {Aaij, R. and others},
    collaboration = {LHCb collaboration},
    title = "{A measurement of the CP asymmetry difference in $\Lambda_{c}^{+} \to pK^{-}K^{+}$ and $p\pi^{-}\pi^{+}$ decays}",
    eprint = {1712.07051},
    archivePrefix = {arXiv},
    primaryClass = {hep-ex},
    reportNumber = {CERN-EP-2017-316, LHCB-PAPER-2017-044},
    doi = {10.1007/JHEP03(2018)182},
    journal = {JHEP},
    volume = {03},
    pages = {182},
    year = {2018}
}

@article{LHCb:2018roe,
    author = {Aaij, R. and others},
    collaboration = {LHCb collaboration},
    title = "{Physics case for an LHCb Upgrade II - Opportunities in flavour physics, and beyond, in the HL-LHC era}",
    eprint = {1808.08865},
    archivePrefix = {arXiv},
    primaryClass = {hep-ex},
    reportNumber = {LHCB-PUB-2018-009, CERN-LHCC-2018-027},
    year = {2018}
}

@article{Bordone:2025cde,
    author = "Bordone, Marzia and Cornella, Claudia and Davighi, Joe",
    title = "{Precision Tests in $b\to s\ell^+\ell^-$ ($\ell=e,\mu$) at FCC-ee}",
    eprint = "2503.22635",
    archivePrefix = "arXiv",
    primaryClass = "hep-ph",
    reportNumber = "CERN-TH-2025-065, ZU-TH 22/25",
    doi = "10.1140/epjc/s10052-025-14696-8",
    journal = "Eur. Phys. J. C",
    volume = "85",
    number = "9",
    pages = "995",
    year = "2025"
}

@article{AlvarezCartelle:2025mtx,
    author = {Alvarez Cartelle, P. and Kenzie, M. and Mangrulkar, R. and Wiederhold, A. R. and Wood, E.},
    title = "{Prospects of searches for invisible $B$-meson decays at FCC-ee}",
    eprint = {2508.04471},
    archivePrefix = {arXiv},
    primaryClass = {hep-ex},
    doi = {10.17181/q9yx5-9cx10},
    year = {2025},
}

@article{LHCb:2018qsd,
    author = {Aaij, R. and others},
    collaboration = {LHCb collaboration},
    title = "{Measurement of Angular and CP Asymmetries in $D^0\to\pi^+\pi^-\mu^+\mu^-$ and $D^0\to K^+K^-\mu^+\mu^-$ decays}",
    eprint = {1806.10793},
    archivePrefix = {arXiv},
    primaryClass = {hep-ex},
    reportNumber = {CERN-EP-2018-162, LHCb-PAPER-2018-020, LHCB-PAPER-2018-020},
    doi = {10.1103/PhysRevLett.121.091801},
    journal = {Phys. Rev. Lett.},
    volume = {121},
    number = {9},
    pages = {091801},
    year = {2018}
}

@article{LHCb:2021yxk,
    author = {Aaij, R. and others},
    collaboration = {LHCb collaboration},
    title = "{Angular Analysis of $D^0 \to \pi^+\pi^-\mu^+\mu^-$ and $D^0 \to K^+K^-\mu^+\mu^-$ Decays and Search for $CP$ Violation}",
    eprint = {2111.03327},
    archivePrefix = {arXiv},
    primaryClass = {hep-ex},
    reportNumber = {LHCb-PAPER-2021-035, CERN-EP-2021-212},
    doi = {10.1103/PhysRevLett.128.221801},
    journal = {Phys. Rev. Lett.},
    volume = {128},
    number = {22},
    pages = {221801},
    year = {2022}
}

@article{Asner:2008nq,
    author = {Asner, D. M. and others},
    title = "{Physics at BESIII}",
    eprint = {0809.1869},
    archivePrefix = {arXiv},
    primaryClass = {hep-ex},
    reportNumber = {IHEP-PHYSICS-REPORT-BES-III-2008-001},
    journal = {Int. J. Mod. Phys. A},
    volume = {24},
    pages = {S1--794},
    year = {2009}
}

@article{BESIII:2021slf,
    author = {Ablikim, M. and others},
    collaboration = {BESIII collaboration},
    title = "{Search for the decay $D^{0} \to \pi^{0} \nu \bar{\nu}$}",
    eprint = {2112.14236},
    archivePrefix = {arXiv},
    primaryClass = {hep-ex},
    doi = {10.1103/PhysRevD.105.L071102},
    journal = {Phys. Rev. D},
    volume = {105},
    number = {7},
    pages = {L071102},
    year = {2022}
}

@article{FCC:2025lpp,
    author = {Benedikt, M. and others},
    collaboration = {FCC collaboration},
    title = "{Future Circular Collider Feasibility Study Report: Volume 1, Physics, Experiments, Detectors}",
    eprint = {2505.00272},
    archivePrefix = {arXiv},
    primaryClass = {hep-ex},
    reportNumber = {CERN-FCC-PHYS-2025-0002},
    doi = {10.17181/CERN.9DKX.TDH9},
    year = {2025},
}

@article{FCC:2025jtd,
    author = {Benedikt, M. and others},
    collaboration = {FCC collaboration},
    title = "{Future Circular Collider Feasibility Study Report: Volume 3, Civil Engineering, Implementation and Sustainability}",
    eprint = {2505.00273},
    archivePrefix = {arXiv},
    primaryClass = {physics.acc-ph},
    reportNumber = {CERN-FCC-ACC-2025-0003},
    doi = {10.17181/CERN.I26X.V4VF},
    year = {2025},
}

@article{FCC:2025uan,
    author = {Zimmermann, F. and others},
    editor = {Benedikt, M.},
    collaboration = {FCC collaboration},
    title = "{Future Circular Collider Feasibility Study Report: Volume 2, Accelerators, Technical Infrastructure and Safety}",
    eprint = {2505.00274},
    archivePrefix = {arXiv},
    primaryClass = {physics.acc-ph},
    reportNumber = {CERN-FCC-ACC-2025-0004},
    doi = {10.17181/CERN.EBAY.7W4X},
    year = {2025},
}

@article{CEPCStudyGroup:2018ghi,
    author = {Dong, M. and others},
    editor = {Guimar{\~a}es da Costa, J. B. and others},
    collaboration = {CEPC Study Group},
    title = "{CEPC Conceptual Design Report: Volume 2 - Physics {\&} Detector}",
    eprint = {1811.10545},
    archivePrefix = {arXiv},
    primaryClass = {hep-ex},
    reportNumber = {IHEP-CEPC-DR-2018-02, IHEP-EP-2018-01, IHEP-TH-2018-01},
    year = {2018},
}

@article{CEPCStudyGroup:2018rmc,
    collaboration = {CEPC Study Group},
    title = "{CEPC Conceptual Design Report: Volume 1 - Accelerator}",
    eprint = {1809.00285},
    archivePrefix = {arXiv},
    primaryClass = {physics.acc-ph},
    reportNumber = {IHEP-CEPC-DR-2018-01, IHEP-AC-2018-01},
    year = {2018},
}

@article{Davidson:2010ew,
    author = {Davidson, N. and Przedzinski, T. and Was, Z.},
    title = "{PHOTOS interface in C++: Technical and Physics Documentation}",
    eprint = {1011.0937},
    archivePrefix = {arXiv},
    primaryClass = {hep-ph},
    reportNumber = {CERN-PH-TH-2010-261, IFJPAN-IV-2010-6},
    doi = {10.1016/j.cpc.2015.09.013},
    journal = {Comput. Phys. Commun.},
    volume = {199},
    pages = {86--101},
    year = {2016}
}

@book{Breiman,
  author = {Breiman, L. and Friedman, J. H. and Olshen, R. A. and Stone, C. J.},
  title = "{Classification and regression trees}",
  publisher = {Wadsworth International Group},
  year = {1984},
  address = {Belmont, California, USA}
}

@article{Roe,
   author = {Roe, B. P. and Yang, H.-J. and Zhu, J. and Liu, Y. and Stancu, I. and McGregor, G.},
   title = "{Boosted decision trees as an alternative to artificial neural networks for particle identification}",
   journal = {Nucl. Instrum. Meth.},
   eprint = {physics/0408124},
   archivePrefix = {arXiv},
   primaryClass = {physics},
   year = {2005},
   volume = {A543},
   pages = {577--584},
   doi = {10.1016/j.nima.2004.12.018}
}

@article{FCCee_ESPPU,
  author = {Lusiani, A. and Wilkinson, G. and Kamenik, J. F. and Monteil, S.},
  title = "{Prospects in Flavour Physics at the FCC}",
  note = {Contribution to the European Strategy for Particle Physics Update 2025--2026, https://doi.org/10.17181/jnzpp-1fw39},
  year = {2025}
}

@article{Gisbert:2020vjx,
  author = {Gisbert, H. and Golz, M. and Mitzel, D. S.},
  title = "{Theoretical and experimental status of rare charm decays}",
  eprint = {2011.09478},
  archivePrefix = {arXiv},
  primaryClass = {hep-ph},
  reportNumber = {DO-TH 20/13, CERN-OPEN-2020-012},
  doi = {10.1142/S0217732321300020},
  journal = {Mod. Phys. Lett. A},
  volume = {36},
  number = {04},
  pages = {2130002},
  year = {2021}
}

@misc{b_to_stautau_FCC,
  author = {Miralles, T. and Monteil, S.},
  doi = {10.17181/d772d-egz40},
  publisher = {CERN},
  title = "{Study of the feasibility of the observation of $B^0 \to K^*(892) \tau^+ \tau^-$ at FCC-ee and related vertex detector performance requirements}",
  year = {2024},
}

@article{b_to_taunu_FCC,
  author = {Zuo, X. and Fedele, M. and Helsens, C. and Hill, D. and Iguro, S. and Klute, M.},
  title = "{Prospects for $B_c^+$ and $B^+ \to \tau^+ \nu_\tau$ at FCC-ee}",
  journal = {Eur. Phys. J. C},
  volume = {84},
  number = {1},
  pages = {87},
  year = {2024},
  doi = {10.1140/epjc/s10052-024-12418-0},
  eprint = "2305.02998",
  archivePrefix = {arXiv}
}

@article{b_to_snunu_FCC,
  author = {Amhis, Y. and Kenzie, M. and Reboud, M. and Wiederhold, A. R.},
    title = "{Prospects for searches of $ b\to s\nu \overline{\nu} $ decays at FCC-ee}",
    eprint = "2309.11353",
    archivePrefix = {arXiv},
    primaryClass = "hep-ex",
    reportNumber = "EOS-2023-04, IPPP/23/51",
    doi = "10.1007/JHEP01(2024)144",
    journal = "JHEP",
    volume = "01",
    pages = "144",
    year = "2024"
}

@article{Bause:2020xzj,
  author = {Bause, R. and Gisbert, H. and Golz, M. and Hiller, G.},
  title = "{Rare charm $\boldsymbol{c\to u\,\nu\bar{\nu}}$ dineutrino null tests for $e^+e^-$ machines}",
  eprint = {2010.02225},
  archivePrefix = {arXiv},
  primaryClass = {hep-ph},
  doi = {10.1103/PhysRevD.103.015033},
  journal = {Phys. Rev. D},
  volume = {103},
  number = {1},
  pages = {015033},
  year = {2021}
}

@article{FCC_ee_the_lepton_collider,
  author = {Abada, A. and others},
  doi = {10.1140/epjst/e2019-900045-4},
  journal = {Eur. Phys. J. Spec. Top.},
  number = {2},
  pages = {261--623},
  title = {{FCC-ee: The Lepton Collider}},
  volume = {228},
  year = {2019}
}

@article{Belle2_Physics_Book,
  author = {{Belle II collaboration}},
  doi = {10.1093/ptep/ptz106},
  journal = {Prog. Theor. Exp. Phys.},
  number = {12},
  pages = {123C01},
  title = "{The Belle II Physics Book}",
  volume = {2019},
  eprint = "1808.10567",
  archivePrefix = "arXiv",
  primaryClass = "hep-ex",
  year = {2019},
  note = "[Erratum: PTEP 2020, 029201 (2020)]"
}

@article{IDEA_detector,
  author = {{IDEA Study Group}},
  title = "{The IDEA detector concept for FCC-ee}",
  eprint = {2502.21223},
  archivePrefix = {arXiv},
  primaryClass = {physics.ins-det},
  year = {2025},
}

@article{Delphes_reference,
  author = {{DELPHES 3 collaboration}},
  doi = {10.1007/JHEP02(2014)057},
  journal = {JHEP},
  number = {2},
  pages = {57},
  eprint = "1307.6346",
  archivePrefix = "arXiv",
  primaryClass = "hep-ex",
  title = "{DELPHES 3: a modular framework for fast simulation of a generic collider experiment}",
  volume = {2014},
  year = {2014}
}

@article{Pythia8_reference,
  author = {Sj{\"o}strand, T. and Ask, S. and Christiansen, J. R. and Corke, R. and Desai, N. and Ilten, P. and Mrenna, S. and Prestel, S. and Rasmussen, C. O. and Skands, P. Z.},
  title = "{An introduction to PYTHIA 8.2}",
  journal = {Comput. Phys. Commun.},
  volume = {191},
  pages = {159--177},
  year = {2015},
  doi = {10.1016/j.cpc.2015.01.024},
  url = {https://www.sciencedirect.com/science/article/pii/S0010465515000442}
}

@article{EvtGen_reference,
      author         = "Lange, D. J.",
      title          = "{The EvtGen particle decay simulation package}",
      booktitle      = "{Proceedings, 7th International Conference on \B physics
                        at hadron machines (BEAUTY 2000): Maagan, Israel,
                        September 13-18, 2000}",
      journal        = "Nucl. Instrum. Meth. A",
      volume         = "462",
      year           = "2001",
      pages          = "152",
      doi            = "10.1016/S0168-9002(01)00089-4",
      SLACcitation   = "%%CITATION = NUIMA,A462,152;%%"
}

@article{Lisovyi_2016,
  author = {Lisovyi, M. and Verbytskyi, A. and Zenaiev, O.},
  title = {Combined analysis of charm-quark fragmentation-fraction measurements},
  journal = {Eur. Phys. J. C},
  eprint = "1509.01061",
  archivePrefix = "arXiv",
  primaryClass = "hep-ex",
  volume = {76},
  number = {7},
  pages = {},
  year = {2016},
  doi = {10.1140/epjc/s10052-016-4246-y},
  url = {http://dx.doi.org/10.1140/epjc/s10052-016-4246-y}
}

@article{Bedeschi_2022,
  author = {Bedeschi, F. and Gouskos, L. and Selvaggi, M.},
  title = {Jet flavour tagging for future colliders with fast simulation},
  journal = {Eur. Phys. J. C},
  eprint = "2202.03285",
  archivePrefix = "arXiv",
  primaryClass = "hep-ex",
  volume = {82},
  number = {7},
  year = {2022},
  doi = {10.1140/epjc/s10052-022-10609-1},
  url = {http://dx.doi.org/10.1140/epjc/s10052-022-10609-1}
}

@article{Belle:2010ouj,
    author = "Petric, M. and others",
    collaboration = "Belle",
    title = "{Search for leptonic decays of $D^0$ mesons}",
    eprint = "1003.2345",
    archivePrefix = "arXiv",
    primaryClass = "hep-ex",
    doi = "10.1103/PhysRevD.81.091102",
    journal = "Phys. Rev. D",
    volume = "81",
    pages = "091102",
    year = "2010"
}

@article{PDG2024,
    author = "Navas, S. and others",
    collaboration = "Particle Data Group",
    title = "{Review of particle physics}",
    doi = "10.1103/PhysRevD.110.030001",
    journal = "Phys. Rev. D",
    volume = "110",
    number = "3",
    pages = "030001",
    year = "2024"
}

@article{FREUND1997119,
  title = {A Decision-Theoretic Generalization of On-Line Learning and an Application to Boosting},
  journal = {Journal of Computer and System Sciences},
  volume = {55},
  number = {1},
  pages = {119-139},
  year = {1997},
  issn = {0022-0000},
  doi = {https://doi.org/10.1006/jcss.1997.1504},
  url = {https://www.sciencedirect.com/science/article/pii/S002200009791504X},
  author = {Freund, Y. and Schapire, R. E.}
}

@misc{Belle2DecayFile,
  collaboration = {Belle II collaboration},
  howpublished = {\url{https://github.com/belle2/basf2/blob/main/decfiles/dec/DECAY_BELLE2.DEC}},
  note         = {Accessed: 2025-02-18}
}

@inproceedings{Punzi:2003bu,
    author    = "Punzi, G.",
    title     = "{Sensitivity of searches for new signals and its optimization}",
    booktitle = "{Statistical Problems in Particle Physics, Astrophysics, and Cosmology (PHYSTAT 2003)}",
    year      = 2003,
    eprint    = "physics/0308063",
    archivePrefix = "arXiv",
    primaryClass  = "physics.data-an",
    SLACcitation  = "%%CITATION = PHYSICS/0308063;%%"
}

@article{Golz:2021imq,
    author = "Golz, M. and Hiller, G. and Magorsch, T.",
    title = "{Probing for new physics with rare charm baryon ($\Lambda_c, \Xi_c \Omega_c$) decays}",
    eprint = "2107.13010",
    archivePrefix = "arXiv",
    primaryClass = "hep-ph",
    reportNumber = "DO-TH 21/13",
    doi = "10.1007/JHEP09(2021)208",
    journal = "JHEP",
    volume = "09",
    pages = "208",
    year = "2021"
}

@article{Bause:2020auq,
    author = "Bause, R. and Gisbert, H. and Golz, M. and Hiller, G.",
    title = "{Lepton universality and lepton flavor conservation tests with dineutrino modes}",
    eprint = "2007.05001",
    archivePrefix = "arXiv",
    primaryClass = "hep-ph",
    reportNumber = "DO-TH 20/07",
    doi = "10.1140/epjc/s10052-022-10113-6",
    journal = "Eur. Phys. J. C",
    volume = "82",
    number = "2",
    pages = "164",
    year = "2022"
}

@article{Gutsche:2013pp,
    author = "Gutsche, T. and Ivanov, M. A. and Korner, J. G. and Lyubovitskij, V. E. and Santorelli, P.",
    title = "{Rare baryon decays $\Lambda_b \to \Lambda {l^{+}l^{-}} (l=e, \mu, \tau)$ and $\Lambda_b \to \Lambda\gamma$: differential and total rates, lepton- and hadron-side forward-backward asymmetries}",
    eprint = "1301.3737",
    archivePrefix = "arXiv",
    primaryClass = "hep-ph",
    reportNumber = "DSF-2012-6 (NAPOLI), MZ-TH-12-50 (MAINZ), DSF-2012-6-(NAPOLI), MZ-TH-12-50-(MAINZ)",
    doi = "10.1103/PhysRevD.87.074031",
    journal = "Phys. Rev. D",
    volume = "87",
    pages = "074031",
    year = "2013"
}

@article{Lin:2025cmn,
    author = "Lin, W. and Huang, X.-E. and Cheng, S. and Yao, D.-L.",
    title = "{Semileptonic Decays of $D \to \rho l^+ \nu$ and $D_{(s)} \to K^\ast l^+ \nu$ from Light-Cone Sum Rules}",
    eprint = "2505.01329",
    archivePrefix = "arXiv",
    primaryClass = "hep-ph",
    year = "2025"
}

@article{BESIII:2018qmf,
    author = "Ablikim, M. and others",
    collaboration = "BESIII collaboration",
    title = "{Observation of $D^+ \to f_0(500) e^+\nu_e$ and Improved Measurements of $D \to\rho e^+\nu_e$}",
    eprint = "1809.06496",
    archivePrefix = "arXiv",
    primaryClass = "hep-ex",
    doi = "10.1103/PhysRevLett.122.062001",
    journal = "Phys. Rev. Lett.",
    volume = "122",
    number = "6",
    pages = "062001",
    year = "2019"
}

@article{Golz:2022alh,
    author = "Golz, M. and Hiller, G. and Magorsch, T.",
    title = "{Pinning down $|\Delta c|=|\Delta u|=1$ couplings with rare charm baryon decays}",
    eprint = "2202.02331",
    archivePrefix = "arXiv",
    primaryClass = "hep-ph",
    reportNumber = "DO-TH 21/32",
    doi = "10.1140/epjc/s10052-022-10302-3",
    journal = "Eur. Phys. J. C",
    volume = "82",
    number = "4",
    pages = "357",
    year = "2022"
}

@article{Blake:2017une,
    author = "Blake, T. and Kreps, M.",
    title = "{Angular distribution of polarised $\Lambda_b$ baryons decaying to $\Lambda \ell^+\ell^-$}",
    eprint = "1710.00746",
    archivePrefix = "arXiv",
    primaryClass = "hep-ph",
    reportNumber = "EOS-2017-02",
    doi = "10.1007/JHEP11(2017)138",
    journal = "JHEP",
    volume = "11",
    pages = "138",
    year = "2017"
}

@article{Meinel:2017ggx,
    author = "Meinel, S.",
    title = "{$\Lambda_c \to N$ form factors from lattice QCD and phenomenology of $\Lambda_c \to n \ell^+ \nu_\ell$ and $\Lambda_c \to p \mu^+ \mu^-$ decays}",
    eprint = "1712.05783",
    archivePrefix = "arXiv",
    primaryClass = "hep-lat",
    reportNumber = "RBRC-1262, RBRC-1262",
    doi = "10.1103/PhysRevD.97.034511",
    journal = "Phys. Rev. D",
    volume = "97",
    number = "3",
    pages = "034511",
    year = "2018"
}

@article{Adolph:2020ema,
    author = "Adolph, N. and Brod, J. and Hiller, G.",
    title = "{Radiative three-body $D$-meson decays in and beyond the standard model}",
    eprint = "2009.14212",
    archivePrefix = "arXiv",
    primaryClass = "hep-ph",
    reportNumber = "DO-TH 20/11",
    doi = "10.1140/epjc/s10052-021-08832-3",
    journal = "Eur. Phys. J. C",
    volume = "81",
    number = "1",
    pages = "45",
    year = "2021"
}

@article{Lee:1992ih,
    author = "Lee, Clarence L. Y. and Lu, Ming and Wise, Mark B.",
    title = "{$B_{l4}$ and $D_{l4}$ decay}",
    reportNumber = "CALT-68-1771",
    doi = "10.1103/PhysRevD.46.5040",
    journal = "Phys. Rev. D",
    volume = "46",
    pages = "5040--5048",
    year = "1992"
}

@article{Fajfer:2023tkp,
    author = "Fajfer, S. and Solomonidi, E. and Vale Silva, L.",
    title = "{S-wave contribution to rare $D^0\to\pi^+\pi^-\ell^+\ell^-$ decays in the standard model and sensitivity to new physics}",
    eprint = "2312.07501",
    archivePrefix = "arXiv",
    primaryClass = "hep-ph",
    doi = "10.1103/PhysRevD.109.036027",
    journal = "Phys. Rev. D",
    volume = "109",
    pages = "3",
    year = "2024"
}

@article{BaBar:2013thi,
    author = "Lees, J. P. and others",
    collaboration = "BaBar collaboration",
    title = "{Measurement of the $D^{*+}(2010)$ meson width and the $D^{*+}(2010) - D^0$ mass difference}",
    eprint = "1304.5657",
    archivePrefix = "arXiv",
    primaryClass = "hep-ex",
    reportNumber = "BABAR-PUB-13-004, SLAC-PUB-15424",
    doi = "10.1103/PhysRevLett.111.111801",
    journal = "Phys. Rev. Lett.",
    volume = "111",
    number = "11",
    pages = "111801",
    year = "2013"
}

@article{BaBar:2010vmf,
    author = "del Amo Sanchez, P. and others",
    collaboration = "BaBar collaboration",
    title = "{Analysis of the $D^+ \to K^- \pi^+ e^+ \nu_e$ decay channel}",
    eprint = "1012.1810",
    archivePrefix = "arXiv",
    primaryClass = "hep-ex",
    reportNumber = "BABAR-PUB-10-021, SLAC-PUB-14329",
    doi = "10.1103/PhysRevD.83.072001",
    journal = "Phys. Rev. D",
    volume = "83",
    pages = "072001",
    year = "2011"
}

@article{Das:2014sra,
    author = "Das, D. and Hiller, G. and Jung, M. and Shires, A.",
    title = "{The $ \overline{B}\to \overline{K}\pi \ell \ell $ and $ {\overline{B}}_s\ \to \overline{K}K\ell \ell $ distributions at low hadronic recoil}",
    eprint = "1406.6681",
    archivePrefix = "arXiv",
    primaryClass = "hep-ph",
    reportNumber = "DO-TH-14-10, QFET-2014-09",
    doi = "10.1007/JHEP09(2014)109",
    journal = "JHEP",
    volume = "09",
    pages = "109",
    year = "2014"
}

@article{Melikhov:2000yu,
    author = "Melikhov, D. and Stech, B.",
    title = "{Weak form-factors for heavy meson decays: An Update}",
    eprint = "hep-ph/0001113",
    archivePrefix = "arXiv",
    reportNumber = "HD-THEP-00-01",
    doi = "10.1103/PhysRevD.62.014006",
    journal = "Phys. Rev. D",
    volume = "62",
    pages = "014006",
    year = "2000"
}

@article{Gisbert:2024kob,
    author = "Gisbert, H. and Hiller, G. and Suelmann, D.",
    title = "{Effective field theory analysis of rare $\Delta c = \Delta u = 1$ charm decays}",
    eprint = "2410.00115",
    archivePrefix = "arXiv",
    primaryClass = "hep-ph",
    doi = "10.1007/JHEP12(2024)102",
    journal = "JHEP",
    volume = "12",
    pages = "102",
    year = "2024"
}

@article{deBoer:2017que,
    author = "de Boer, S. and Hiller, G.",
    title = "{Rare radiative charm decays within the standard model and beyond}",
    eprint = "1701.06392",
    archivePrefix = "arXiv",
    primaryClass = "hep-ph",
    reportNumber = "DO-TH-16-20, DO-TH 16/20, QFET-2016-18",
    doi = "10.1007/JHEP08(2017)091",
    journal = "JHEP",
    volume = "08",
    pages = "091",
    year = "2017"
}

@article{Grzadkowski:2010es,
    author = "Grzadkowski, B. and Iskrzynski, M. and Misiak, M. and Rosiek, J.",
    title = "{Dimension-Six Terms in the Standard Model Lagrangian}",
    eprint = "1008.4884",
    archivePrefix = "arXiv",
    primaryClass = "hep-ph",
    reportNumber = "IFT-9-2010, TTP10-35",
    doi = "10.1007/JHEP10(2010)085",
    journal = "JHEP",
    volume = "10",
    pages = "085",
    year = "2010"
}

@article{deBoer:2015boa,
    author = "de Boer, S. and Hiller, G.",
    title = "{Flavor and new physics opportunities with rare charm decays into leptons}",
    eprint = "1510.00311",
    archivePrefix = "arXiv",
    primaryClass = "hep-ph",
    reportNumber = "DO-TH-15-10, QFET-2015-25",
    doi = "10.1103/PhysRevD.93.074001",
    journal = "Phys. Rev. D",
    volume = "93",
    number = "7",
    pages = "074001",
    year = "2016"
}

@article{BaBar:2011ouc,
    author = "Lees, J. P. and others",
    collaboration = "BaBar collaboration",
    title = "{Searches for Rare or Forbidden Semileptonic Charm Decays}",
    eprint = "1107.4465",
    archivePrefix = "arXiv",
    primaryClass = "hep-ex",
    reportNumber = "BABAR-PUB-11-008, SLAC-PUB-14482",
    doi = "10.1103/PhysRevD.84.072006",
    journal = "Phys. Rev. D",
    volume = "84",
    pages = "072006",
    year = "2011"
}

@article{LHCb:2025bfy,
    author = "Aaij, R. and others",
    collaboration = "LHCb collaboration",
    title = "{Search for resonance-enhanced CP and angular asymmetries in the $\Lambda_c^+ \to p\mu^+\mu^-$ decay at LHCb}",
    eprint = "2502.04013",
    archivePrefix = "arXiv",
    primaryClass = "hep-ex",
    reportNumber = "LHCb-PAPER-2024-051, CERN-EP-2024-340",
    doi = "10.1103/PhysRevD.111.L091102",
    journal = "Phys. Rev. D",
    volume = "111",
    number = "9",
    pages = "L091102",
    year = "2025"
}

@article{Feldmann:2011xf,
    author = "Feldmann, T. and Yip, M. W. Y.",
    title = "{Form factors for $\Lambda_b \to \Lambda$ transitions in the  soft-collinear effective theory}",
    eprint = "1111.1844",
    archivePrefix = "arXiv",
    primaryClass = "hep-ph",
    reportNumber = "IPPP-11-70, DURHAM-IPPP-11-70, DCPT-11-140",
    doi = "10.1103/PhysRevD.85.014035",
    journal = "Phys. Rev. D",
    volume = "85",
    pages = "014035",
    year = "2012",
    note = "[Erratum: Phys.Rev.D 86, 079901 (2012)]"
}

@article{BESIII:2024mgg,
    author = "Ablikim, M. and others",
    collaboration = "BESIII collaboration",
    title = "{Observation of a rare beta decay of the charmed baryon with a Graph Neural Network}",
    eprint = "2410.13515",
    archivePrefix = "arXiv",
    primaryClass = "hep-ex",
    doi = "10.1038/s41467-024-55042-y",
    journal = "Nature Commun.",
    volume = "16",
    number = "1",
    pages = "681",
    year = "2025"
}

@article{Angelescu:2020uug,
    author = "Angelescu, A. and Faroughy, D. A. and Sumensari, O.",
    title = "{Lepton Flavor Violation and Dilepton Tails at the LHC}",
    eprint = "2002.05684",
    archivePrefix = "arXiv",
    primaryClass = "hep-ph",
    reportNumber = "ZU-TH 02/20",
    doi = "10.1140/epjc/s10052-020-8210-5",
    journal = "Eur. Phys. J. C",
    volume = "80",
    number = "7",
    pages = "641",
    year = "2020"
}

@article{BESIII:2015tql,
    author = "Ablikim, M. and others",
    collaboration = "BESIII collaboration",
    title = "{Study of Dynamics of $D^0 \to K^- e^+ \nu_{e}$ and $D^0\to\pi^- e^+ \nu_{e}$ Decays}",
    eprint = "1508.07560",
    archivePrefix = "arXiv",
    primaryClass = "hep-ex",
    doi = "10.1103/PhysRevD.92.072012",
    journal = "Phys. Rev. D",
    volume = "92",
    number = "7",
    pages = "072012",
    year = "2015"
}

@article{FermilabLattice:2022gku,
    author = "Bazavov, Alexei and others",
    collaboration = "Fermilab Lattice, MILC",
    title = "{D-meson semileptonic decays to pseudoscalars from four-flavor lattice QCD}",
    eprint = "2212.12648",
    archivePrefix = "arXiv",
    primaryClass = "hep-lat",
    reportNumber = "MIT-CTP/5513, FERMILAB-PUB-22-943-T",
    doi = "10.1103/PhysRevD.107.094516",
    journal = "Phys. Rev. D",
    volume = "107",
    number = "9",
    pages = "094516",
    year = "2023"
}

@article{Fuentes-Martin:2020lea,
    author = "Fuentes-Martin, Javier and Greljo, Admir and Martin Camalich, Jorge and Ruiz-Alvarez, Jos{\'e} David",
    title = "{Charm physics confronts high-p$_{T}$ lepton tails}",
    eprint = "2003.12421",
    archivePrefix = "arXiv",
    primaryClass = "hep-ph",
    reportNumber = "CERN-TH-2020-047, ZU-TH 07/20",
    doi = "10.1007/JHEP11(2020)080",
    journal = "JHEP",
    volume = "11",
    pages = "080",
    year = "2020"
}

@article{BESIII:2017ylw,
    author = "Ablikim, M. and others",
    collaboration = "BESIII collaboration",
    title = "{Analysis of $D^+\to\bar K^0e^+\nu_e$ and $D^+\to\pi^0e^+\nu_e$ semileptonic decays}",
    eprint = "1703.09084",
    archivePrefix = "arXiv",
    primaryClass = "hep-ex",
    doi = "10.1103/PhysRevD.96.012002",
    journal = "Phys. Rev. D",
    volume = "96",
    number = "1",
    pages = "012002",
    year = "2017"
}
\addcontentsline{toc}{section}{References}

\end{document}